\documentclass[10pt,letterpaper]{article}
\usepackage[utf8]{inputenc}
\usepackage{amsmath}
\usepackage{amssymb}
\usepackage{graphicx}
\usepackage{epstopdf}
\usepackage{hyperref}
\usepackage{subcaption}
\usepackage[percent]{overpic}
\usepackage{pdfpages}
\usepackage[normalem]{ulem}
\usepackage[left]{lineno}
\usepackage[top=0.85in,left=2.75in,footskip=0.75in]{geometry}
\usepackage[object=vectorian]{pgfornament} 
\usepackage{vwcol}

\raggedright
\usepackage[aboveskip=1pt,labelfont=bf,labelsep=period,justification=raggedright,singlelinecheck=off]{caption}

\makeatletter
\renewcommand{\@biblabel}[1]{\quad#1.}
\makeatother

\usepackage{lastpage,fancyhdr,graphicx}
\usepackage{epstopdf}
\fancyheadoffset[L]{2.25in}
\fancyfootoffset[L]{2.25in}
\begin{document}
\vspace*{0.2in}

\begin{flushleft}
{\Large
\textbf\newline{A political spectrograph: High-resolution examinations of the United States' ideological landscape} 
}
\newline
\\
David Sabin-Miller\textsuperscript{1*},
Mary C. McGrath\textsuperscript{2},
Marisa C. Eisenberg\textsuperscript{1,3}
\\
\bigskip
\textbf{1} Center for the Study of Complex Systems, University of Michigan, Ann Arbor, MI, USA
\\
\textbf{2} Department of Political Science, Northwestern University, Evanston, IL, USA
\\
\textbf{3} Department of Epidemiology, School of Public Health, University of Michigan, Ann Arbor, MI, USA
\\
\bigskip
* Corresponding Author\\
E-mail: dasami@umich.edu

\end{flushleft}

\begin{abstract}
    The concept of ideology is central to political discourse and dynamics, and is often cast as falling primarily on a one-dimensional scale from ``left-wing/liberal" to ``right-wing/conservative", but the validity of this simple quantitative treatment is uncertain. Here we investigate and compare various high-resolution measures of ideology, both internal (self-identification and policy-stance agreements) and external (estimating the ideological position of political opinion statements). We find strong consistency between internal measures, although policy-stance agreement ideology yields a systematically centralizing and liberalizing portrait relative to more abstract ``liberal/conservative" measures. More remarkably, we find that external assessments of ideology, while noisy, are largely consistent across observers, even for highly dissonant ideas and regardless of speaker identity markers. This supports the use of these responses as meaningful, comparable quantities, which general members of the public reliably project from the abstract space of political thought onto a shared one-dimensional domain. 
\end{abstract}


\section{Introduction}

What does political ideology look like in the public mind? Despite the ostensible importance of ideology for structuring political conflict and political organization in the abstract, there is little clear agreement about what constitutes ideology in the context of American politics---about what political ideology means, what it means to the public, and whether it even exists outside a narrow segment of the population.

For decades the dominant view within political science was that ideology does not exist in the public mind—that, with the exception of political elites and some politically sophisticated members of the general public, American voters do not hold anything resembling a political ideology (e.g., \cite{converse2006nature, kinder2017neither}).

Within the general population, party identity and liberal/conservative identification are more aligned than at any time in recent history (\cite{pew2017divide, dimock2014political,levendusky2009partisan, mason2018uncivil}). Debate remains over how to interpret this alignment in the public---whether a public devoid of ideological thinking is simply mirroring cues from ideologically-sorted partisan elites, or if the increasing and increasingly party-aligned ideological self-identification is based in something more reflective of underlying political preferences (e.g., \cite{Simas2023}).

Conceptions of how representation functions and what democracy consists of are tightly linked to this question of ideology among the public. A democracy could operate with a purely non-ideological public. But if underlying ideological principles do exist among the American public, a presumption that the public is “innocent of ideology” \cite{converse2006nature} could deter efforts to uncover those principles and diminish the quality of representation.

As Kalmoe \cite{kalmoe2020uses} notes, scholarship on ideology in the American public has tended to arrive at one of two apparently contradictory conclusions. In contrast to the ideological innocence view of public opinion “signifying nothing ideological for most,” (\cite{kalmoe2020uses}, p. 789), the maximal-dispositional paradigm holds that liberal/conservative ways of viewing the world are deeply ingrained and pervasive. Jost and colleagues write that “Left-right orientations permeate people’s public and private lives” (\cite{jost2023psychology} p. 57). Hibbing et al. \cite{hibbing2014differences} go even further, positing that a liberal/conservative outlook is not only “universal” but “probably in our DNA” (p. 298).

The observations about ideology among the American public from these different schools of thought are not incompatible. The predictive power of ideological self-identification that the maximal-dispositional view tends to rely on does not in itself show that a liberal-conservative framework plays an active role in shaping political attitudes. And non-conformity of political attitudes to the liberal-conservative framework does not in itself indicate that the public is non-ideological. 

A persistent minority of scholars have raised and reiterated the point that a one-dimensional liberal-conservative framework is by no means the only form of ideological thinking possible (\cite{Lane1973, Morgan2017, chong1993people, carmines2015new}). Indeed, a number of apparently conflicting observations about ideology among the American public are accommodated when “ideological” is not treated as synonymous with liberal/conservative, and when allowing for different uses of the liberal-conservative framework by political elites, political sophisticates, and other members of the public. 

While a small portion of the population—the highly-attuned `political sophisticates' identified by Kalmoe and others—may use the liberal-conservative framework to arrive at their views, this does not imply that the rest of the population is non-ideological. Rather, the degree of non-identification, instability, and inconsistency in the public’s use of the liberal-conservative framework could arise if this framework does not reflect the principles by which members of the public form their political views, but rather a foreign language that can be learned and employed for communication. With the aid of elite partisan-ideological sorting, this language has become more intelligible over the past several decades, leading to increased self-identification on the liberal-conservative spectrum (\cite{abramowitz2008polarization, halliez2021examining}), increased strength of liberal/conservative identification in predicting political attitudes (\cite{kozlowski2021issue}); and greater temporal stability now than in the past (\cite{gries2017does}). 

We present evidence that, far from exhibiting ideological innocence, voters exhibit a nuanced, consistent, and shared understanding of the liberal-conservative ideological spectrum--but that public opinion also deviates from the liberal-conservative framework in systematic ways. We argue that this combination of findings suggest that members of the public make use of the liberal-conservative spectrum as a \emph{meaningful signal}. Rather than a generative framework they employ to form their political views, the public can learn to translate their political views into the language of this spectrum constructed by political elites---and, aided in deciphering this language by elite partisan-ideological sorting, they have increasingly done so.

In an initial study and replication, we find that—although public opinion departs in important ways from the liberal-conservative framework, supporting the idea that a considerable portion of the public does not consist of liberal-conservative ideologues—rather than ideological innocence, people show a remarkably consistent, shared, and relatively internalized understanding of this spectrum. Though it is outside the scope of this article to investigate, the consistency and accuracy with which people are able to communicate in this foreign/elite language suggests there may well be a set of underlying principles, shared within subsets of the population, that people refer to when forming political opinions.

\newgeometry{margin = 0.85in}
\fancyfootoffset[L]{0.5in}
\fancyfootoffset[R]{0.9in}
\textwidth 6.25in 
\textheight 9in

The consistency of this one-dimensional projection is a factor of considerable importance for mathematical modeling of ideological dynamics (often under the broader categories of ``opinion dynamics" or ``sociophysics," e.g.~\cite{galam2008sociophysics, porter2016dynamical, sirbu2017opinion, sabin2020pull, kan2023adaptive}). These models have generally utilized abstract, binary or one-dimensional ``opinion" variables to explore the implications of theoretical influence environments, without necessarily connecting to real-world data due to the practical difficulties of observing and interpreting relevant (often internal/individually-interpreted) quantities. 

In this study, we present a multi-faceted evaluation of individuals' sense of their own ideology and their ideological perception of political statements. We hope this work may serve as a bridge for these interested parties to understand the degree of consistency and variability inherent in taking this crucial step of contention with empirical data, and that with such grounding, future efforts may beget theory-experiment feedback loops aimed at elucidating these fascinating and powerful societal dynamics.

\section{Background}
Converse's seminal 1964 study of ideology in the American public began with a broad definition of ideology as “a configuration of ideas and attitudes in which the elements are bound together by some form of constraint or functional interdependence” (\cite{converse2006nature}, p. 3). However, finding that ``the liberal-conservative continuum… was almost the only dimension of the sort that occurred empirically” in their interviews with members of the public, Converse narrowed the primary focus of the study to whether political views among the public reflected this continuum. Seeing little evidence that they did, he concluded that most of the American public was “remarkably innocent” with regard to “the familiar belief systems that, in view of their historical importance, tend most to attract the sophisticated observer” (\cite{converse2006nature}, p. 66).

Recent scholarship continues Converse's `ideological innocence' debate. Kinder and Kalmoe \cite{kinder2017neither} revisit and update Converse’s project, finding this thesis to hold for the modern (2017) American public. Kalmoe \cite{kalmoe2020uses} reiterates that only the most politically sophisticated citizens hold views that conform to the liberal-conservative ideological framework. In contrast, John Jost, Jonathan Haidt, and others working from a psychologically oriented perspective, see pervasive influence of the liberal-conservative framework on people’s thinking as well as indications that liberal-conservative outlooks are connected to dispositional traits and psychological needs (e.g., \cite{graham2009liberals}). Jost et al. \cite{jost2023psychology} hold that “most people do have political preferences (including beliefs, opinions, and values) that can be understood fruitfully in left-right terms, whether they realize it or not.” 

Another line of research emphasizes the symbolic nature of liberal-conservative identification among the public. Conover \& Feldman \cite{conover1981origins}, noting that research on ideology within the public “has tended to ignore—or perhaps take for granted—the meaning of liberal/conservative self-identifications,'' argue that “ideological identifications constitute more a symbolic than issue-oriented link to the political world.” In a similar vein, Ellis and Stimson \cite{ellis2009symbolic} contend that an identity-based symbolic ideological self-placement should be considered as distinct from an operational, issue-based ideology—noting that when separating these two measures, the American public consistently exhibits issue-based/operational ideological preferences that are more liberal than their general ideological self-placement. 

Other authors have pushed against the ideological innocence thesis on its own terms. Gries \cite{gries2017does} argues that problems with traditional measures have obscured that liberal-conservative ideology among the public is ``not only temporally stable and internally reliable, but also powerfully structures
sociopolitical attitudes" (p. 133). Simas \cite{Simas2023} shows that newer data and improved measurement reveal operational ideology to play an important role in political views for a larger segment of the American population than previous literature has granted.  

Our argument that the mass public use the liberal-conservative spectrum as a meaningful signal rather than a generative ideology is consistent with a number of observations about American public opinion. It accommodates the widespread evidence that, in a variety of respects, a single liberal-conservative spectrum does not adequately describe political views in the American public (see, e.g., \cite{carmines2012political, klar2014multidimensional, feldman2014understanding} among others on dimensionality; \cite{treier2009nature} on ordinality; \cite{grossmann2016asymmetric} on asymmetry) as well as the observation that voters exhibit a nuanced understanding of what the spectrum represents \cite{goggin2020goes, Hare2015}; that deviations from this spectrum are not due to a lack of awareness about ‘what goes with what’ \cite{Groenendyk2023}; and that preferences that appear associated with liberal-conservative identification are in fact driven by issue preferences \cite{ahler2018delegate}. Willingness and ability to locate oneself on the liberal-conservative ideological spectrum has increased as elite partisan-ideological sorting has made the language of the liberal-conservative spectrum more intelligible \cite{lupton2020values, halliez2021examining}. 

This is not to suggest that people very often employ whatever ideology---the “shared and systematic beliefs about how the world does and should work” \cite{gries2017does}---they might have. With elite partisan-ideological sorting, the language of liberal-conservative ideology alongside the binary of partisanship make it very easy to achieve a good-enough guess when it comes to the coarse political choices—e.g., selecting between two candidates—that comprise the entirety of political decision-making for the vast majority of the public. We also do not suggest that most people know what principles define the liberal/conservative framework, but rather that, with increased sorting and exposure, the public has gained greater clarity on “what goes with what” within that framework, even if they do not know or care  what the throughline is. This ignorance does not mean they are non-ideological, but that they are not liberal/conservative in the strictest sense: they make no use of these principles to inform their views. To them, the liberal-conservative spectrum is merely a signaling language they need to learn if they want to convey their views to elites (whether through voting or other means) in the manner afforded to them by nature of democratic politics.

\subsection{Terminology and Survey Overview}

In this section, we provide a brief overview of the terminology used in this study, as well as a basic overview of the survey we conducted. For detailed study methods including exact wordings, please see Section~\ref{Sec:Methods}. 

\begin{itemize}
    \item \textbf{General ideological self-placement (General ideology)}: $[-50,50]$ ideological self-placement. 
    \item \textbf{Issue-by-issue ideology}: $[-50,50]$ ideological self-placement on each of thirteen politically relevant issues.
    \item \textbf{Average issue-by-issue ideology}: average of issue-by-issue ideology values for one individual.
    \item \textbf{Policy-stance agreement ideology}: $[-50,50]$ agreement with each of 10 major policy positions/attitudes, where agreement for liberal positions is sign-flipped to align with other ideology measures.
    \item \textbf{Average policy-stance agreement ideology}: average of policy-agreement ideology values for one individual.
\end{itemize}

For interpretation of the magnitude of these numbers: the ideology sliders had descriptive labels at scores of approximately 
\begin{itemize}
    \item $\pm 48$ Extremely Conservative/ Extremely Liberal
    \item $\pm 32$ Very Conservative/Very Liberal
    \item $\pm 16$ Somewhat Conservative/Somewhat Liberal
    \item $0$ Unsure/Centrist
\end{itemize}

\noindent On agreement-type sliders, those same numbers corresponded to
\begin{itemize}
    \item $\pm 48$ Emphatically Agree/Vehemently Disagree
    \item $\pm 32$ Strongly Agree/Strongly Disagree
    \item $\pm 16$ Somewhat Agree/Somewhat Disagree
    \item $0$ Neither Agree Nor Disagree
\end{itemize}
The non-standard, emotionally charged language on the slider endpoints was part of an intentional attempt to expand the resolved response space by reserving the endpoints for more extreme views and emotions (see Methods for more details).

A pilot study was conducted with participants (N=296) recruited from Mechanical Turk (N=166), a large midwestern university (N = 90), and volunteers (N = 40). The primary study (N=508) was conducted with participants recruited through Prolific, and collected to provide a sample representative of the U.S.~population on age, gender, race$/$ethnicity, and party identification. Figures in the main text are based on this second, nationally-representative sample, with corresponding figures for the initial study in Appendix \ref{sec:mv_replication}. 

\section{Results}

\subsection{Internal Ideology Assessments} \label{sec:internal_ideos}
\setcounter{subsection}{1} 

\begin{figure*} [h!]
    \centering
    \begin{overpic}
        [width=1.1\textwidth]{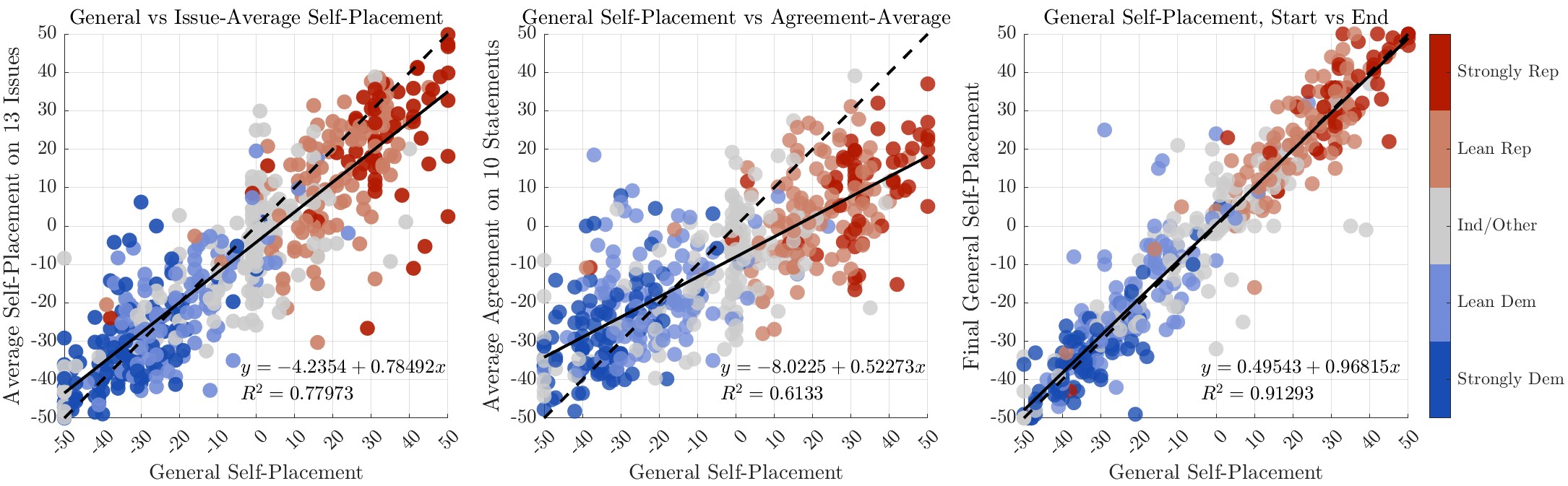}
        \put(4.5,27){\large{\textbf{a}}}
        \put(35,26.8){\large{\textbf{b}}}
        \put(65.7,27){\large{\textbf{c}}}
    \end{overpic}
    \caption{\textbf{Comparing Ideology Measures}. General ideological self-placement (at survey start) on the political ideology spectrum ($x$ axis) plotted against three other measures to assess consistency and relative bias.  \textbf{a)} Comparison to average of self-placement on 13 prominent political issues. \textbf{b)} Comparison to average policy-stance agreement ideology. We see a systematic centralizing and liberalizing effect of the agreement-based metric (best-fit slope of $0.52$, intercept $-8.0$), along with a weaker, but still substantial fit ($R^2 = 0.6133$) based on respondents' general ideology.
    \textbf{c)} For ``null" comparison, general ideology is seen to be very consistent with itself across the length of the survey, despite potential re-contextualization of the political environment from the intervening stimuli---the average absolute deviation between measures of general self-placement was $5.25$ on the 100-point scale, serving as an upper bound on individuals' inherent response deviation.}
    \label{fig:ideo_scatter}
\end{figure*}

First, we examine different measures of individuals' ideology in comparison: their overall, ``general" self-reported sense of their own ideology versus more granular measures. These more granular metrics take two different forms: self-placement on a liberal-conservative scale for individual issues, and policy-agreement. We find several noteworthy effects, as summarized in Figure \ref{fig:ideo_scatter}. 

\subsubsection{Average issue-by-issue ideology is largely consistent with general ideology.} 
We find that average issue self-placement largely agrees with general ideological self-placement, albeit with a slight moderating bias, skewing slightly towards liberalizing respondents, as seen in Fig \ref{fig:ideo_scatter}a. This suggests that general self-placement is a fairly accurate reflection of individuals' more granular issue-based positions as rated on the same individually-interpreted ideological scale.

\begin{figure*}[h!] 

    \centering
    \includegraphics[width= 0.33\textwidth]{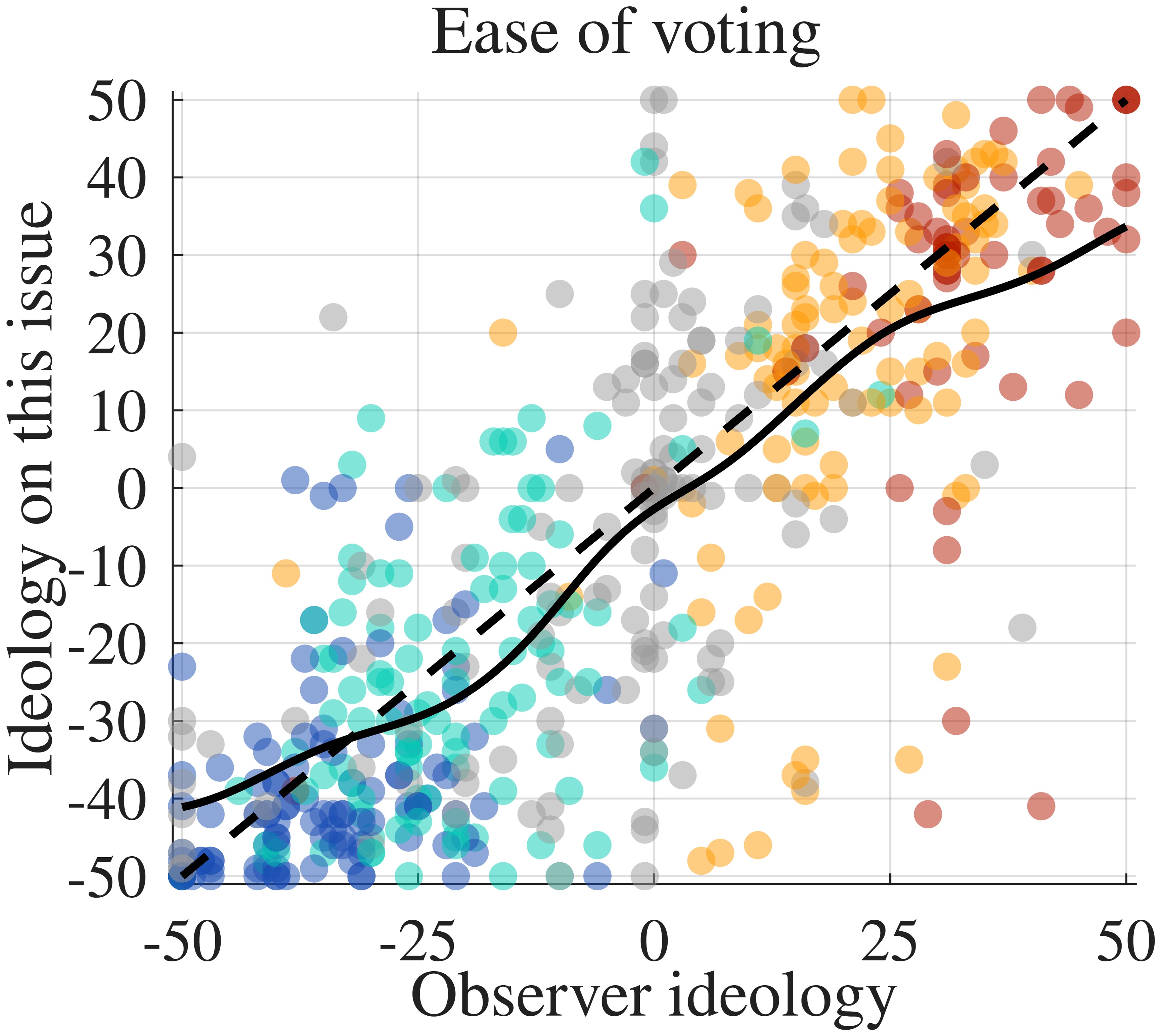}
    \raisebox{0.1\height}{\includegraphics[width= 0.095\textwidth]{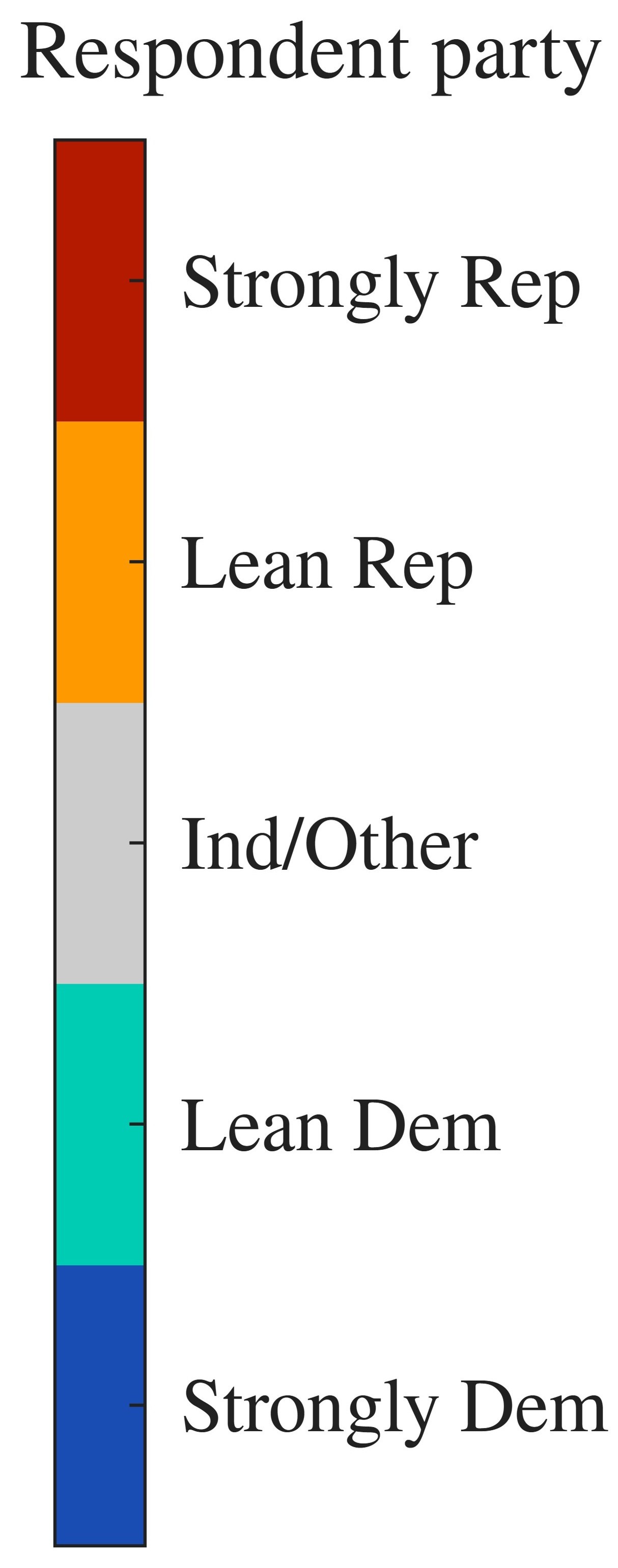}}
    \\
    \includegraphics[width= 0.26\textwidth]{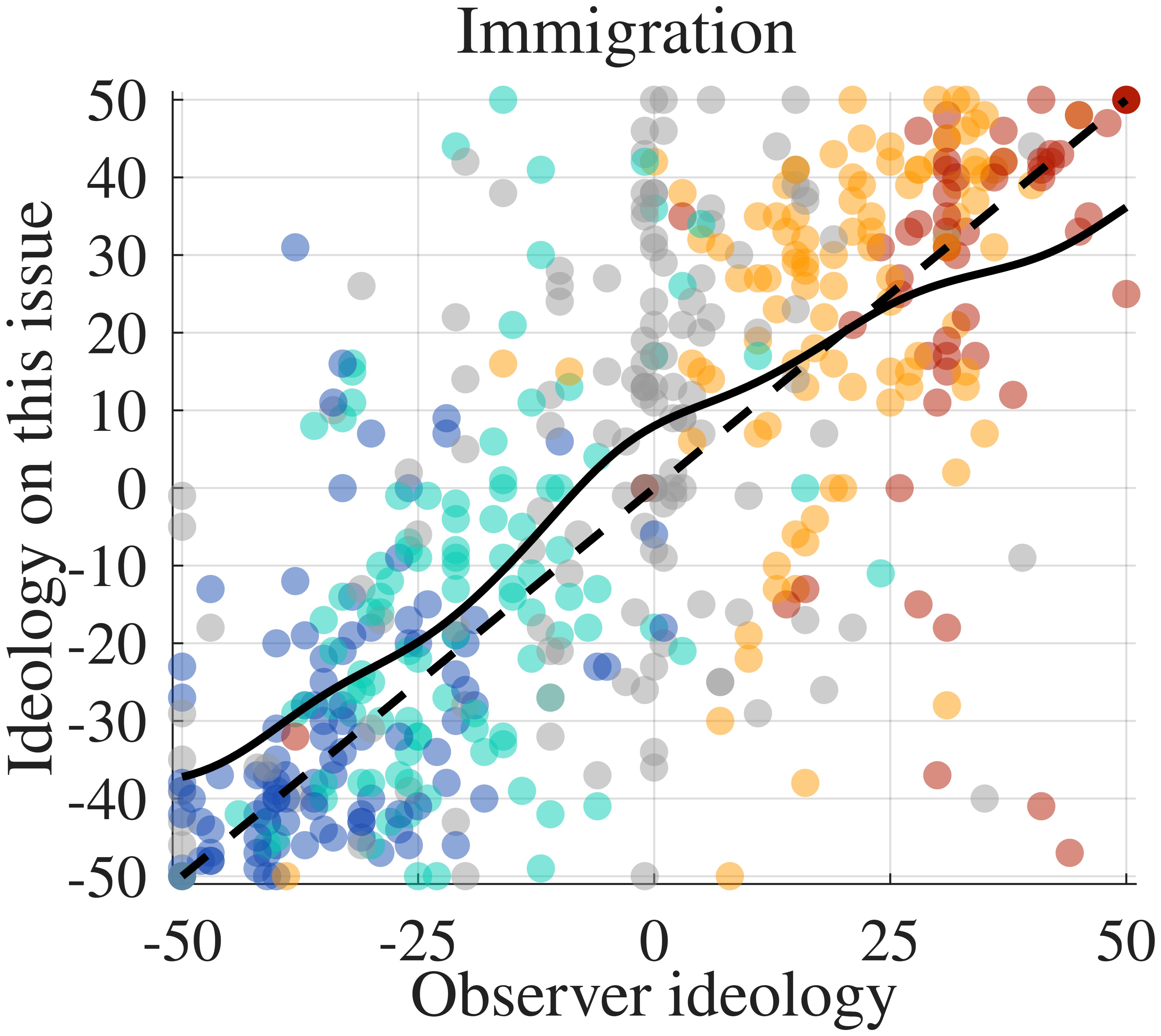}
    \includegraphics[width= 0.26\textwidth]{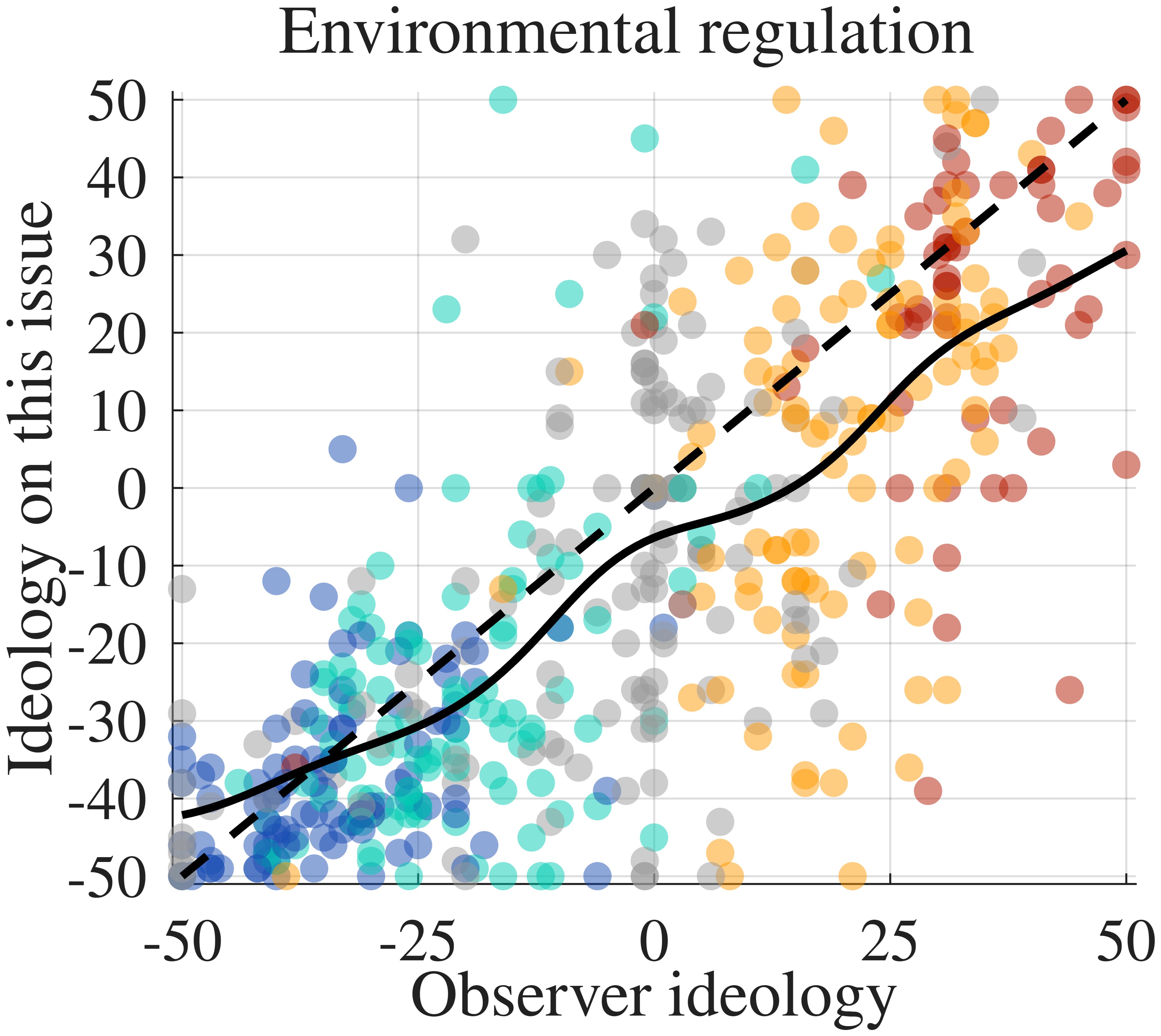}
    \includegraphics[width= 0.26\textwidth]{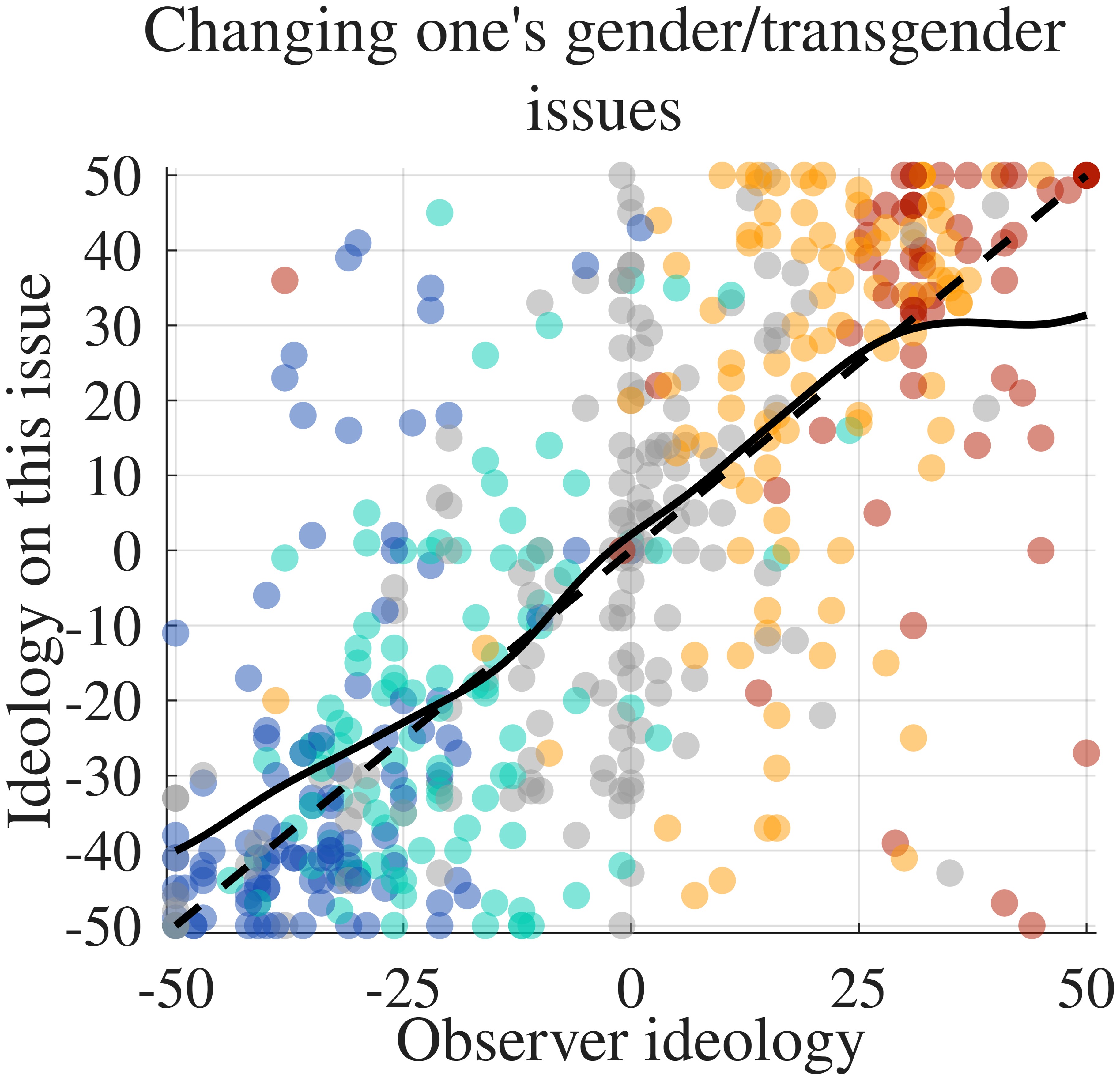}
    \includegraphics[width= 0.26\textwidth]{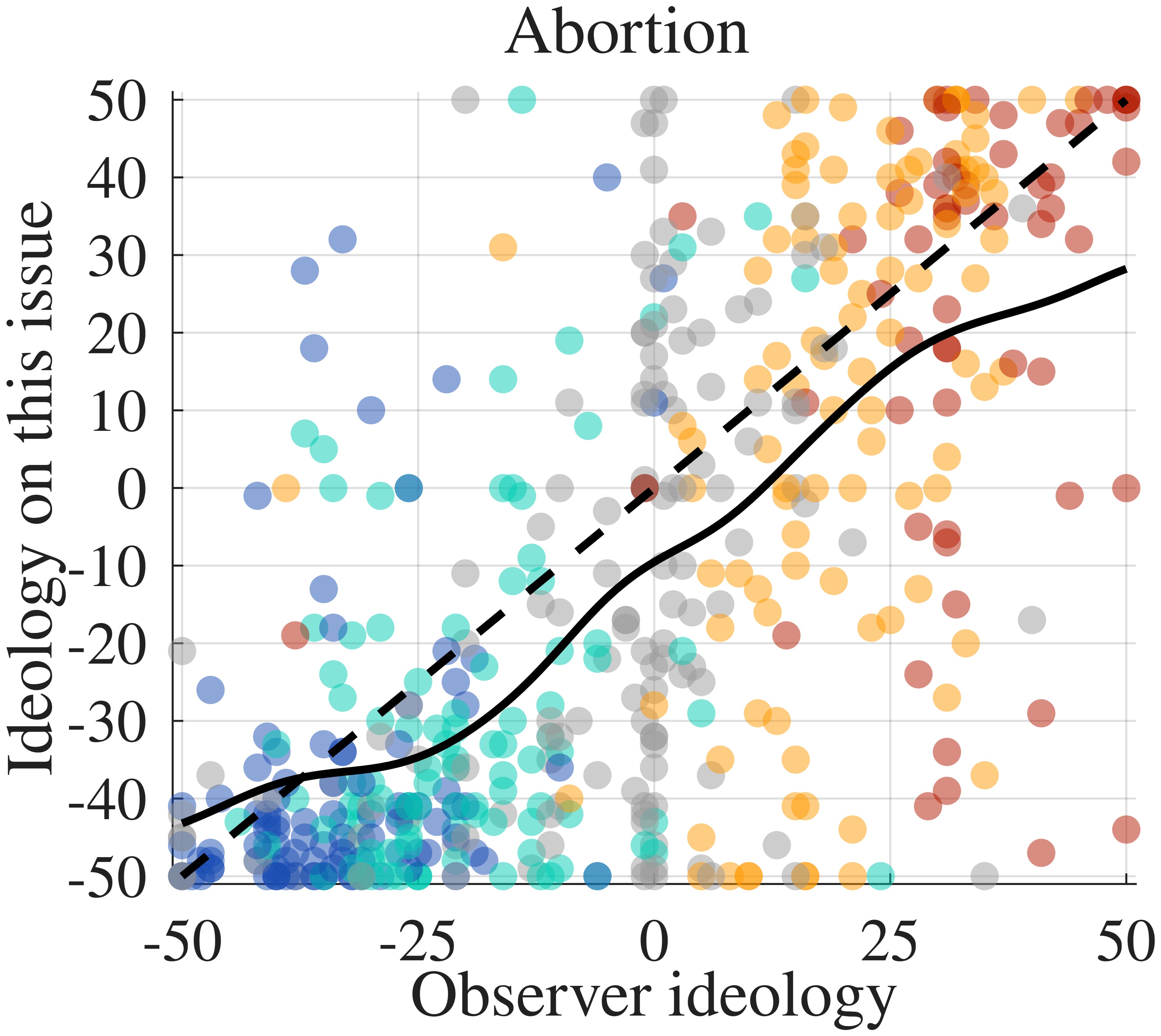}
    \\
    \vspace{2mm}
    \includegraphics[width= 0.26\textwidth]{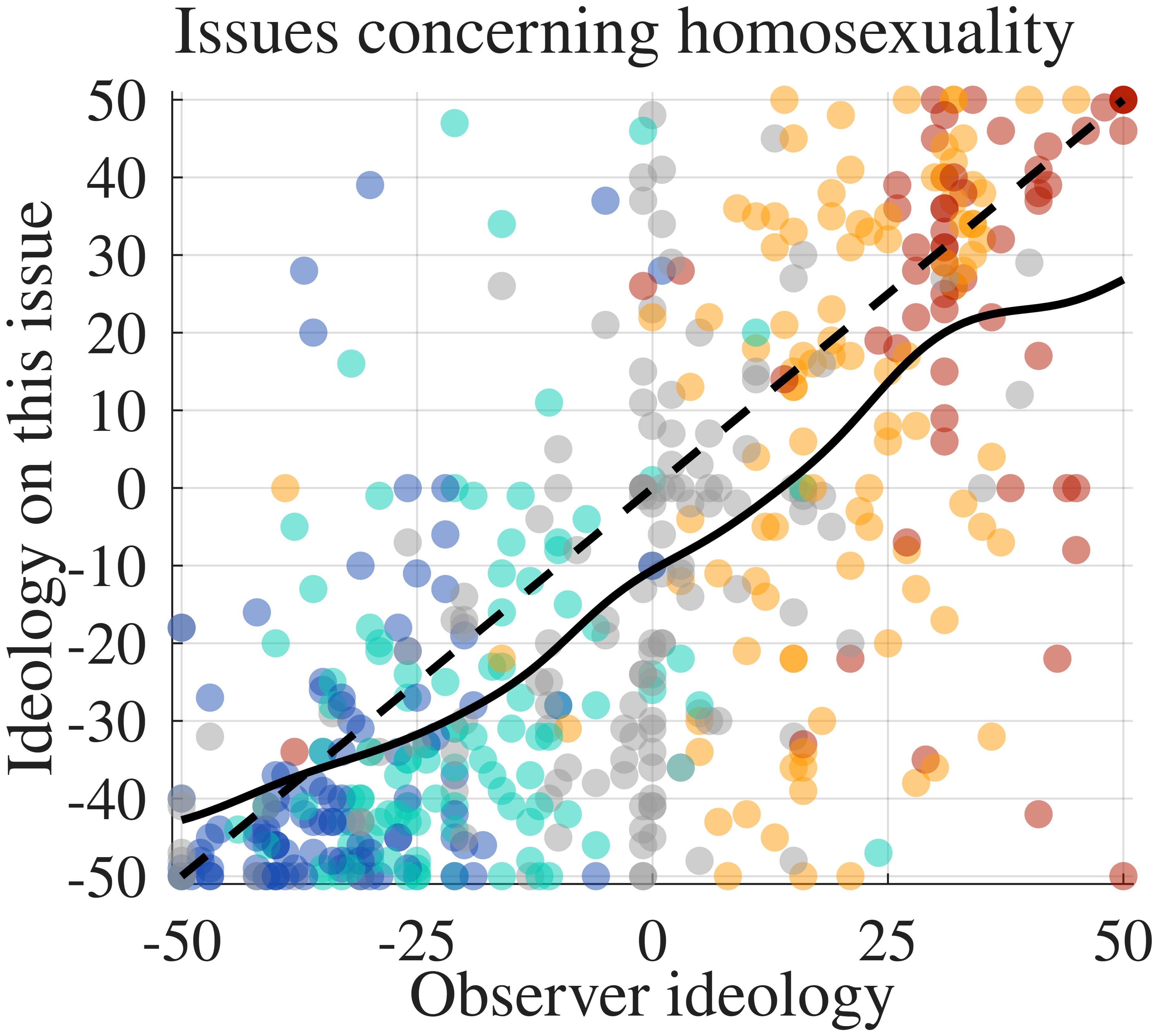}
    \includegraphics[width= 0.26\textwidth]{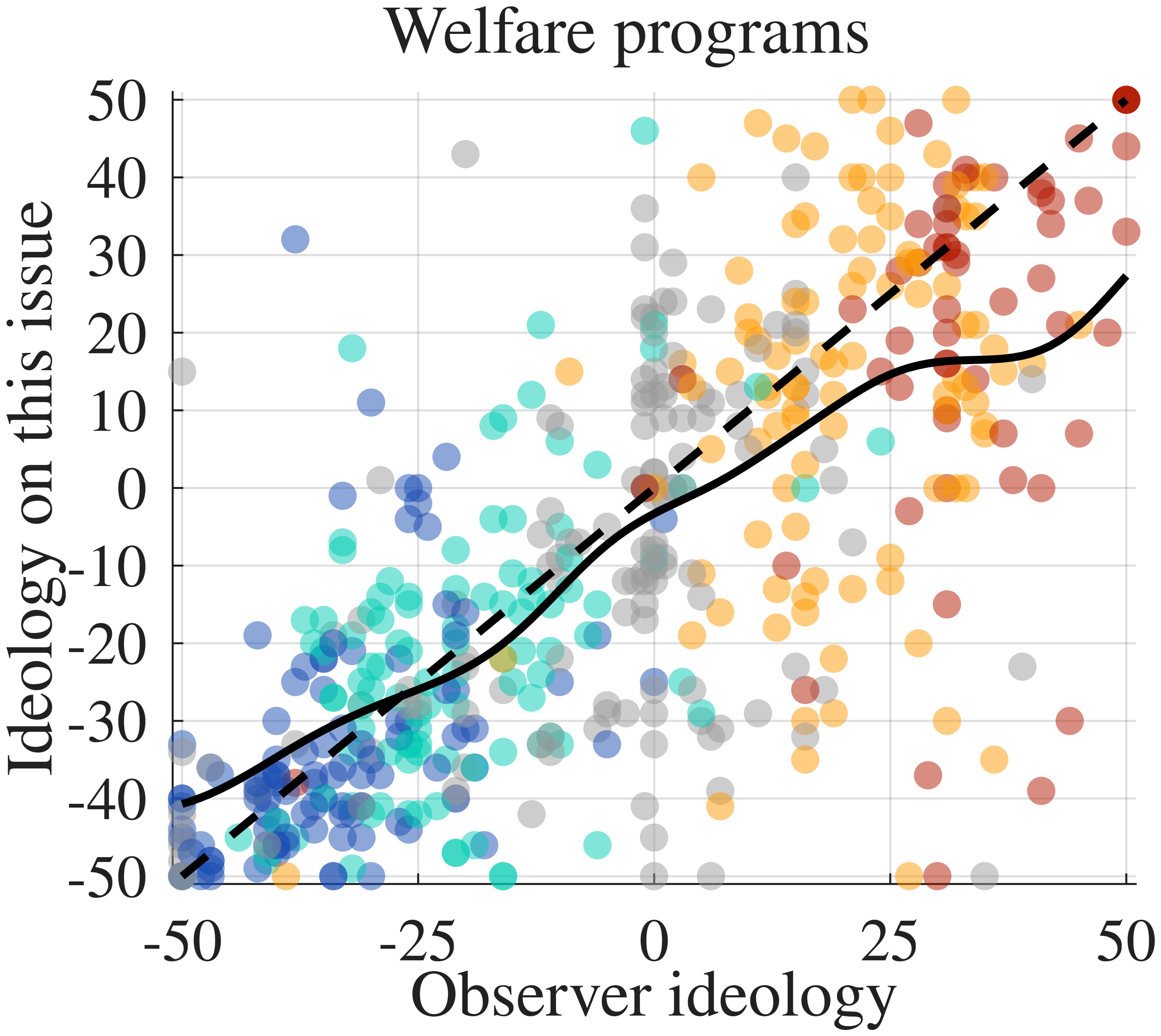}
    \includegraphics[width= 0.26\textwidth]{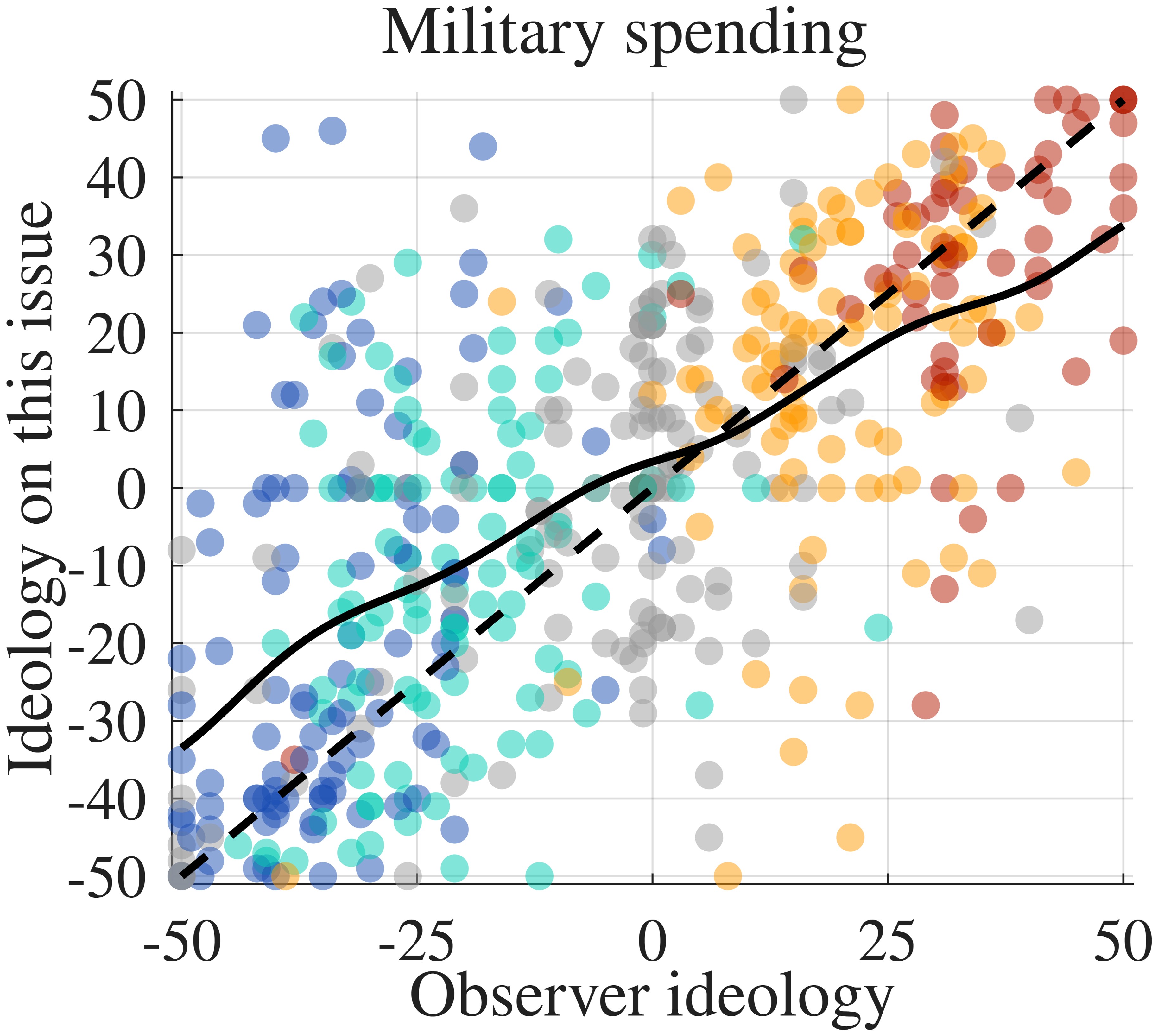}
    \includegraphics[width= 0.26\textwidth]{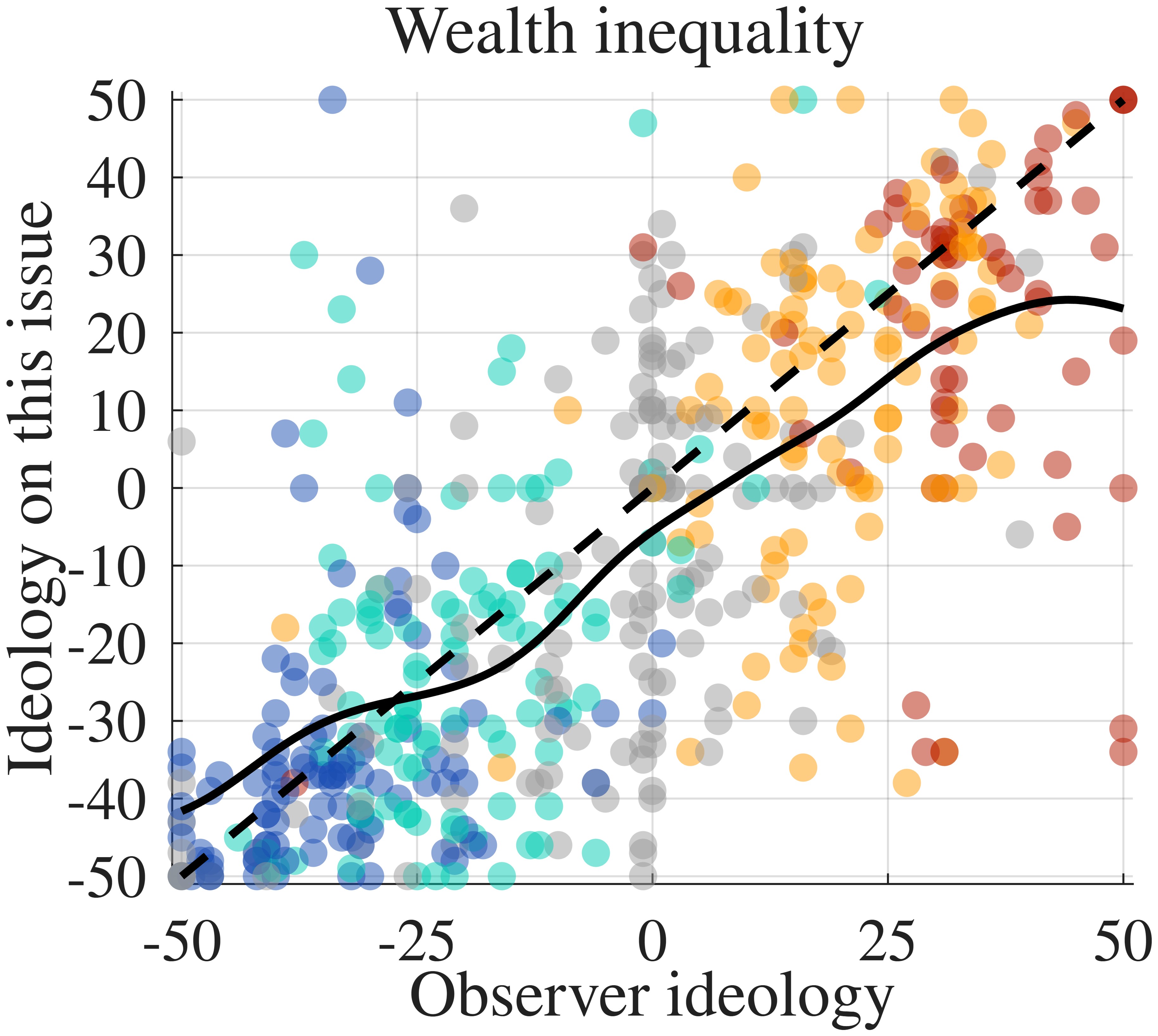}
    \\
    \vspace{2mm}
    \includegraphics[width= 0.26\textwidth]{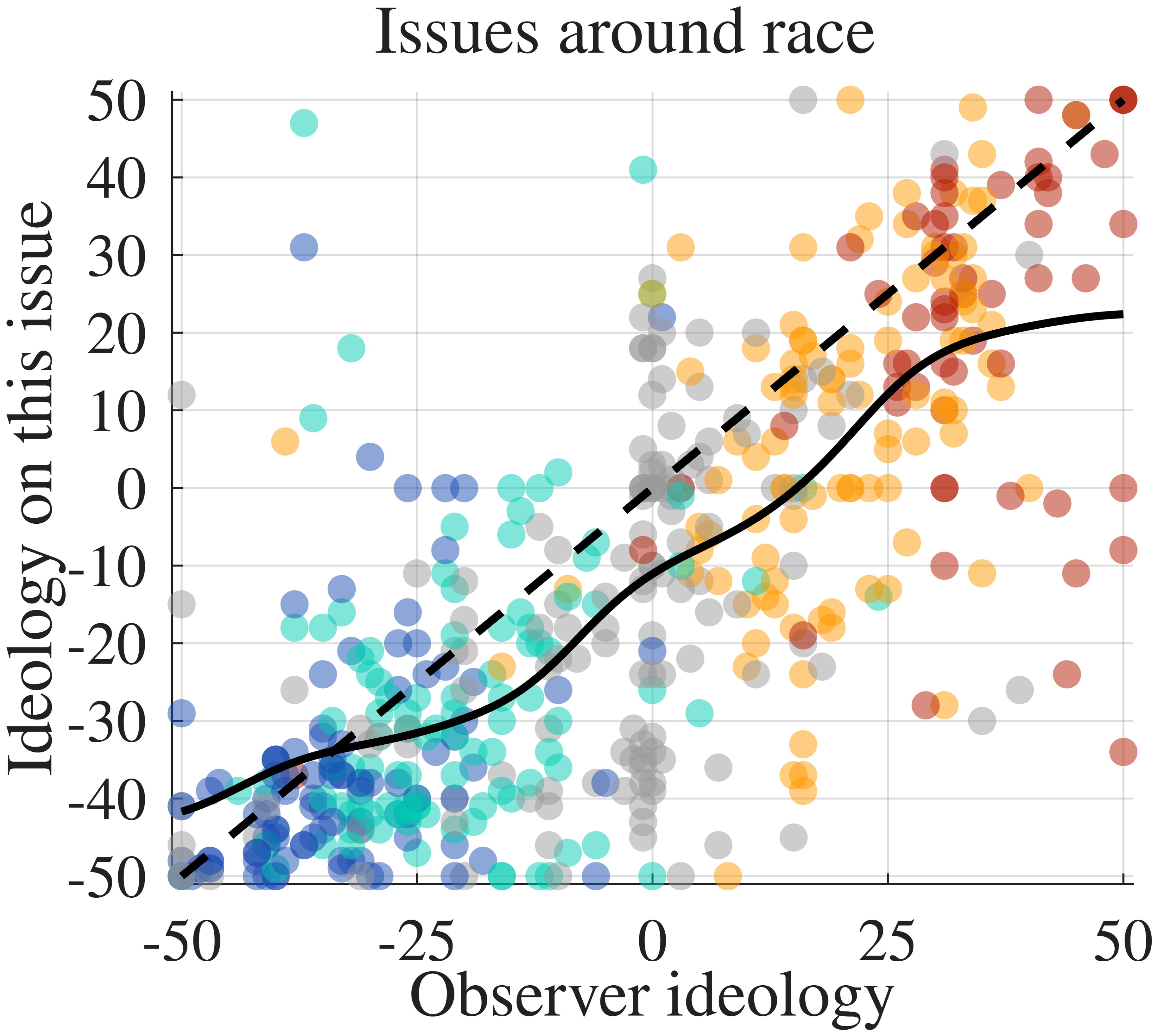}
    \includegraphics[width= 0.26\textwidth]{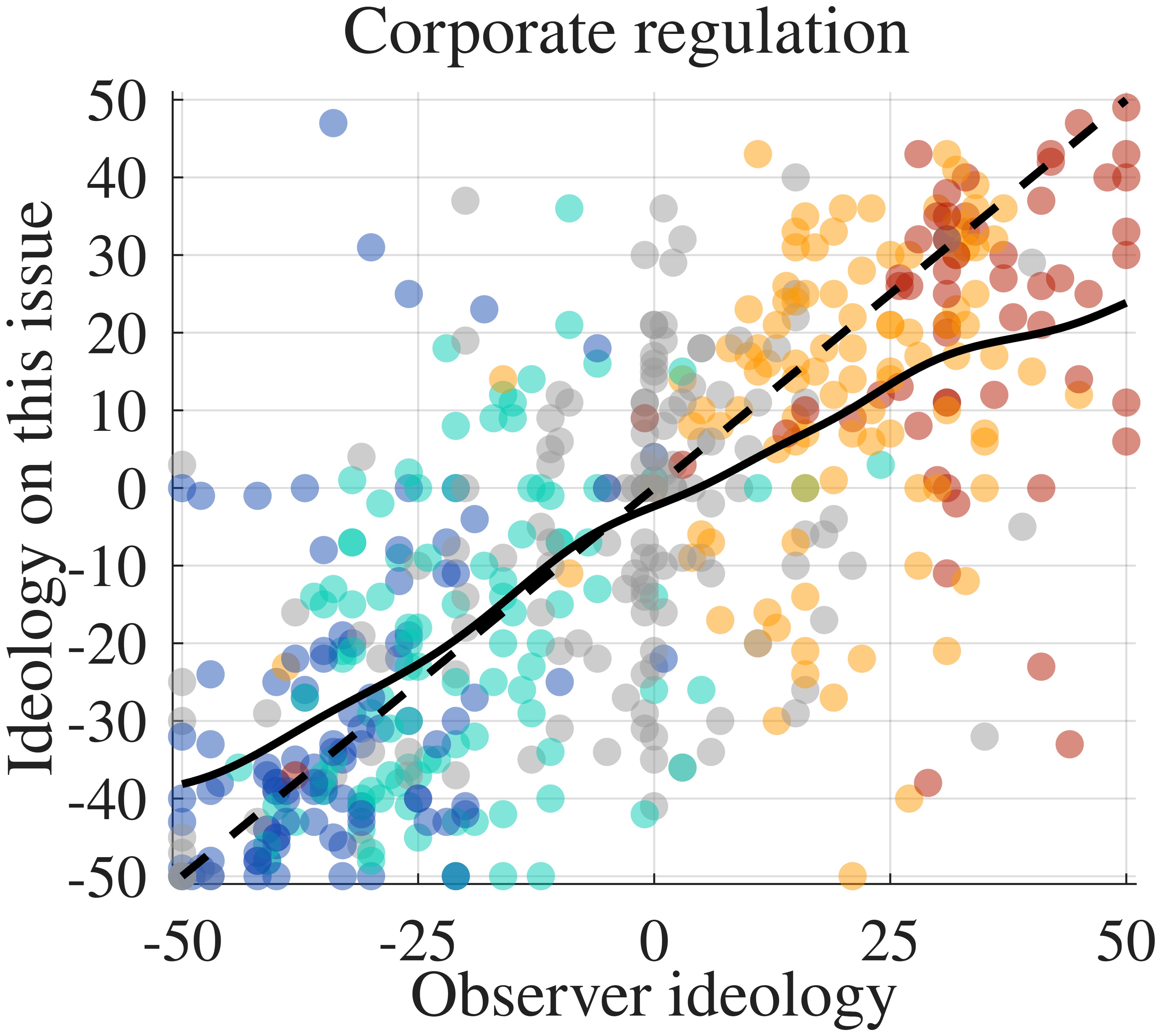}
    \includegraphics[width= 0.26\textwidth]{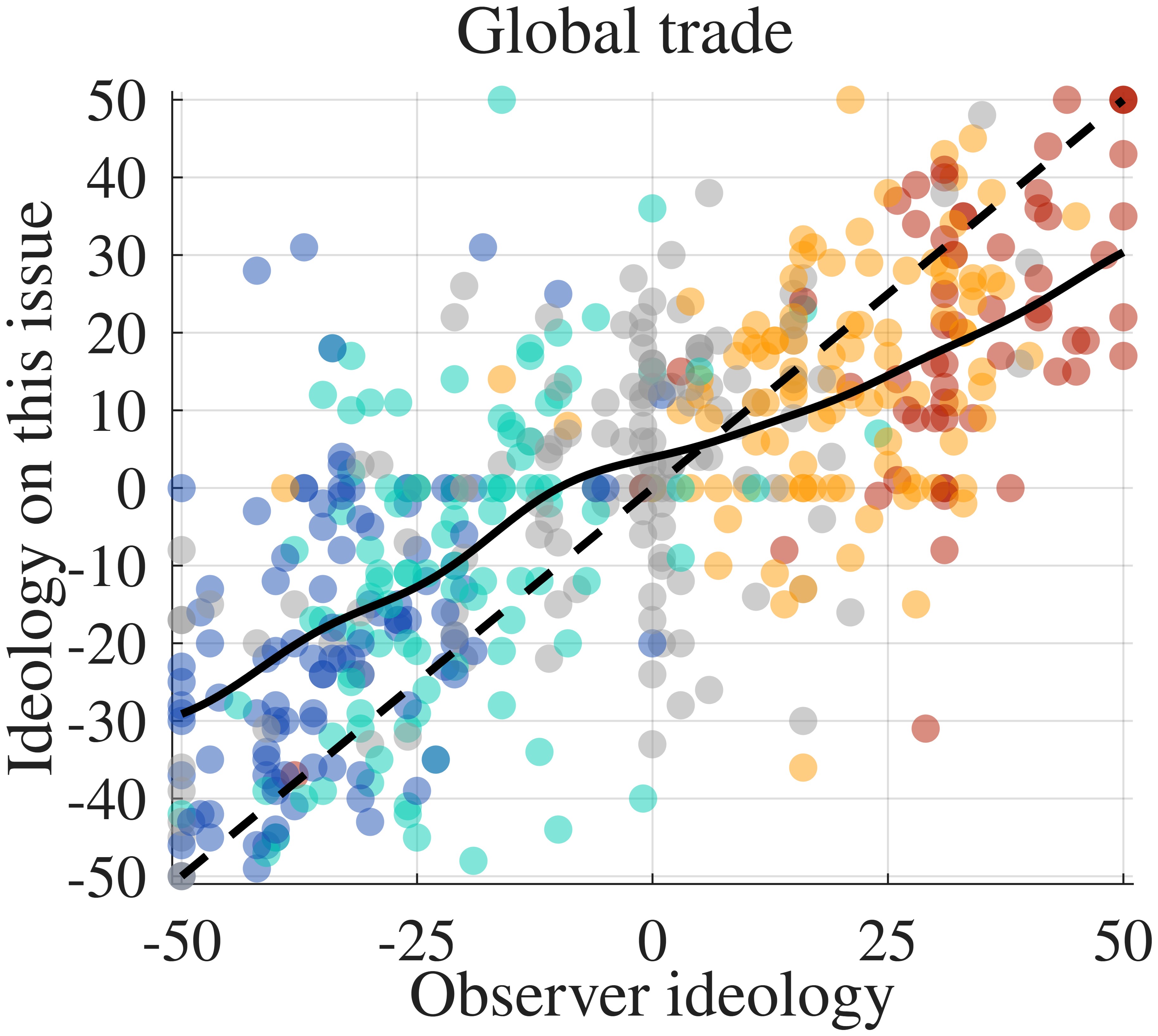}
    \includegraphics[width= 0.26\textwidth]{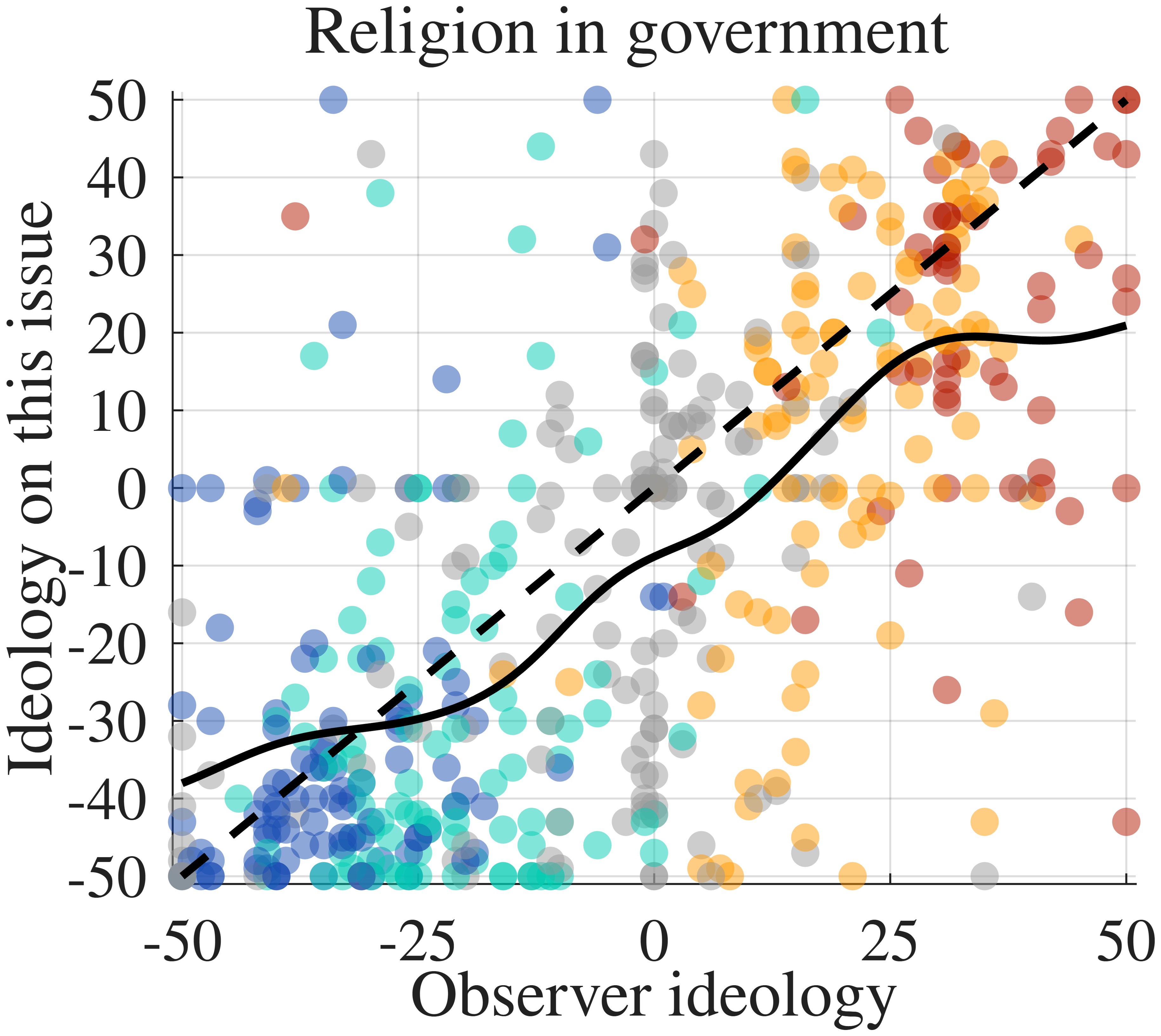}
    
    \caption{\textbf{Issue Self-Placement.} Full response data for ideological self-placement on each ``major issue," plotted against each respondent's general ideology. Black trend curves are Gaussian-weighted moving average ($\sigma = 7$). Issues are ordered by decreasing polarization (far-left/far-right mean difference). Dots colored by respondent party affiliation, with additional hue variation to aid distinguishing despite dot transparency.
    }
    \label{fig:13_major_selfplace_scatters}
\end{figure*}

Fig \ref{fig:13_major_selfplace_scatters} shows the ideological self-placement data for each individual  issue. Overall, these issue-ideology scores follow the general-ideology-consonant diagonal, with some slight liberalization at the conservative end of the spectrum. 

\subsubsection{Average policy-agreement ideology is largely predicted by general ideology, but is liberally biased relative to general ideology for right-leaning participants.}

\begin{figure*} [h!]
    \centering
    \begin{overpic}
        [width=.54\textwidth]{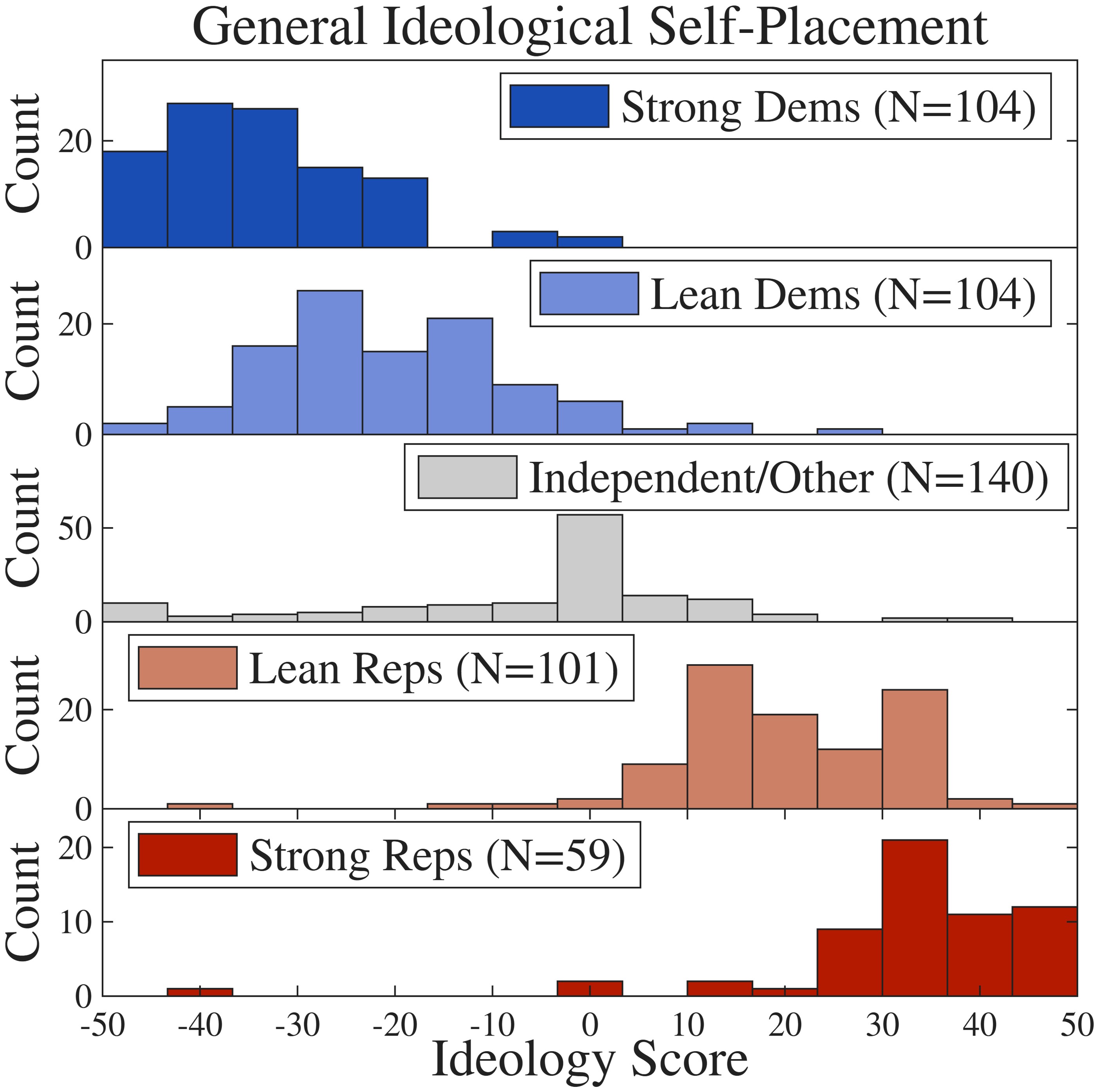}
        \put(4,95){\large{\textbf{a}}}
    \end{overpic}
    \begin{overpic}
        [width=.54\textwidth]{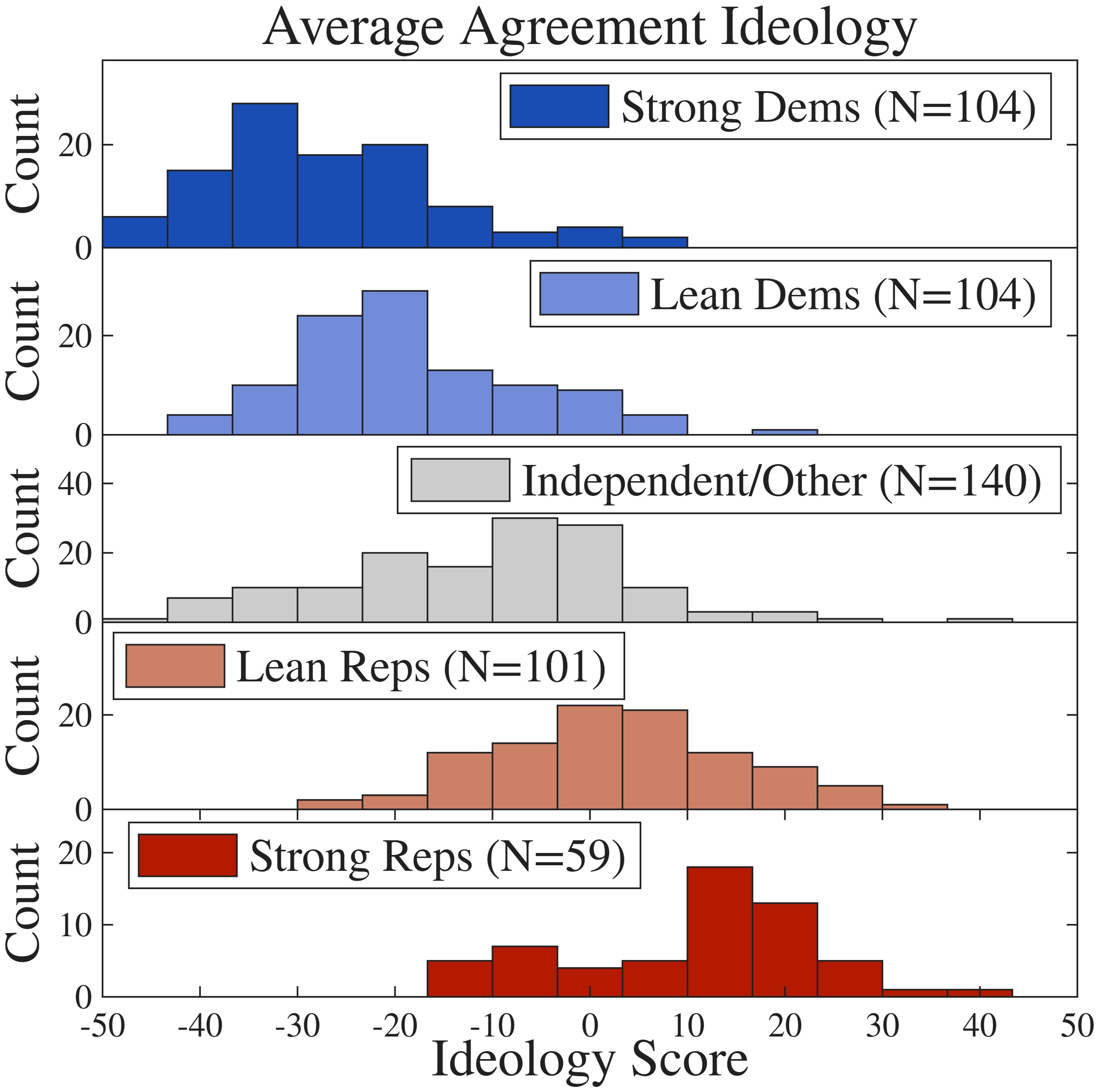}
        \put(4,95){\large{\textbf{b}}}
    \end{overpic}
    \\
    \vspace{2mm}
    \begin{overpic}
        [width=.54\textwidth]{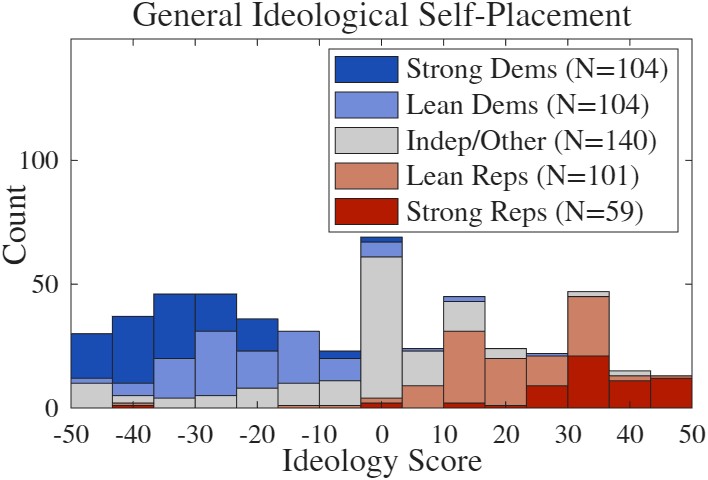}
        \put(5,64){\large{\textbf{c}}}
    \end{overpic}
    \begin{overpic}
        [width=.54\textwidth]{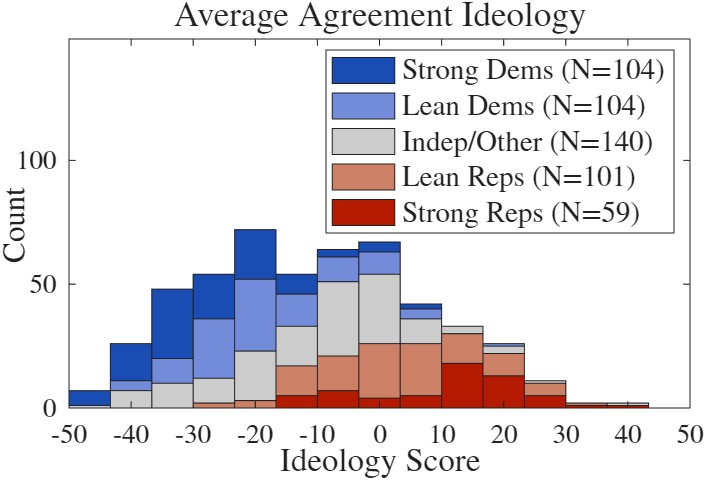}
        \put(5,64){\large{\textbf{d}}}
    \end{overpic}
    \caption{\textbf{Ideology Histograms:} Comparison of overall self-placement ideology (left panels) and policy-agreement average ideology (right panels), to illustrate the distribution of ideology for each party-identity group (top panels) and the overall ideology distribution of the population (bottom panels) by these two measures. 
    Top panels are visually normalized with differing $y$ axes to show the shape and position of each party-identity group, rather than their magnitude, while the bottom panels aggregate the bars to show the overall envelope of the sampled population while preserving party breakdown at each ideology level. \textbf{a)} General self-placement results in a nearly symmetric set of distributions, distinctly ordered by party identity, evenly filling the ideology space. \textbf{b)} Agreement ideology skews distributions central and liberal, though the distributions are still well-ordered by party.
     \textbf{c)} General ideology shows a more spread-out distribution which fills the whole ideological domain. \textbf{d)} Agreement ideology shows a more condensed distribution, slightly left of center.}
    \label{fig:ideo_hists}
\end{figure*}

General self-placement is also highly predictive of ideology as measured by average policy agreement. As seen in Fig \ref{fig:ideo_scatter}b, a large amount (over 61\%) of the variance in the policy-agreement ideology is accounted for by overall self-placement. However, the resulting distributions are systematically biased such that individuals generally register as more moderate than their general self-placement would suggest. This is seen in the slightly flatter-than-diagonal slope in Fig \ref{fig:ideo_scatter}b, with the largest differences for right-leaning participants. This liberalizing tilt to ideology as measured by policy-stance agreement reproduces a durable finding that U.S.~public opinion appears more liberal when people are asked about policy specifics than when asked to place their ideology in general terms (see e.g., \cite{ellis2012ideology, mason2018ideologues, grossmann2016asymmetric}). This effect on is clearly visible in Fig \ref{fig:ideo_hists}.

\begin{figure*} [h!]
    \centering
    \includegraphics[width=0.26\textwidth]{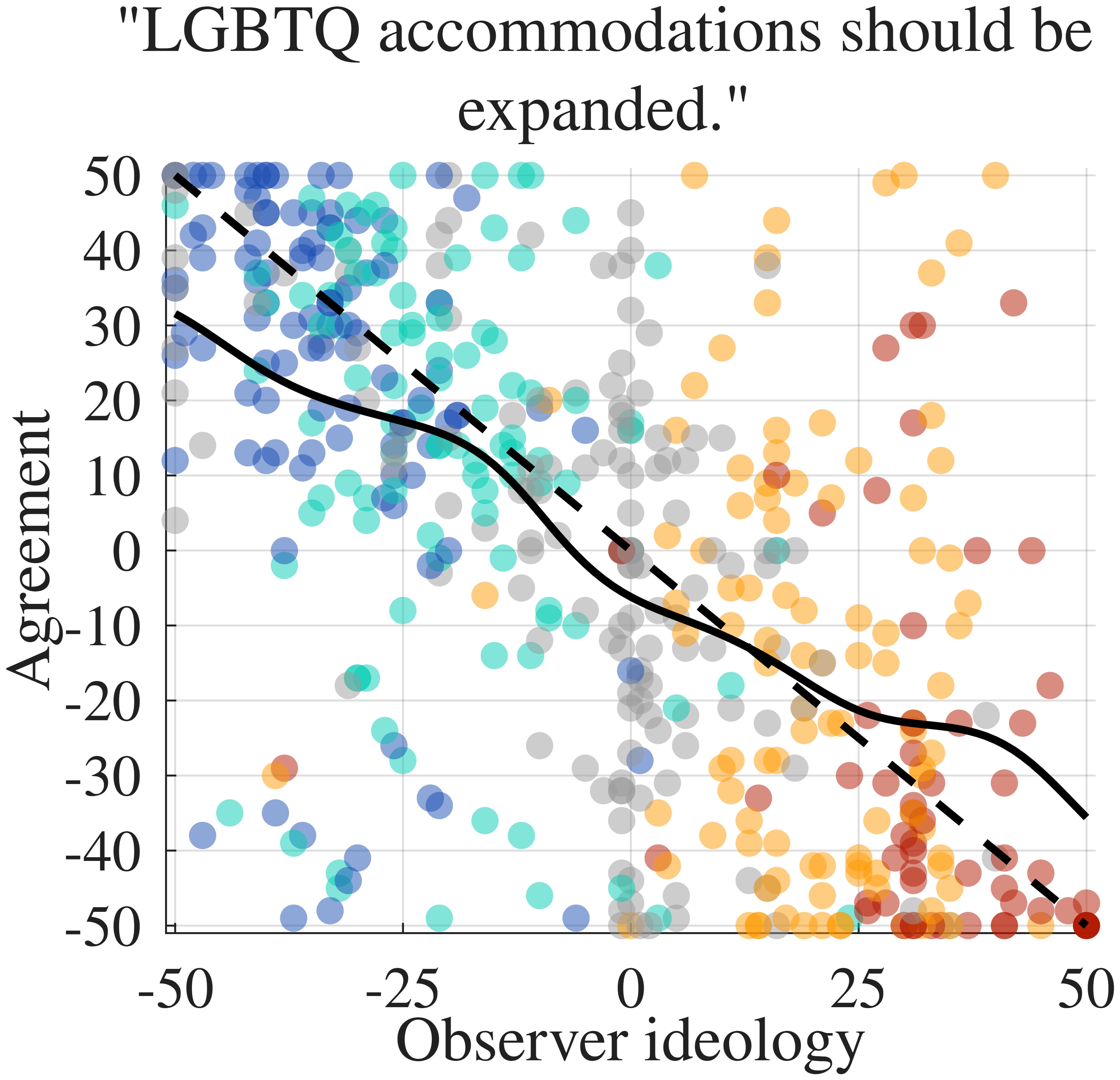}
    \includegraphics[width=0.26\textwidth]{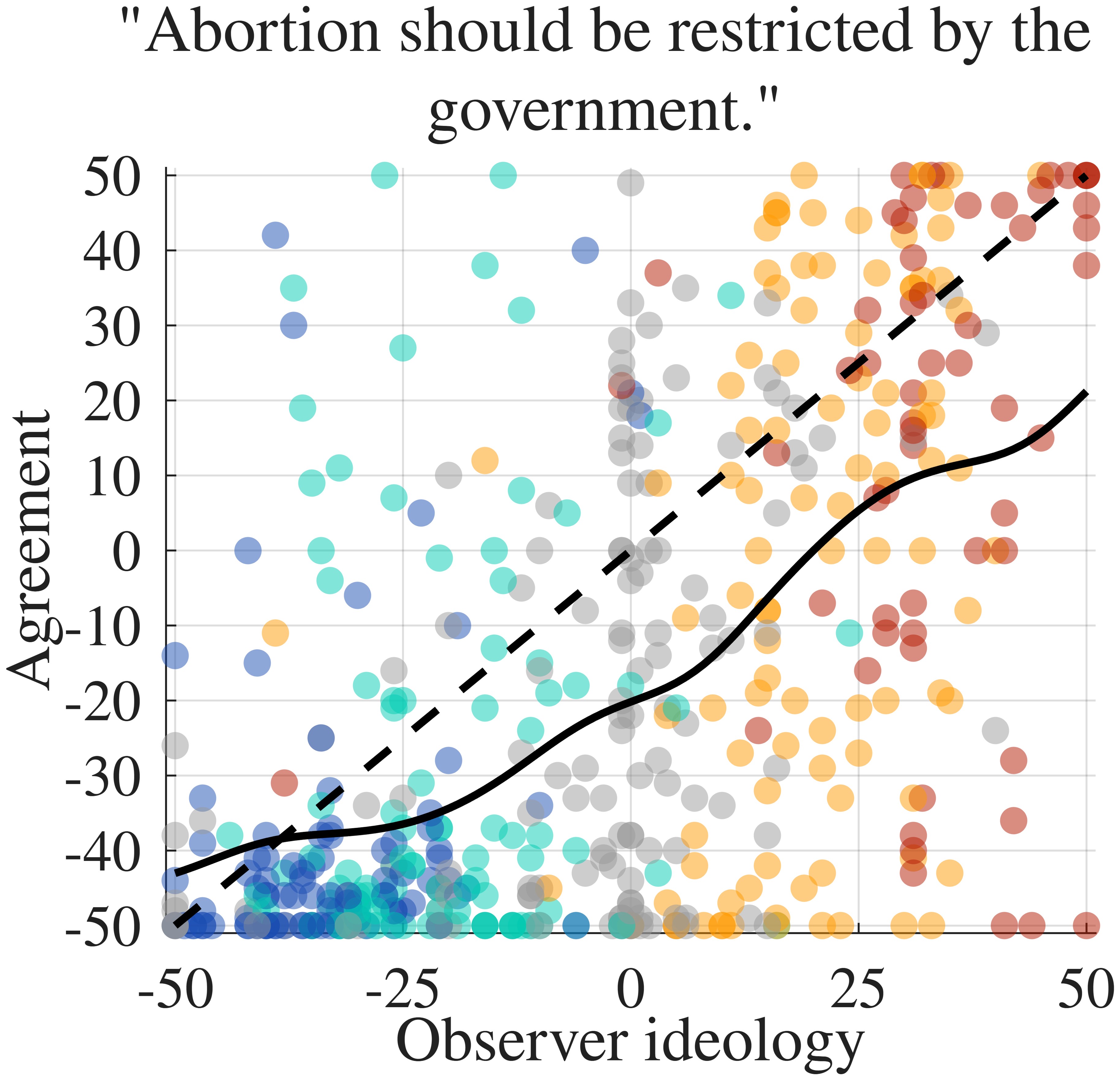}
    \includegraphics[width=0.26\textwidth]{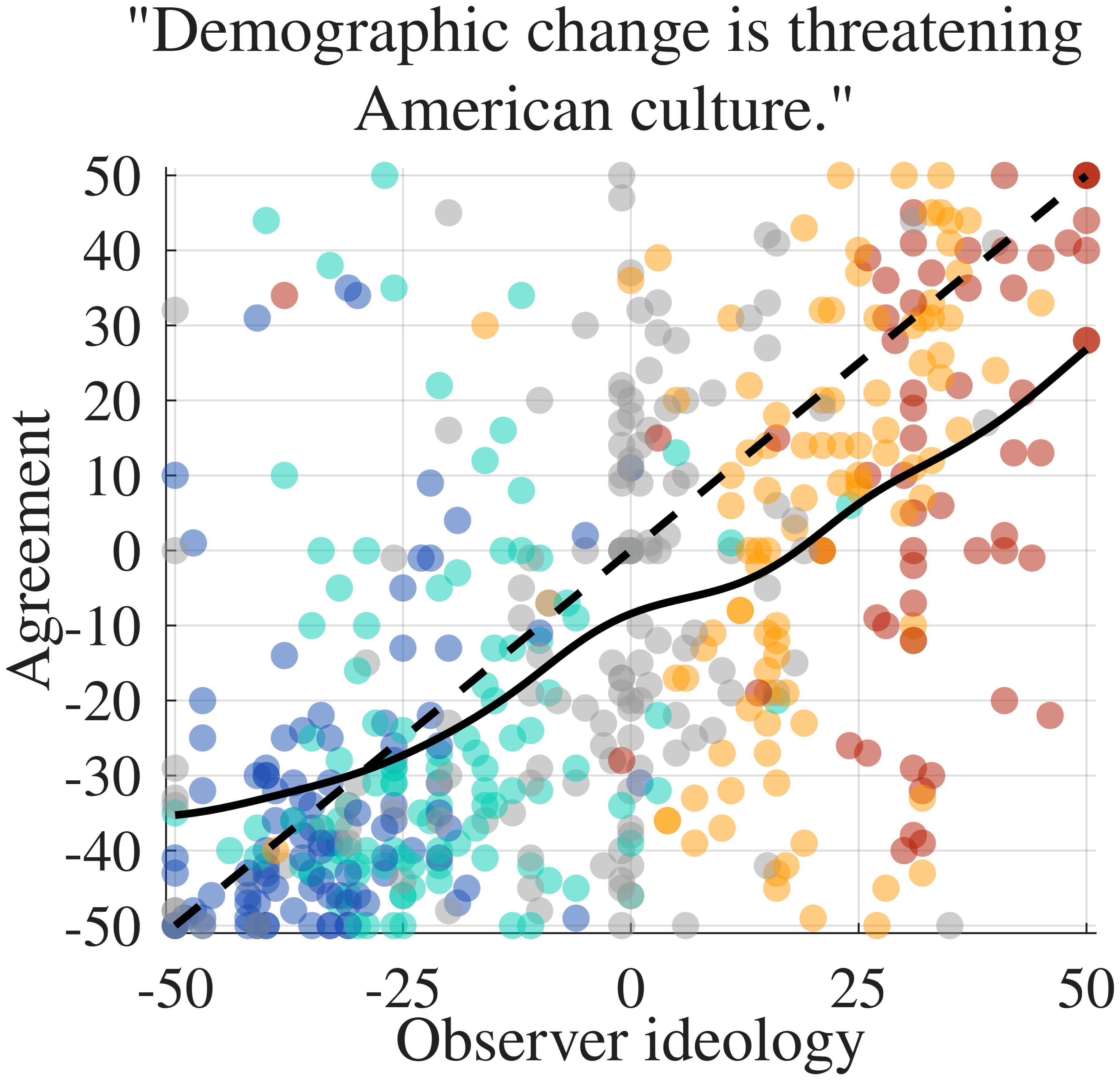}
    \raisebox{0.05\height}{\includegraphics[width= 0.09\textwidth]{jpgs/colorbar_short.jpg}}
    \\
    \vspace{2mm}
    \includegraphics[width=0.26\textwidth]{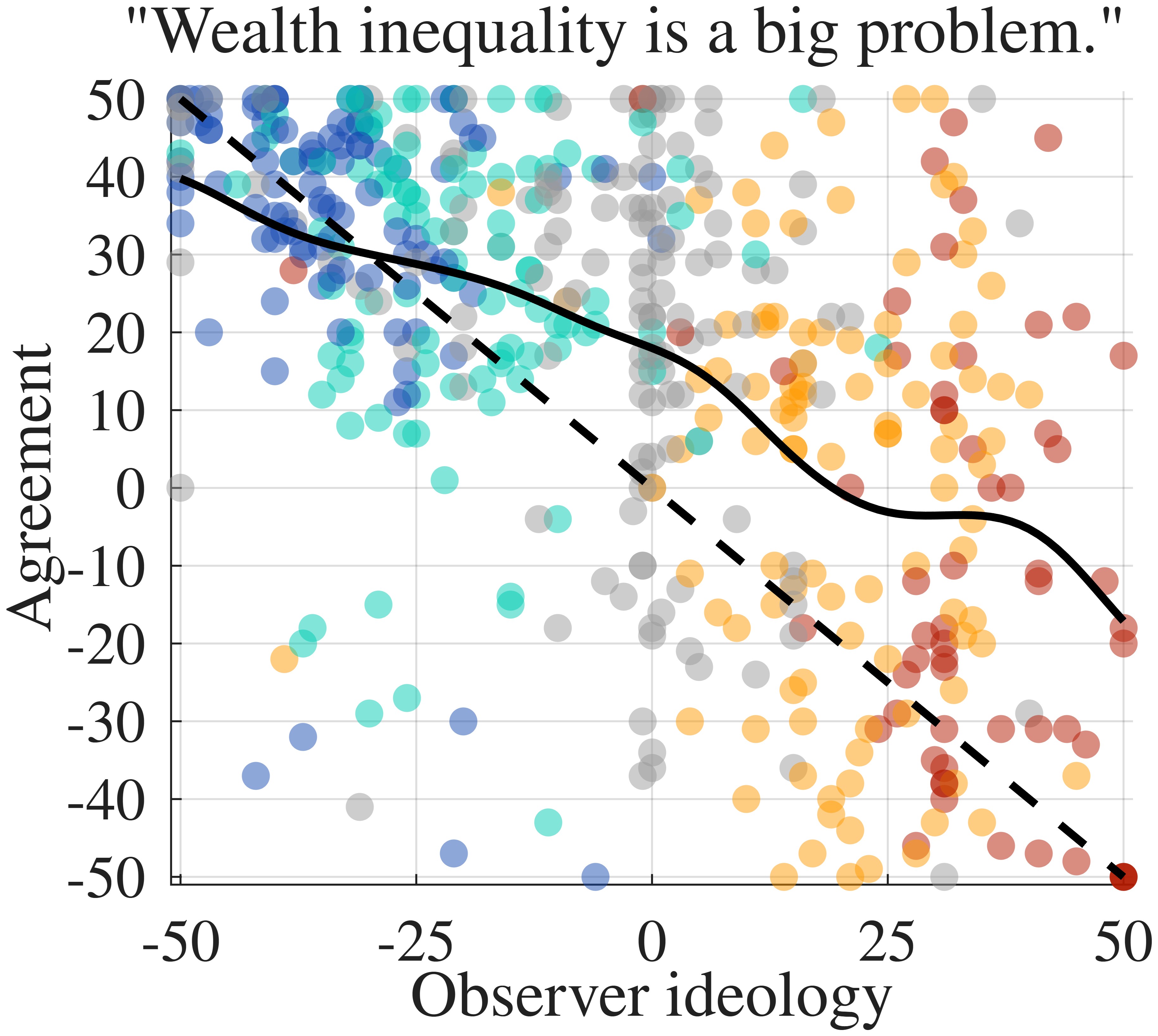}
    \includegraphics[width=0.26\textwidth]{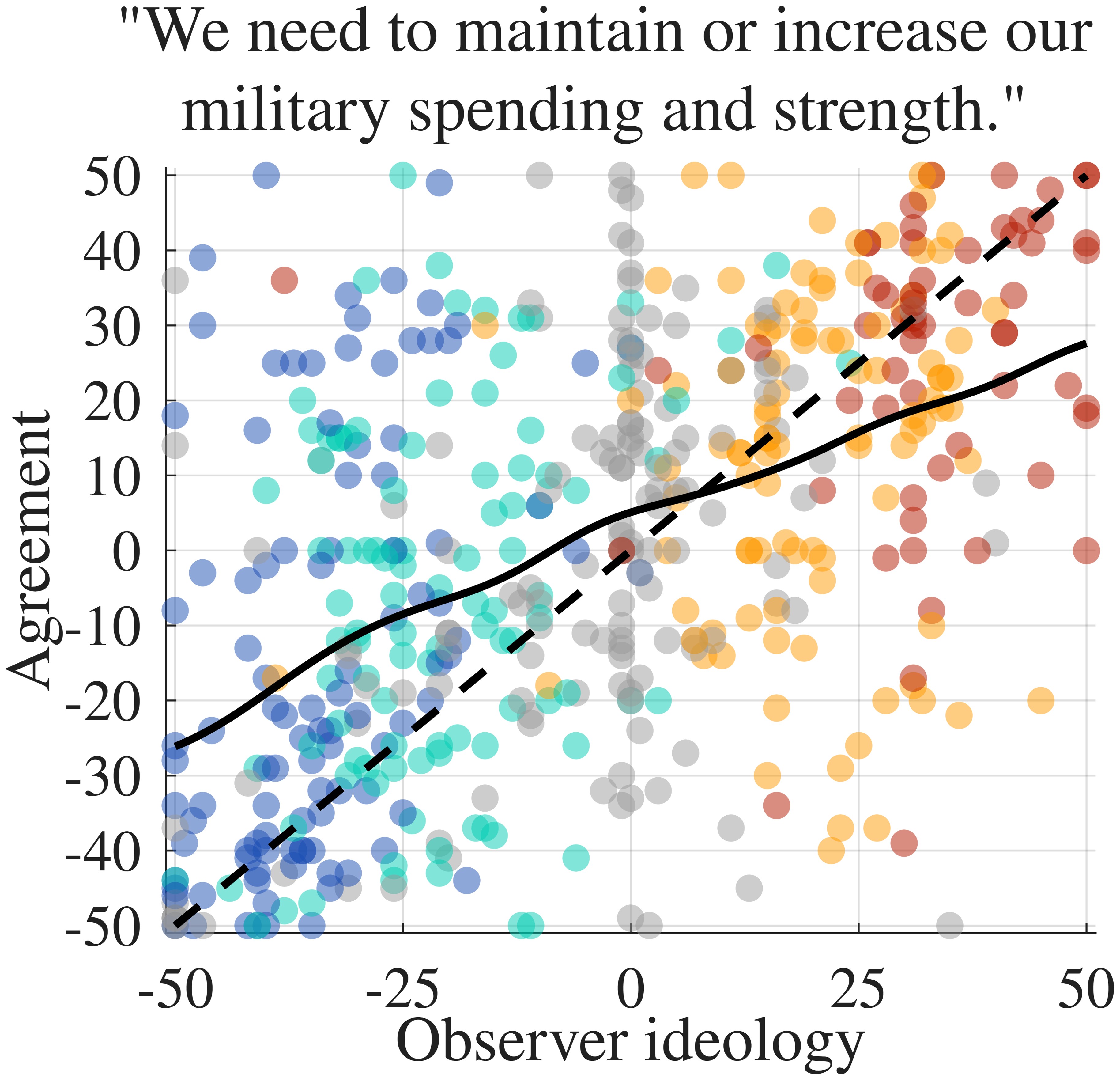}
    \includegraphics[width=0.26\textwidth]{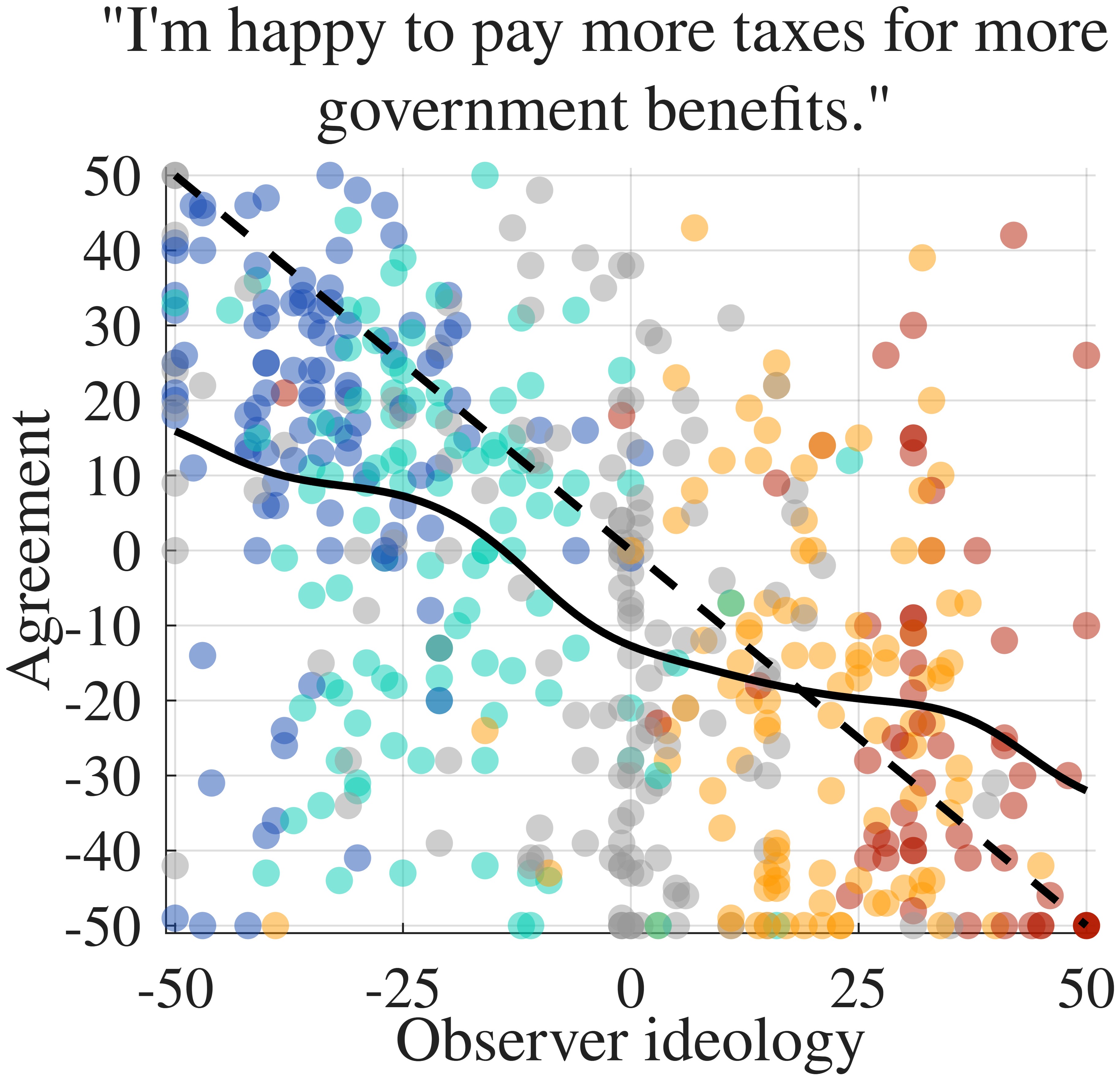}
    \includegraphics[width=0.26\textwidth]{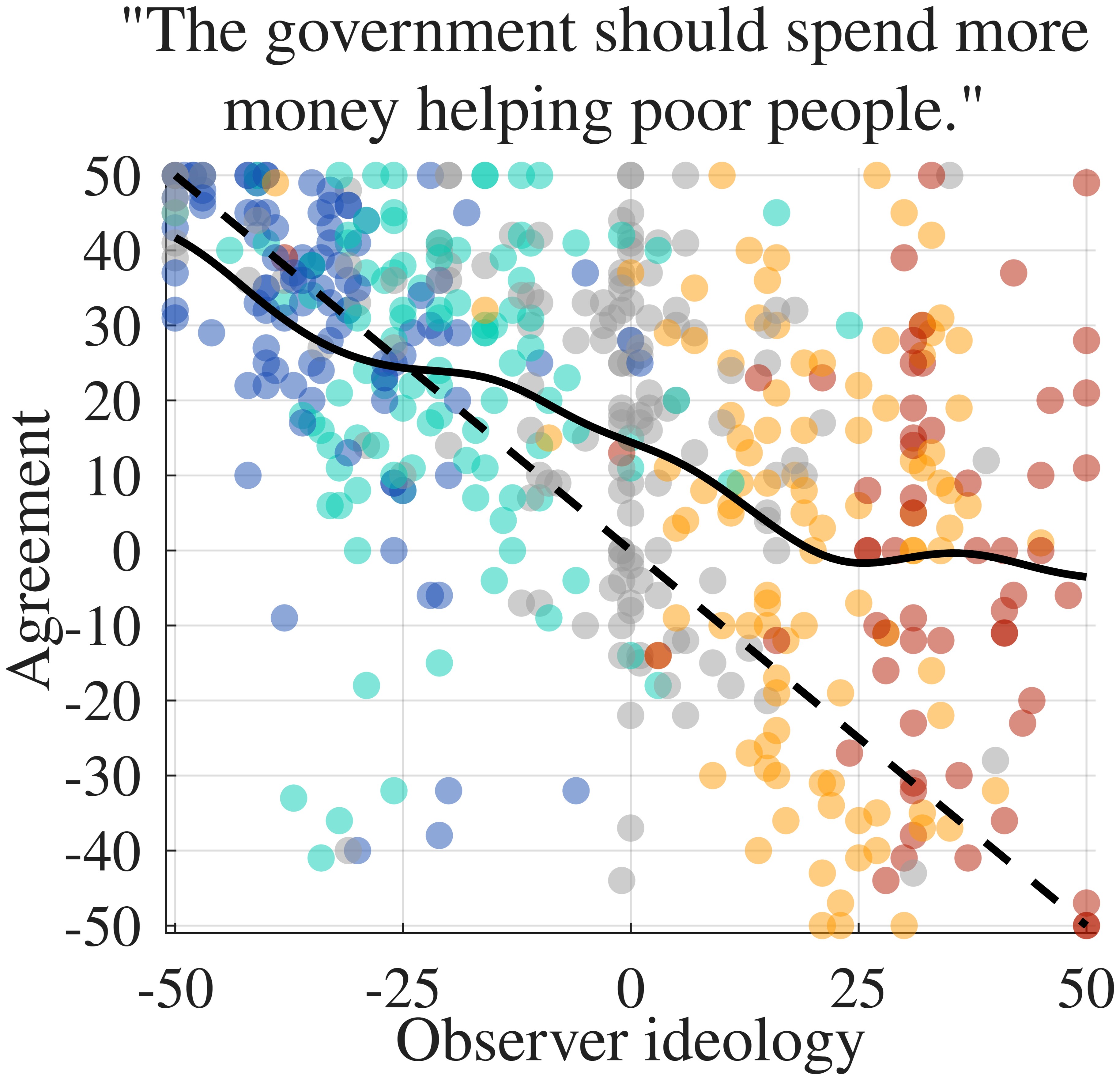}
    \\
    \vspace{2mm}
    \includegraphics[width=0.26\textwidth]{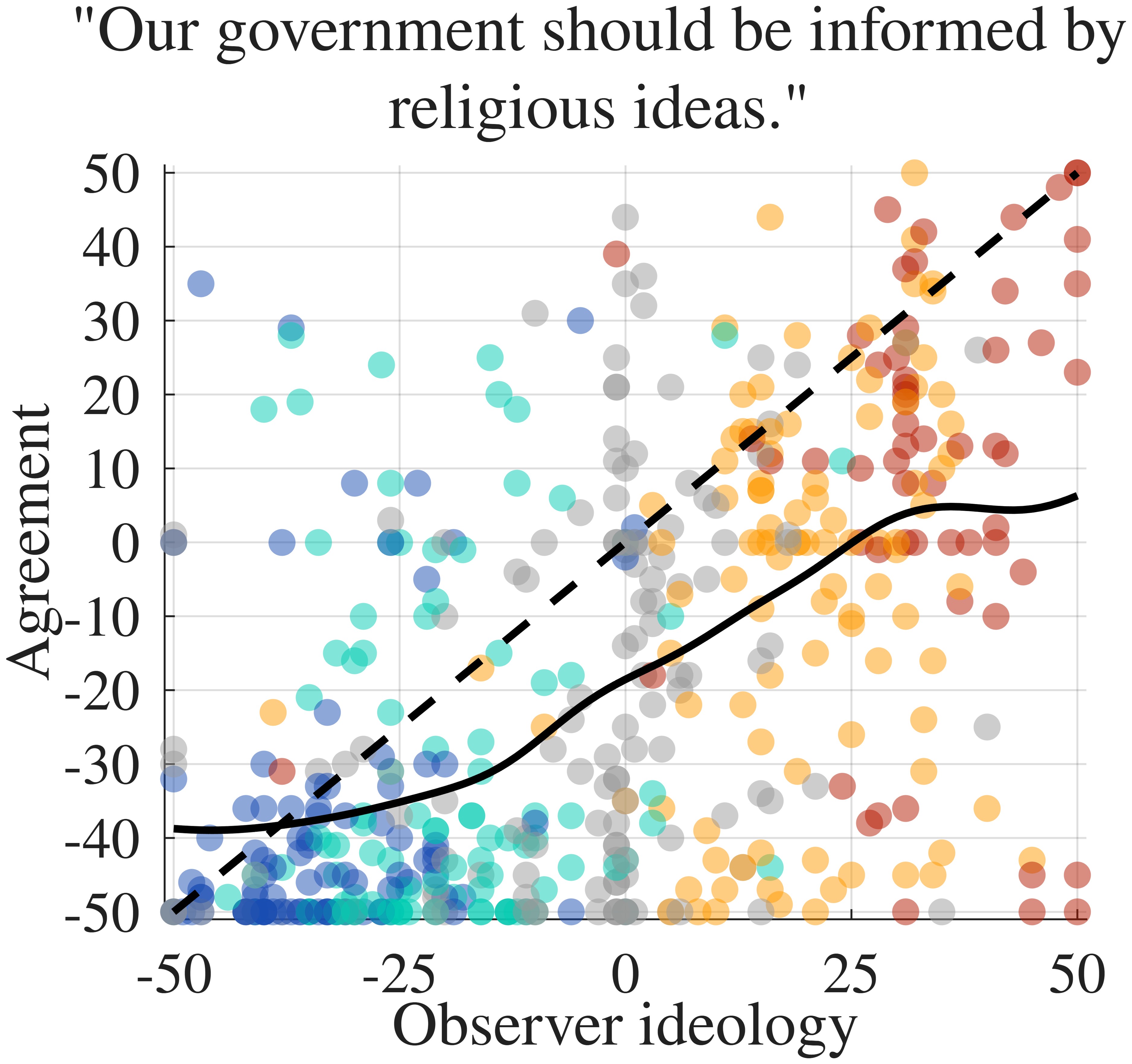}
    \includegraphics[width=0.26\textwidth]{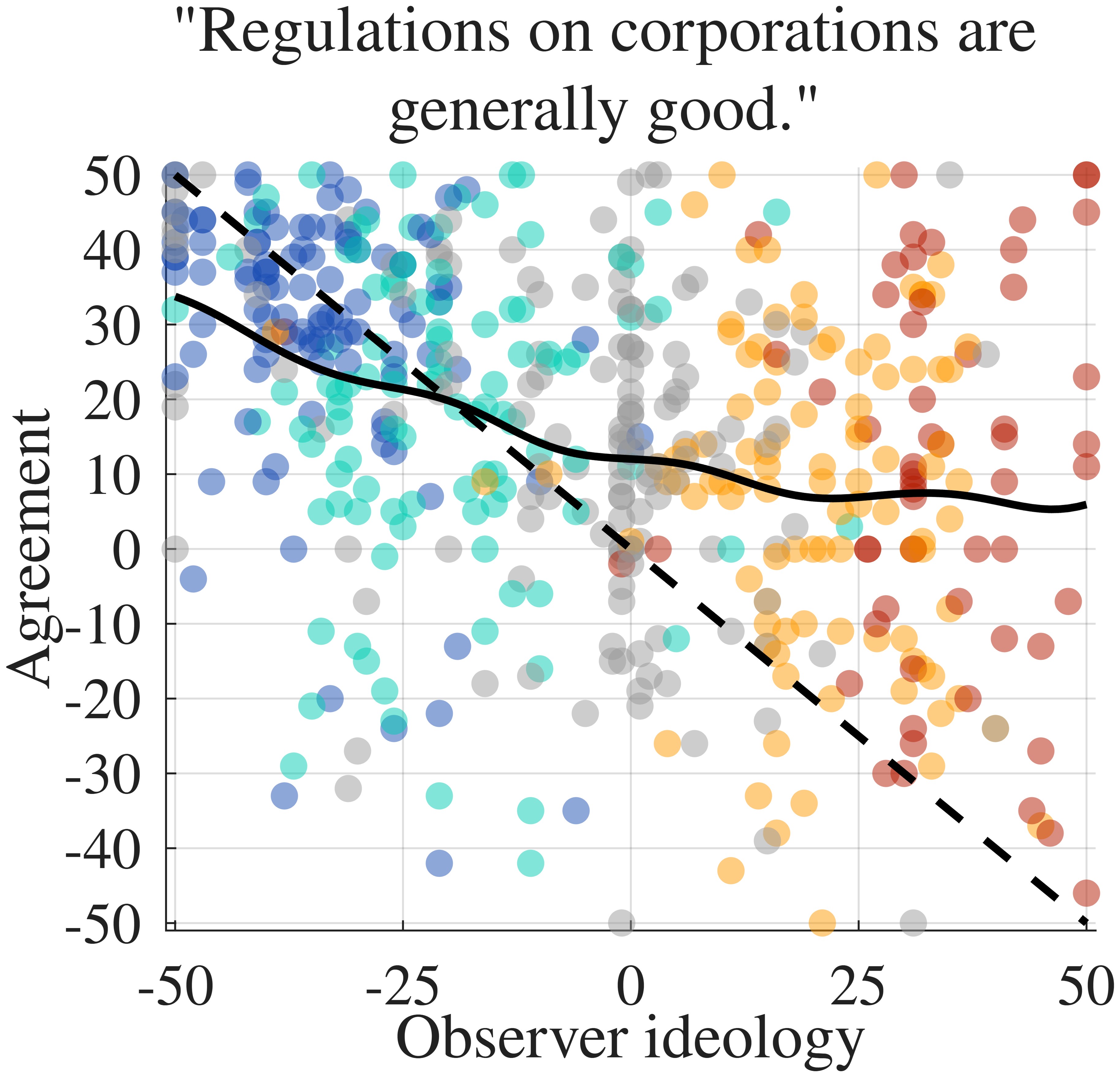}
    \includegraphics[width=0.26\textwidth]{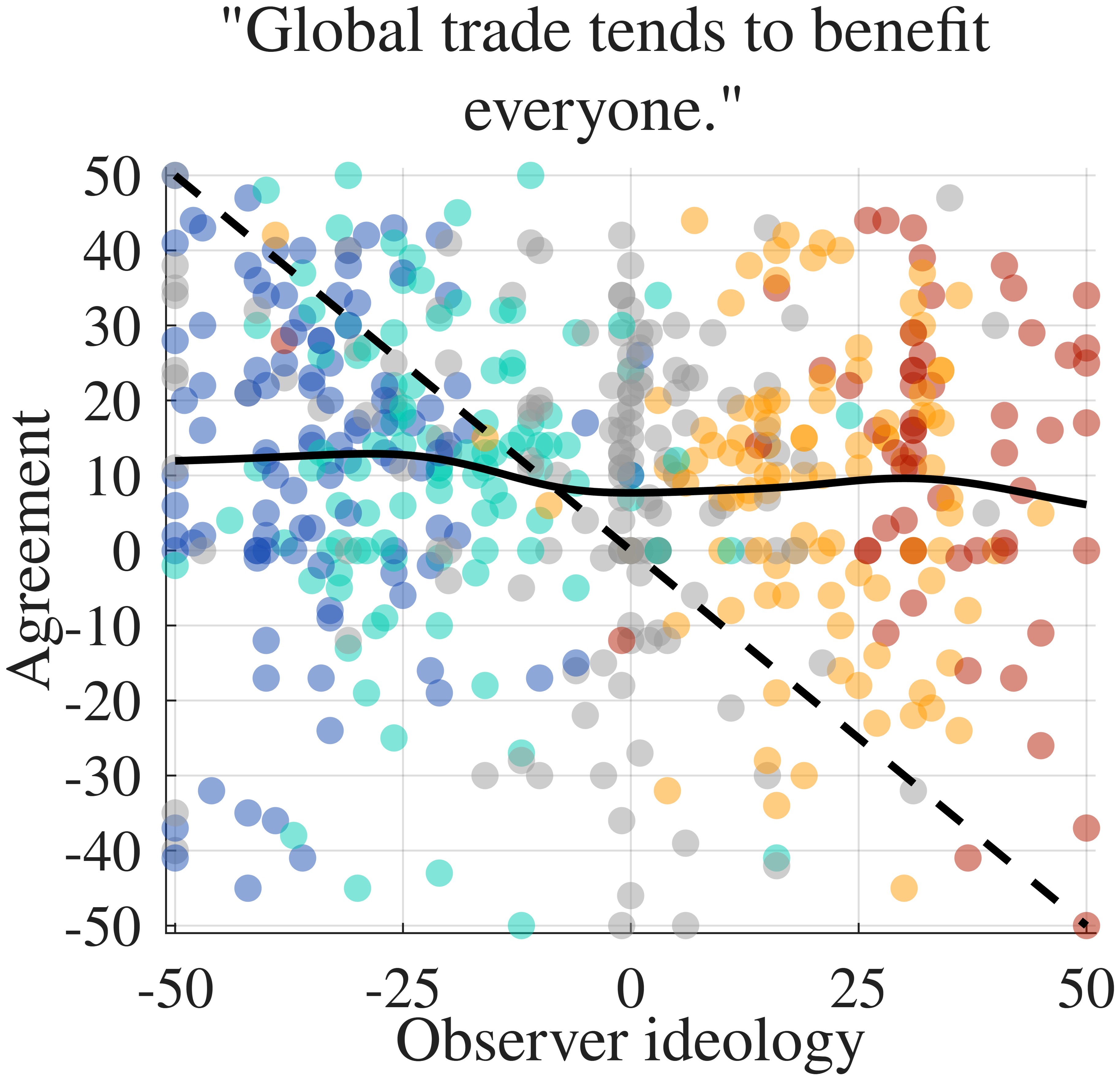} 
    \caption{\textbf{Policy-stance agreement:} 
    Agreement ideology responses by policy area, ordered from most polarized to least polarized (difference between far-left/far-right means) and plotted along with their $\sigma=7$ Gaussian-weighted moving average---compare to Fig \ref{fig:13_major_selfplace_scatters}. 
   }
\label{fig:10_major_agreement_scatters}
\end{figure*}

As Fig \ref{fig:10_major_agreement_scatters} shows, individual policy stances exhibit distinctly different trends from one another, in at least two cases deviating considerably from ``diagonal'' patterns (i.e., agreement-ideology equal to general ideology). 
The location of the curves near or below 0 on the topics of government assistance for the poor and religion in government indicate a broad, if tempered, agreement with the ``liberal" position (favoring more government assistance for the poor, against government informed by religion). 

Abortion, demographic cultural alarm, wealth inequality, assistance for the poor, religiosity in government, and corporate regulation all share a pattern of ``half-indifference" whereby liberals are tightly clustered but conservatives are quite widely spread. Interestingly, while the issues of corporate regulation and especially global trade show relatively flat trends when evaluated by policy-stance, indicating broad agreement with the liberal position across self-identified general ideology, these issues show a pattern fairly close to the diagonal when respondents are asked to use the ideological self-placement scale, as shown in Fig \ref{fig:13_major_selfplace_scatters}. These deviations from the diagonal, particularly amongst conservatives, explain the overall tilt of average policy-agreement ideology from the general ideology diagonal, seen in Fig \ref{fig:ideo_scatter}b. 

Our results suggest that while an even weighting of policy-agreement stances can account for a substantial degree of individuals' self-identification on a general liberal-conservative ideological scale, it may also obscure important variation at the level of individual policy stances.

\subsubsection{Overall ideology is mostly consistent when measured at the beginning and end of the survey.}
To investigate the inherent noise and individual precision of these measurements, we compared respondents' general ideological self-placement at the start and end of the survey (Fig \ref{fig:ideo_scatter}c), and found it to be quite consistent over this short, but ideologically intensive interval: individuals exhibited an average absolute difference of $5.25$ (median absolute deviation $4$, root-mean-squared deviation $\sigma = 8.08$). Rather than a true longitudinal effect (which others have investigated, e.g., \cite{sears1999evidence, berry2007measurement, freeze2016static, zwicker2020persistent}), we intend this comparison to primarily serve as an estimate of the ``fuzziness" of these individually-interpreted quantities on a fine-resolution domain such as ours, to provide important groundwork for future stochastic mathematical-modeling approaches. We consider that on such a near-continuous domain, any measurements are representatives of a narrow but inherently uncertain distribution (due to both psychological quantitative imprecision and indifference to precise slider positioning), whose width we seek.

\subsection{External Ideology Assessments and Reactions}
Next, we present individuals' assessments of political opinion statements, evaluating external stimuli from an ideological standpoint. We established a pool of 68 statements of political opinion---30 ``liberal," 30 ``conservative," and 8 ``centrist," with varying levels of extremity. Participants were shown a random sample of 30 statements from among those 68, and were asked where they would place that statement on the liberal/conservative axis, with a 100-point slider. An overview of the results are shown in Fig \ref{fig:multiscatters} (and in alternate forms in Appendix \ref{sec:alternate_additional_figures}). 
(Full statement list and all results are available alongside this publication.)

\begin{figure*}[h!] 
    \centering
    \begin{overpic}[width=1.1\textwidth]{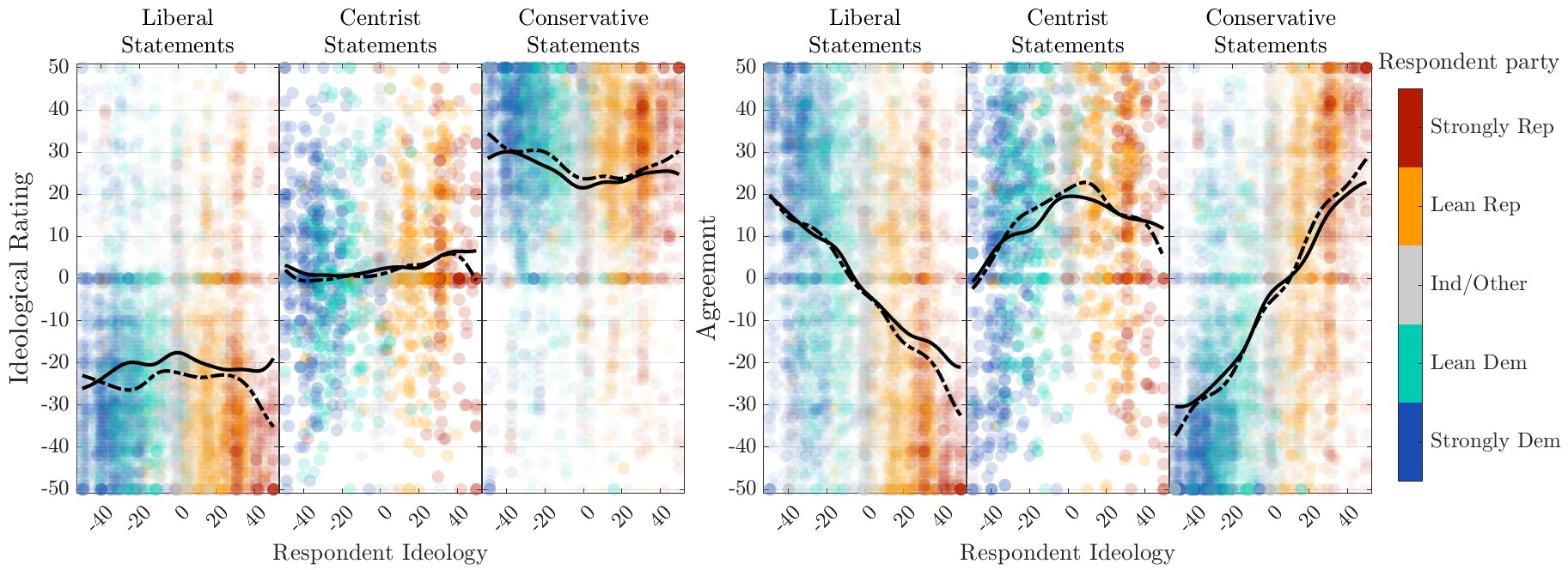}
    \put(5,33.5){\large{\textbf{a}}}
    \put(48,33.5){\large{\textbf{b}}}
    \end{overpic}
    \caption{\textbf{External Assessments:} \textbf{(a)} Average ideological estimation and \textbf{(b)} acceptance of statements as a function of general ideology, divided by statement pool (left/center/right panels). Gaussian-weighted moving average trendlines ($\sigma=7$) are shown for unmarked (solid) and marked (dot-dashed) treatment conditions. \textbf{(a)} We see overall ``flat" trends for the three pools' ideological positions, indicating a general universality of the abstract ideological scale (there remains differences in interpretation between individuals for any given statement---see Fig \ref{fig:assessment_stdev_hist} in Appendix \ref{sec:alternate_additional_figures}---but there does not appear to be systematic observer-ideological bias in those differences). \textbf{(b)} We see clear ``linear" trends in agreement for the liberal and conservative statement pools, and a centrally-peaked trend for the centrist statement pool. 
    \textbf{Comparing trend lines:} We see little, if any, impact of speaker party-identity information in most cases. A slight liberalizing effect on liberal-statement estimation was not replicated in the other sample (see Figs B5 and \ref{fig:mv_assessment_stdev_hist} in Appendix \ref{sec:mv_replication}).
     }
     \label{fig:multiscatters}
\end{figure*}

\subsubsection{Individuals have a largely shared understanding of the ideological spectrum.}
\label{sec:shared_external_ideo}
Respondents show a nearly-identical view of the spectrum as a whole, regardless of ideology or party affiliation: the primary signature in Fig \ref{fig:multiscatters}a is the ``flatness" of ideological ratings within each statement pool. 

However, it is important to clarify that this lack of systematic bias does \textit{not} imply a lack of variance in ideological estimation for each opinion statement: the average standard deviation of ideological ratings of any particular statement was $\mu_\sigma = 19.64$ (distribution visible in Fig \ref{fig:assessment_stdev_hist}a). Thus, the takeaway should not be that political opinion statements have a single agreed-upon ideological value, but rather that (1) they have a particular \textit{distribution} of ideological interpretations that all individuals draw from that does not vary with their own ideology or party affiliation, and (2) the locations of the bulk of these distributions follow a distinct liberal/centrist/conservative ordering that is widely agreed-upon (again regardless of own ideology). See Fig \ref{fig:assessment_stdev_hist} for more details about individual variation.

\subsubsection{Individuals show systematic, symmetric partisan patterns of agreement.}
As Fig \ref{fig:multiscatters}b shows, respondents exhibited a systematic pattern of agreement for each statement pool. The clear emergence of this trend (which is robust to our alternate-sample replication, as seen in Fig B5) may inspire future quantitative exploration of the underlying psychological patterns at work in aggregate underneath these results. Particular issues for particular observers may clearly deviate from this pattern, and of course phrasing and reasoning are large determining factors. However, all other things being equal, estimating the likely distribution of reactions by observers across the political spectrum based on only ideological distance could be a valuable guide for advocacy and outreach efforts.

For an alternative sense of these agreement distributions, see box-plots in Fig \ref{fig:alternate_multibars}.

\subsubsection{External ideological assessments are largely unaffected by speaker party identity.}
Two versions of the survey were administered randomly which differed by one factor: whether statements were presented with the speaker's supposed political party affiliation, e.g., ``A [Democrat/Independent/Republican] says, \ldots" prefacing each statement, with supposed partisan identity corresponding to which ``statement pool" it came from. One hypothesis this was intended to interrogate is whether the advertisement of party identity would provide a significant in-group bias effect, i.e.~that reactions to statements by members of the opposing party would suffer a significant penalty to sentiment and agreement (and conversely, in-group statements would enjoy a boost). 

We found little evidence of such an effect, as shown in all panels of Fig \ref{fig:multiscatters}---comparing solid and dot-dashed trend curves, corresponding to unmarked- and marked-conditioned responses, respectively. In almost all cases, these trend curves show no significant differences (we break the data down by respondent party in Fig \ref{fig:multibars} of Appendix \ref{sec:alternate_additional_figures} and find most all but one party affiliation have insignificant mean-differences).

\subsection{Political Party Identification}
Respondents were asked, ``Which option best describes your political party affiliation/voting tendency?" with 5 options, ``Strongly Democrat," ``Lean Democrat," ``Independent/Undecided/Other," ``Lean Republican," and ``Strongly Republican."

We found little to no overlap in general ideology between the combined Democrat and Republican categories (see Fig \ref{fig:ideo_hists}, top): out of 231 Republican respondents, only 7 self-identified left of the ideological center, while only 10 out of 392 Democrat respondents self-identified right of center. However, overlap did appear when considering average policy-stance agreement ideology, as seen in Fig \ref{fig:ideo_hists}, bottom. In contrast to our segregated and near-symmetric general ideological self-placement distributions, these policy-stance agreement results show a consistently-liberal Democratic party and an only weakly-conservative Republican party.

\begin{figure*}[h!]
    \centering
    \includegraphics[width=1.1\textwidth]{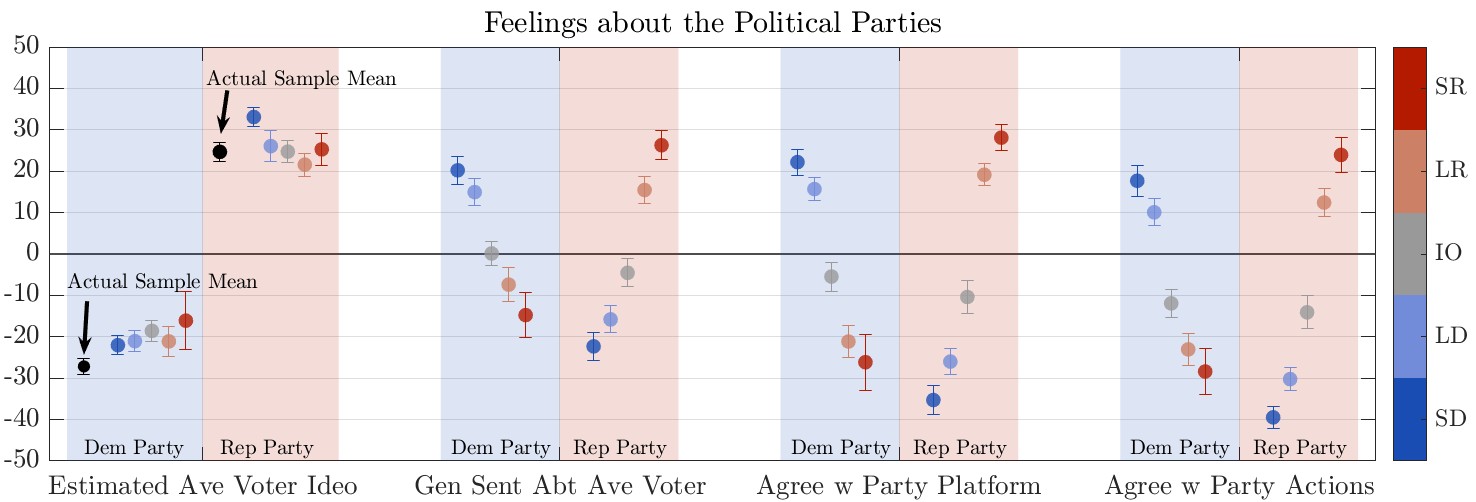}
    \caption{Mean responses to four prompts about the two main political parties (party in question indicated by column shading), broken down by respondent partisanship (dot color). Proceeding from the leftmost column pair, respondents provided \textbf{(1)} their estimation of the ideological position of an average voter for each party, \textbf{(2)} their positive/negative sentiment toward such a person, \textbf{(3)} their agreement with each political party's policy platform, and finally \textbf{(4)} agreement with each party's actions. Vertical scale's meaning is relative to each question (e.g., +32 means ``very conservative" ideologically, ``very positive" sentiment, and ``strongly agree" with party platform or actions). In \textbf{(1)}, along with the estimations of average partisan ideology, the actual ideological means of respondents each party \textit{in our sample} (Strong and Lean combined) are provided in black: Democratic respondents skewed more liberal than all estimates, and Strong Democrat respondents significantly overestimated the conservatism of Republicans. 
    }
    \label{fig:party feelings}
\end{figure*}

In addition, to investigate the perceptions of the political parties themselves, we asked respondents to rate where they believed the average Democrat and Republican voter fell on the ideology scale, their sentiment towards such a person, and their agreement towards the major parties' policy platforms, and finally each party's actions. Mean results for respondents of each affiliation are shown in Fig \ref{fig:party feelings}.

\subsubsection{Strong/Lean/Ind partisan identity well-orders respondent distributions on all ideology measures and reactions.}
In the plots above, respondents are color-coded by party identification. As can be seen via this color-coding, every result above exhibits a strong ordering effect by this 5-way partisan identity, with leaners falling solidly between strong partisans and independents. An important exception to this ordering effect is that external ideological assessments---the placement of statements shown in Fig \ref{fig:multiscatters}, as well as the placement of the average voters shown in the first panel of Fig \ref{fig:party feelings}---appears to be \textit{un}biased by party.

\subsubsection{Individuals are largely accurate at estimating average voter ideology.}

Respondents were generally accurate in estimating the average voter for each party, at least when using our sample's self-report ideology scores as this ``target''---see the leftmost columns of Fig \ref{fig:party feelings}. The exceptions: Strong Democrats significantly overestimated the conservatism of the average Republican, while all groups (especially Strong Republicans) \textit{under}estimated the liberalness of the average Democrat.

\subsubsection{Little clarity on out-party negativity versus in-party positivity.}
In terms of sentiment towards the parties, Democratic respondents (blue dots) showed a pattern of out-party negativity exceeding in-party favorability: relatively muted support of their own party (around $+20$), but vehement opposition to the Republican party (around $-40$). Republican respondents (red dots) showed an up-shifted, more balanced trend with stronger approval of their own party (around $+27$) and less negativity about the Democratic party (around $-27$). Neither Democratic nor Republican respondents show out-party negativity exceeding in-party favorability with respect to average voters in those parties. 

In the replication data-set with Mechanical Turk workers and volunteers (Fig B6), these partisan asymmetries were not present, with all respondents showing the Democratic-respondent pattern of the Prolific data---out-party negativity being much stronger than in-party favorability. 

This mix of results on `out-party hate' versus `in-party love' reflects the wider literature on this topic, as discussed by, e.g. \cite{spinner2024respect, yu2024partisanship}; many observe negative partisanship as stronger, though some find a balanced or reversed pattern \cite{lee2022negative}.

\subsection{Extreme-Value Responses} \label{sec:extreme_responses}
We discouraged respondents from over-using the slider endpoints (with a combination of explicit instructions and nonstandard end-point labels; see Methods), and the resulting frequency with which individuals nonetheless provided extreme answers shows an intriguing pattern. 

\begin{figure*}[h!]
    \centering
    \includegraphics[width=.54\textwidth]{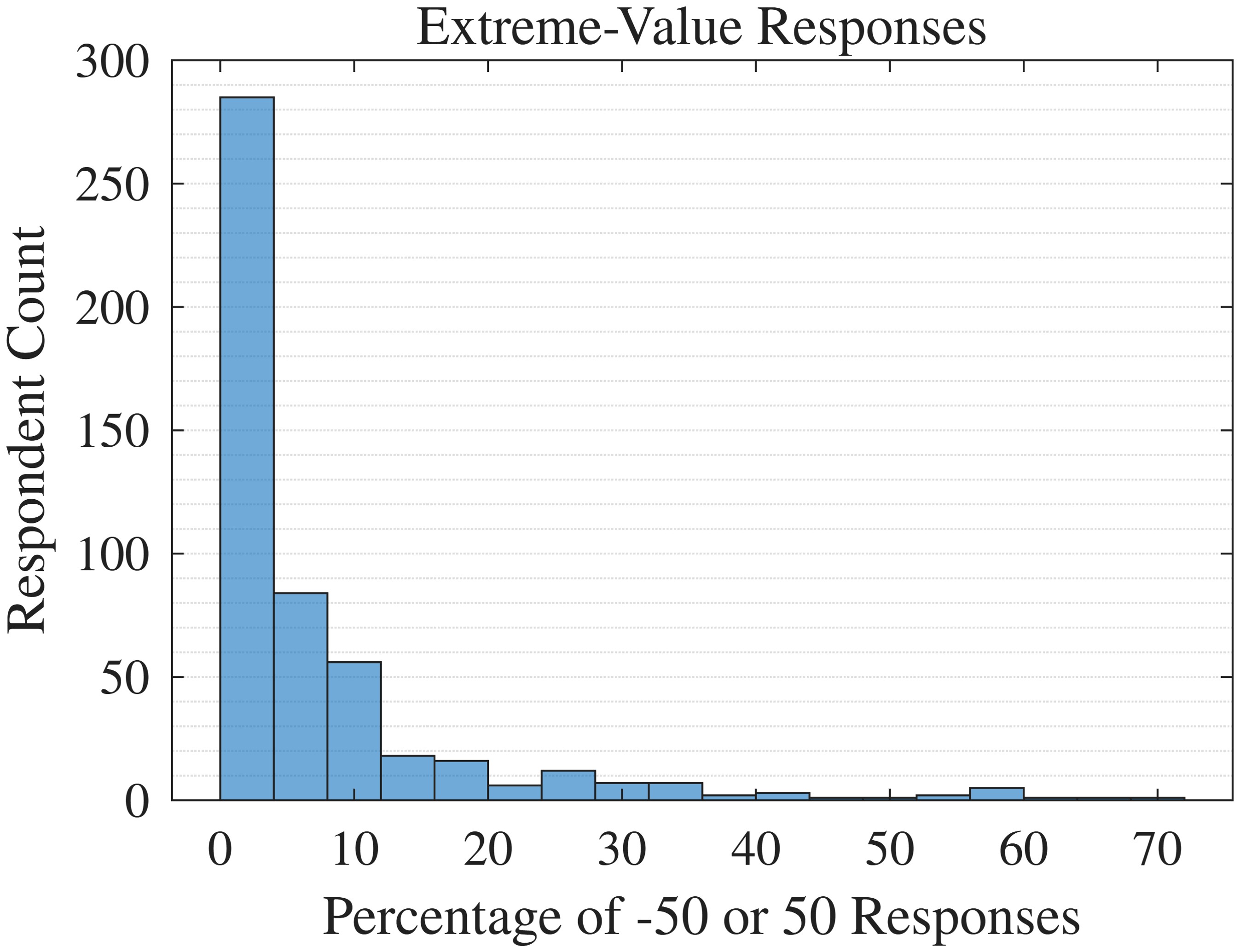}
    \includegraphics[width=.54\textwidth]{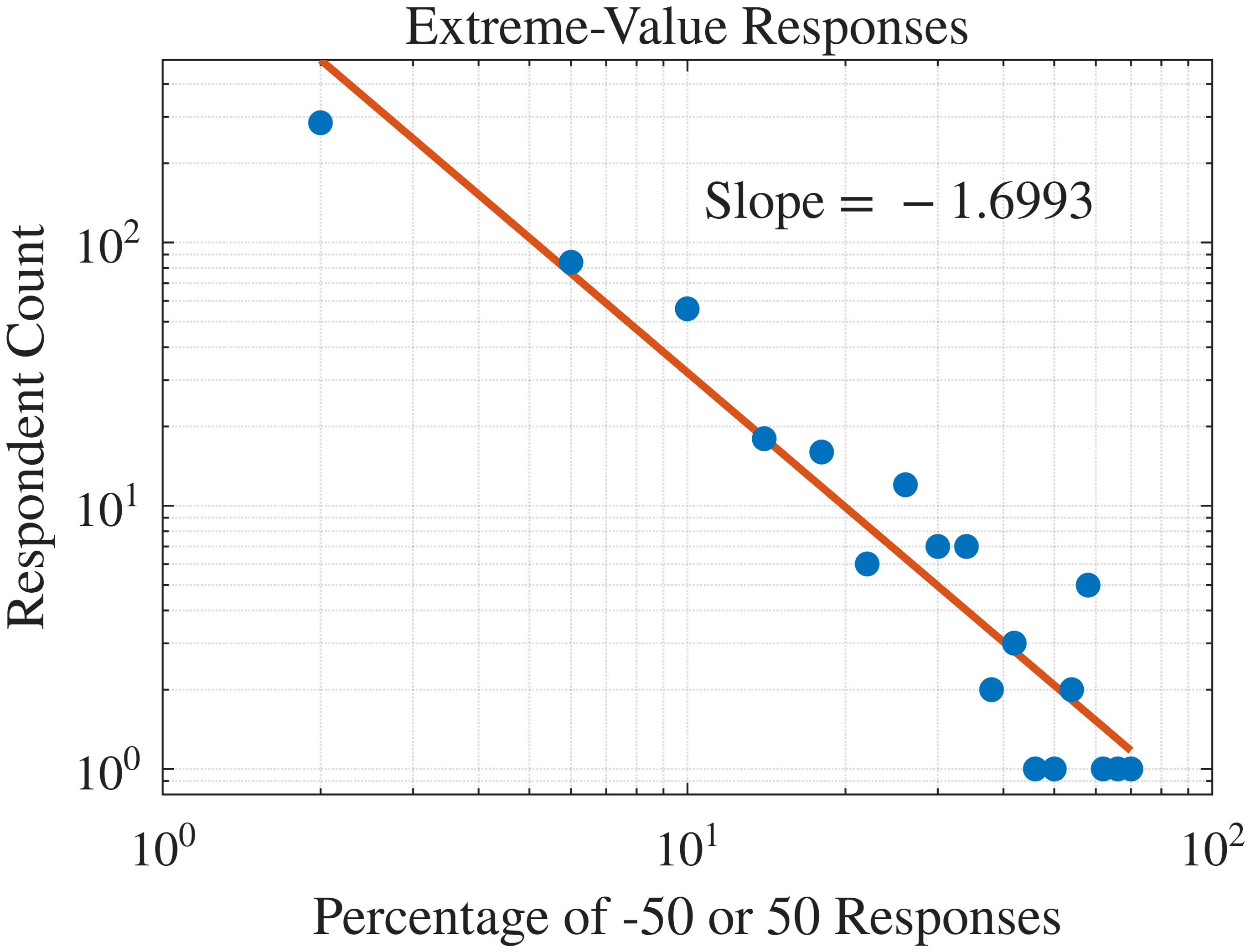}
    \caption{Overall percentage rates of using the endpoints of the slider for external assessments, either for ideological position, agreement, or sentiment. On the log-log plot, we see evidence of a power-law relationship, with an exponent of about $-1.7$.}
    \label{fig:extreme_responses}
\end{figure*}

\subsubsection{The frequency of endpoint-usage rates follows a power-law distribution.}

We tracked what percentage of all external slider responses (ideological estimates, sentiment, and agreement) were at either endpoint---the results are shown in Fig \ref{fig:extreme_responses}. We found that this frequency distribution appeared to follow a power-law relationship with exponent of around $-1.7$. 

\subsubsection{Extreme ideologues use extreme responses more, but outliers exist uniformly throughout.}
\begin{figure}[h!]
    \centering
    \includegraphics[width=.5\columnwidth]{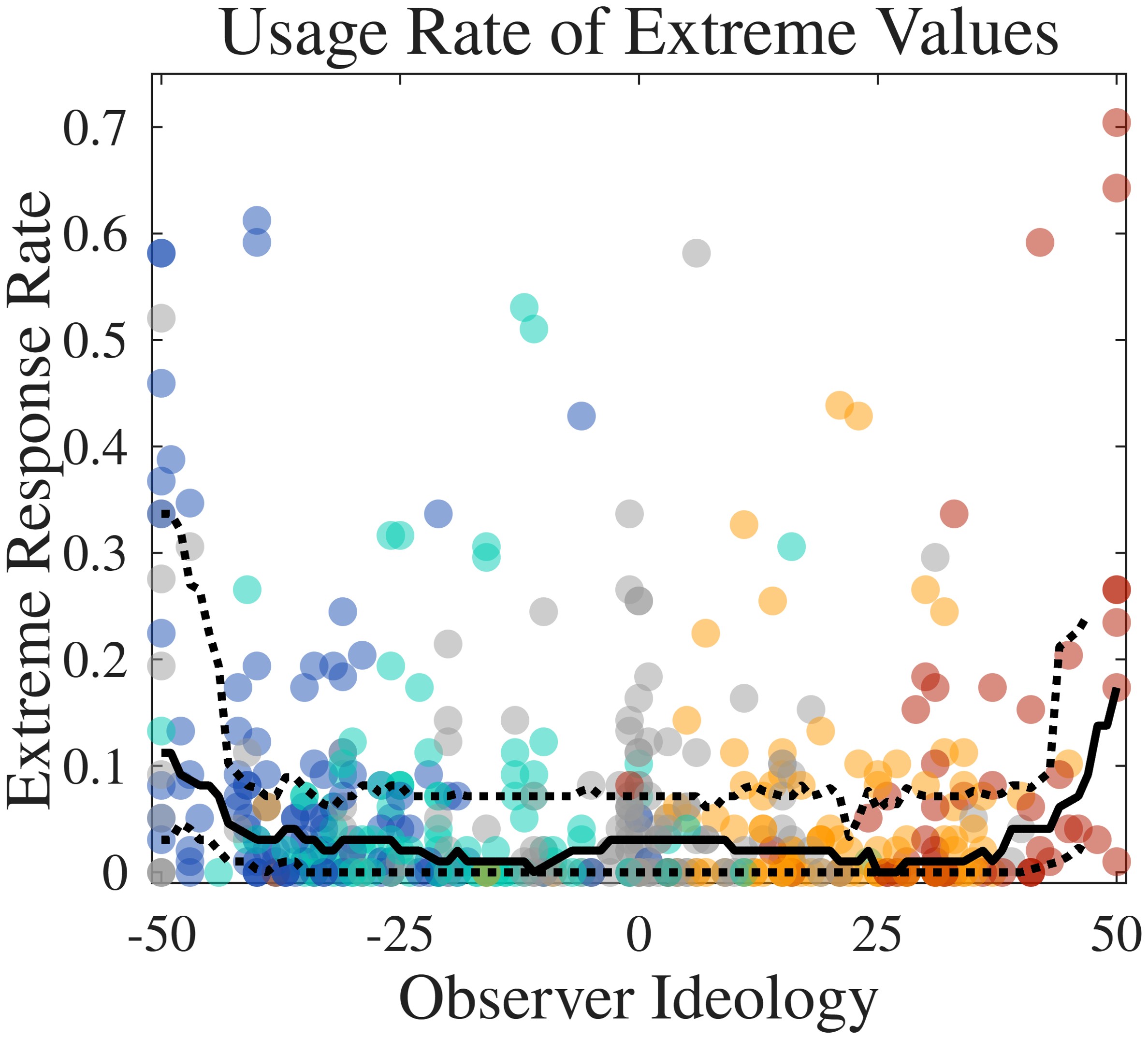}
    \caption{Extreme-value usage rates by ideology, with moving median and 25/75 percentiles (window width = 7) in black. High-usage outliers exist throughout the ideological spectrum, but median rates for non-extreme ideologues are 3\% or lower (moving-average rates, not shown, hover around 6\%). Extreme ideologues, on the other hand, demonstrate a distinctly higher propensity for these extreme judgments.}
    \label{fig:extreme_response_scatter}
\end{figure}

Perhaps unsurprisingly, we found that the individuals who self-identified as ideologically extreme also used extreme values on the other slider responses more often---see Fig \ref{fig:extreme_response_scatter}. However, large extreme-using participants did also exist throughout the ideological spectrum. Overall, 11 out of 508, or approximately 2\% of respondents, used the extreme endpoints more than half the time.

\section{Discussion}\label{sec:discussion}

\medskip\noindent\textit{Summary.} We found that different measures of individuals' political ideology (general self-placement, average issue self-placement, average policy-agreement ideology) broadly agree with one another, though policy agreement is by far the most heterogeneous and issue-dependent. In particular, individuals 
distribute themselves relatively evenly across the political spectrum, while 
policy-agreement ideology puts those same individuals in a single-humped distribution left of ideological center (Figure~\ref{fig:ideo_hists}c and d).

Importantly, we found that the individually interpreted ideological spectrum was remarkably consistent across participants, indicating a collective consensus on a fundamentally abstract construct. This suggests that individuals' own ``projections" of political ideas onto this abstrct, one-dimensional scale are meaningfully comparable across individuals. 

When offered all 5 options transparently, Strong/Lean/Ind party identity well-orders all trends in ideological identification and agreement with policies and statements of political opinion. Attitudes towards political parties were largely symmetric (Figure~\ref{fig:party feelings}). Finally, the prevalence of extreme-responding individuals seems to follow a power law distribution (Figure~\ref{fig:extreme_responses})---being more likely among extreme ideologues but by no means restricted to them (Figure~\ref{fig:extreme_response_scatter}).

Though ostensibly arriving at different conclusions about political ideology among the American public, we view our findings as largely consistent with those of Yeung and Quek \cite{yeung2025self}, who also focus on self-reported measures of ideology. Yeung and Quek find low levels of understanding of textbook definitions of “liberal” and “conservative” in the American public, and show that when asked to place themselves on a scale described only by the theoretical underpinnings of the liberal/conservative framework (social change and active government versus traditional values and limited government) instead of the labels "liberal" and "conservative", respondents self-report as more moderate than when using the traditional ideological measure that presents the labels with no further explanation. These findings regarding the public's relationship to the terminology and theoretical through-lines of liberal/conservative ideology are highly informative, and seem to us consistent with members of the public employing the liberal-conservative framework as a signaling language through which to communicate political views that imperfectly align with elite definitions of the terms.

\medskip\noindent\textit{Agreement vs Sentiment.}
We collected both agreement (``How much do you agree with this statement?") and positive/negative emotional sentiment (``How do you feel about this statement?") throughout the survey, and found that the two tracked each other almost universally---we have displayed only agreement data in the main text for brevity's sake, but the sentiment version of Fig \ref{fig:multiscatters}b is available as Fig \ref{fig:multiscatter_sentiment}, displaying a similar but slightly muted pattern for both datasets. 

\medskip\noindent\textit{Limitations.} 
This survey had several limitations and areas for improvement for similar data-gathering efforts in the future. 

The survey used language ``liberal'' and ``conservative'' for salience and clarity to participants due to their ubiquity in U.S.~political parlance, but these might be alternately described as ``left-wing'' and ``right-wing'' for concordance with international (as opposed to just American) political-spectrum terminology---``liberalism'' indeed has a broader definition outside of the United States, which could leave some more global-politics-minded respondents confused or conflicted. 

Next, while our results indicate that the one-dimensional `political spectrum' is remarkably consistent within and between individuals, our survey did not investigate alternate, higher-dimensional ideology spaces. A survey set up to record two- or higher-dimensional abstract positions (e.g., `political compass'-type economic-redistribution and authoritarian axes) may explain more variance in political identification. However, the increased burden of data collection, questions about assessment when statements may not engage with all axes, and difficulty in visualizing results may be a challenge. 

Additionally, these results are cross-sectional rather than longitudinal, leaving theories of ideological \textit{dynamics} at best indirectly informed. However, the accumulation of this type of high-resolution, transparent quantitative data on political attitudes, reactions, and preferences will gradually strengthen the ability to validate or falsify theories of political ideology and its dynamic changes.

\medskip\noindent\textit{Future work.} This survey was constructed with mathematical modeling in mind, with the goal of interrogating basic consistency of foundational variables like ideology, and seeking broad-stroke, robust patterns that can inform future models of political psychology. We note that the application of unsupervised clustering or other machine-learning analysis to these data may yield some additional ``bottom-up" insights in this regard, though we leave such investigations to future work. We hope this work facilitates further mathematical modeling of the dynamics of political ideology. The apparent robustness of general ideological self-placement lets it serve as a (noisy) variable in nonlinear, complex, and dynamical models which may interrogate novel hypotheses at the psychological (micro) and societal (macro) levels. Theoretical mathematical ``opinion dynamics" and ``sociophysics" models which have relied on abstract opinion variables may iterate with the perspective that contention with sufficiently high-resolution real-world data (for certain generally understood concepts) is indeed possible, and may build those models with such data-accountability in mind.

\section{Methods}\label{Sec:Methods}
This survey was created to elucidate the perceptions and reactions of individuals exposed to political statements of position. For increased resolution and ease of quantitative trend-seeking, all answers except party identification were entered by a 100-point slider. 

Several efforts were made to discourage respondents from over-using the extremes of the response scale, in order to resolve a wider range of true reactions by reserving truly extreme positions and emotions. To this end, first, the following disclaimer preceded the survey: 

``For this survey, please try \textbf{not} to use the extreme values very often---they should represent what you believe to be truly extreme views (e.g.~inclined to drastic action or violence), or highly emotional/zealous mental states.'' 

Second, the reference labels provided along with each slider included non-standard and more emotionally salient language on the ends: 
\begin{itemize}
    \item For ideological placement (e.g.,~``Rate where you think this statement falls on a Liberal/Conservative axis''), the markers were ``Extremely Liberal,'', ``Very Liberal,'' ``Somewhat Liberal,'' ``Unsure/Centrist,'' ``Somewhat Conservative,'' ``Very Conservative,'' and ``Extremely Conservative.''
    \item For agreement (e.g.,~``How much do you agree with this statement?''), the markers were ``Vehemently Disagree,'' ``Strongly Disagree,'' ``Somewhat Disagree,'' ``Unsure/Indifferent,'' ``Somewhat Agree,'' ``Strongly Agree,'' and ``Emphatically Agree.''
    \item For sentiment (e.g.,~``How do you feel about this statement?''), the markers were ``Hatred/Disgust,'' ``Very Negative,'' ``Somewhat Negative,'' ``Indifferent,'' ``Somewhat Positive,'' ``Very Positive,'' and ``Fervent/Impassioned.'' 
\end{itemize}
However, there was no indication of what exact position corresponded to each label, so individuals were encouraged to position sliders smoothly anywhere between these labels. The seven labels were approximately at the locations $\pm 48, \pm 32, \pm 16$, and $0$.

The survey started with an assessment portion, aimed at measuring respondents' ideological position in three different ways, for comparison: self-placement overall, self-placement on thirteen salient political issues, and agreement with a slate of ten broad statements on a similar slate of issues. These measures were compared to assess the accuracy and consistency of self-report with researcher-computed ideological positions. 

The ``general ideological self-placement'' score was their response to the question, ``Below is a scale on which the political views that people might hold are arranged from extremely liberal to extremely conservative. When it comes to politics, where would you place yourself on this scale?''

The thirteen issues for more granular ``issue self-placement": 
\begin{itemize}
    \item Issues around race
    \item Issues concerning homosexuality
    \item Changing one's gender/transgender issues
    \item Abortion
    \item Welfare programs
    \item Military spending
    \item Immigration
    \item Corporate regulation
    \item Global trade
    \item Wealth inequality
    \item Religion in government
    \item Environmental regulation
    \item Ease of voting
\end{itemize}
These were presented in a randomized order, and the ``average self-placement" ideology estimate was the mean of these self-placement scores.

The ten representative statements for ``policy-stance agreement" ideology estimation were:
\begin{itemize}
    \item ``The government should spend more money helping poor people."
    \item ``We need to maintain or increase our military spending and strength."
    \item ``Wealth inequality is a big problem."
    \item ``Our government should be informed by religious ideas."
    \item ``Regulations on corporations are generally good."
    \item ``Abortion should be restricted by the government."
    \item ``LGBTQ accommodations should be expanded."
    \item ``Global trade tends to benefit everyone."
    \item ``I'm happy to pay more taxes for more government benefits."
    \item ``Demographic change is threatening American culture."
\end{itemize}
These statements were likewise presented in a random order. To construct average policy-agreement ideology, the scores for ``liberal-aligned'' statements were sign-flipped and the mean was taken.

Participants were also asked their party affiliation, i.e.~``Which option best describes your political party affiliation/voting tendency?" and the following options:
\begin{itemize}
    \item Strongly Democrat
    \item Lean Democrat
    \item Independent/Undecided/Other
    \item Lean Republican
    \item Strongly Republican
\end{itemize}
This intentionally differs from the standard two-part ANES question, which first asks participants to choose between Democrat/Independent/Republican and then ask (only) the Independents to choose a ``lean'' direction. We believe that offering all options from the start is both simpler and more transparent. 

Respondents then rated where they believed the average Democrat and Republican voter was on the ideology scale, their sentiment towards such a person, and their agreement towards the major parties' policy platforms, and their agreement with each party's \textit{actions}. 

This was followed by the main portion of the survey, where a random sample of thirty out of sixty-eight statements were shown to each participant. Respondents were randomly assigned to a control condition, where the statements were displayed on their own, or a treatment condition, where statements were framed as coming from a speaker of a particular political affiliation (e.g.,~``A Democrat says, `...' ").

The statement pool included thirty ``liberal'' statements, thirty ``conservative'' statements, and eight ``centrist'' statements. This pool of statements was created to represent positions encountered across the political spectrum in late 2022/early 2023, emulating how a politically opinionated person might express their position online or in person. These statements were intended to cover the ideological spectrum as evenly as possible.

\medskip\noindent\textit{Respondent Composition.} \label{sec:respondent_composition}
The respondents whose data are shown in the main text comprised an age/sex/party representative sample gathered through the Prolific platform between May 14, 2024 and May 28, 2024. See Appendix \ref{sec:mv_replication} for the alternate-sample replication of all main-text figures.

\medskip\noindent\textit{Human Subjects Ethics Statement.} \label{sec:human_subjects}
This study was approved as oversight-exempt by the University of Michigan Institutional Review Board, as it was an anonymous survey and no identifiable information was gathered. Regardless, all participants granted written informed consent for their response data to be gathered, analyzed, and shared for academic purposes.


\subsection*{Acknowledgments}
The authors thank Drew Trygstad for his help with preparing the survey and figures, Kirill Truntaev for designing and implementing an interactive website, and Daniel Abrams for his assistance with the original mathematical model which inspired this investigation. Additionally, the authors' sincere gratitude goes out to all volunteers who offered their time and attention to contribute to this endeavor, as well as to the two anonymous reviewers whose helpful comments substantially improved the manuscript.

\bibliography{refs}

\clearpage

\setcounter{figure}{0} 
\renewcommand\thefigure{A\arabic{figure}}    
\appendix

\section{Alternate and Additional Figures} \label{sec:alternate_additional_figures}

Here we offer alternative perspectives on several main-text figures, for transparency and deeper analysis. 
\subsection{Response distributions by political party affiliation}
\begin{figure*}[h!] 
    \centering
     \includegraphics[width=1.1\textwidth]{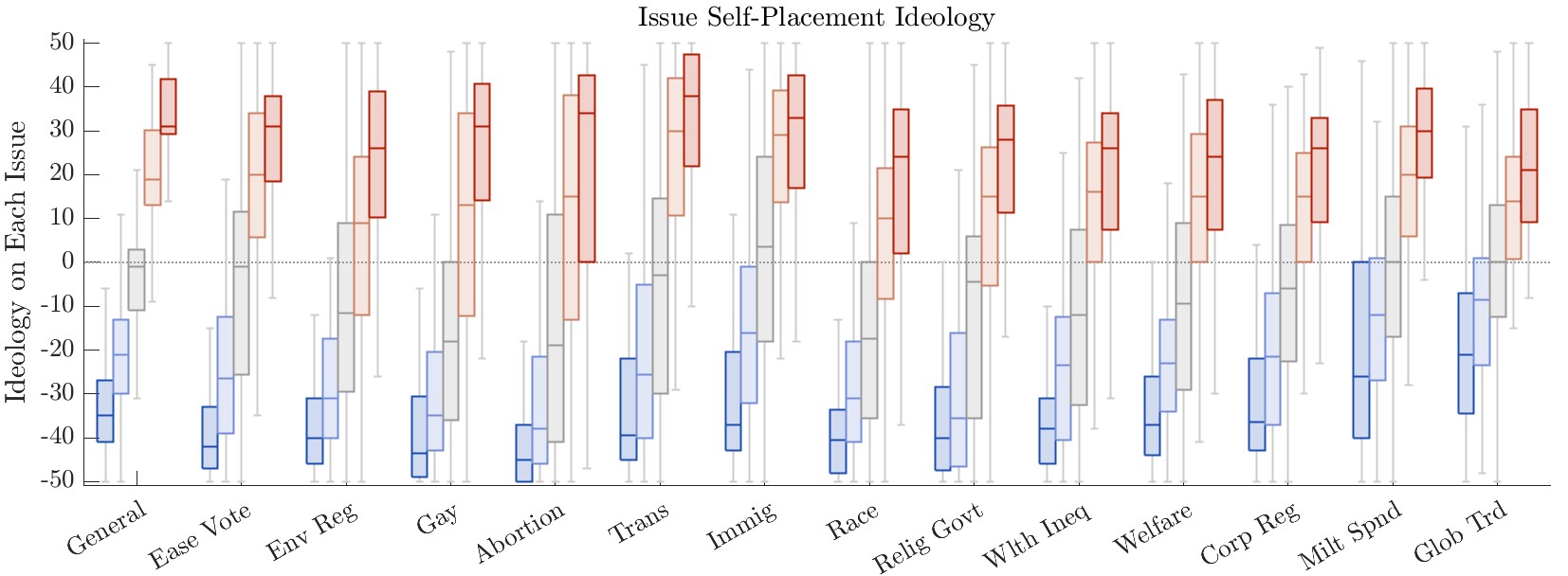}
    \caption{\textbf{Self-Placement Ideology Distributions by Issue and Party.} ``Major Issue Self-Placement" distributions grouped by party, with 13 issues ordered by decreasing polarization (Strong Dem/Strong Rep mean responses---this party-based sorting results in a slightly different ordering than the ideology-based panels in Fig \ref{fig:13_major_selfplace_scatters}). General ideological self-placement distributions by party are also provided on the far left for comparison. 
    }
    \label{fig:13_major_selfplace_boxplots}
\end{figure*}

\begin{figure*}[h!]
    \centering
    \includegraphics[width=1.1\textwidth]{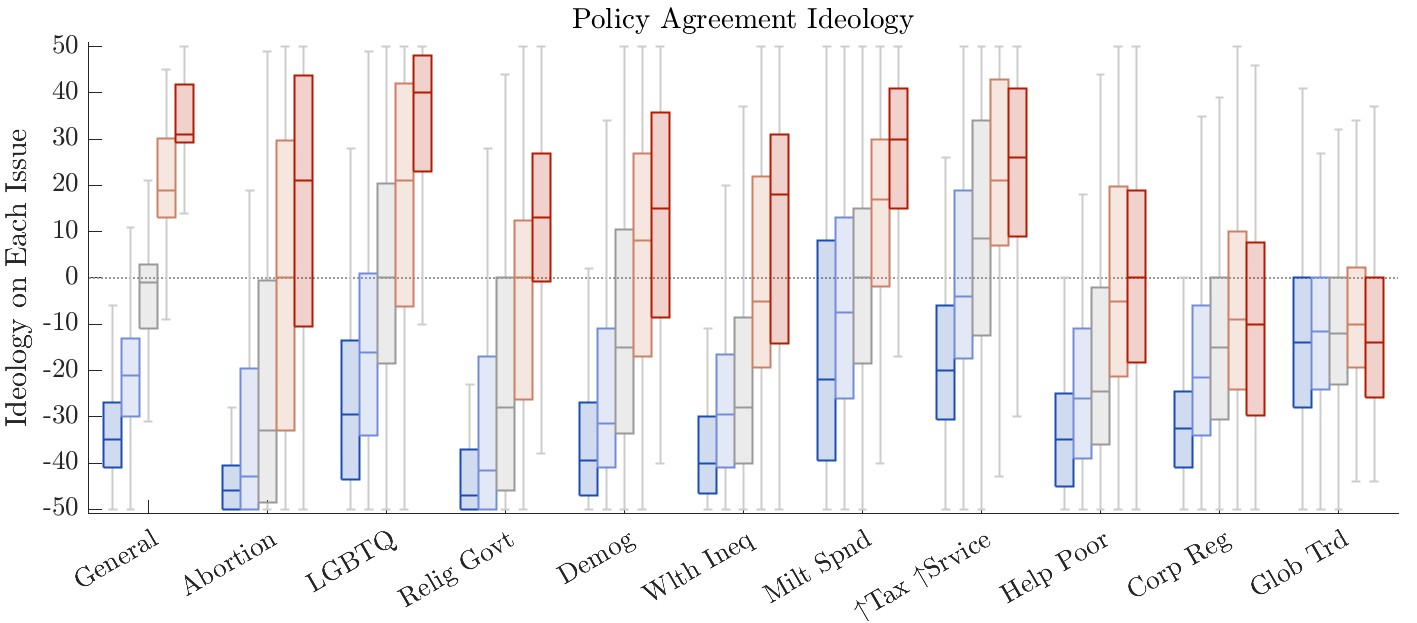}
    \caption{\textbf{Policy-Stance Agreement Ideology Distributions by Issue and Party:} Data from Fig \ref{fig:10_major_agreement_scatters}, divided by respondent party, and ordered from most polarized to least polarized (Strong Dem/Strong Rep mean difference). We can see the more issue-specific patterns manifest again when compared to Fig \ref{fig:13_major_selfplace_boxplots}.
    }
    \label{fig:10_major_agreement_boxplots}
\end{figure*}

\begin{figure*}[h!] 
    \centering
    \begin{overpic}[width=1.1\textwidth]{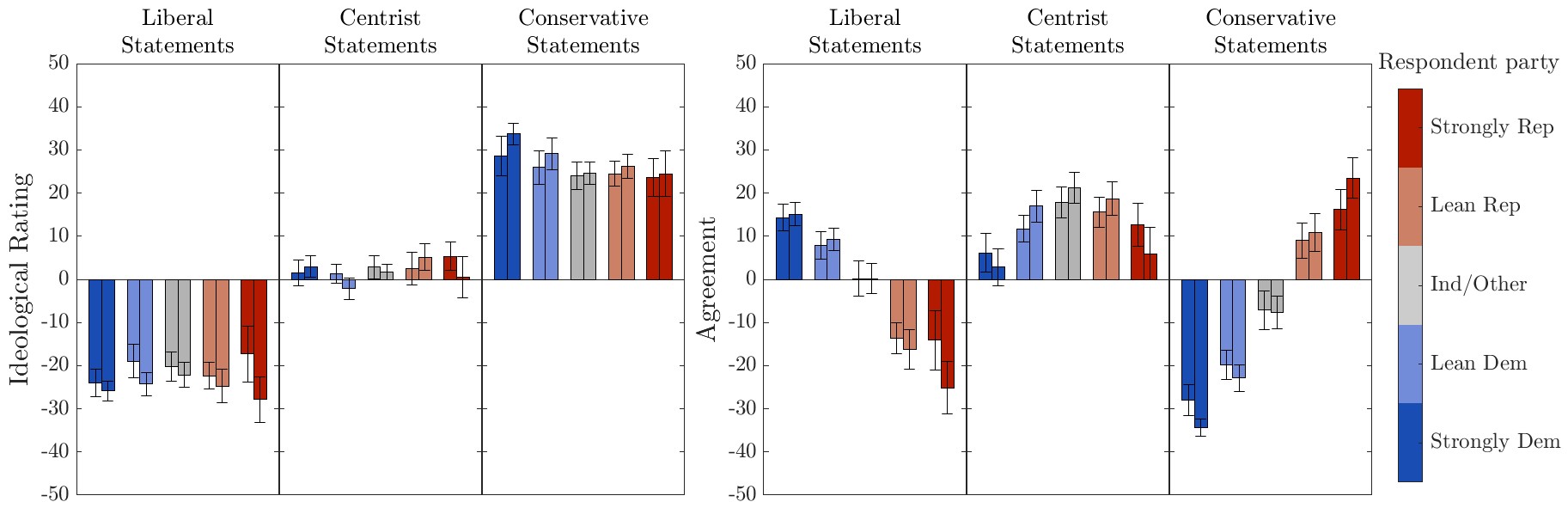}
    \put(5.5,26.5){\large{\textbf{a}}}
    \put(49.5,26){\large{\textbf{b}}}
    \end{overpic}
    \caption{\textbf{Bar-graph view of data in Fig \ref{fig:multiscatters}, by party:} \textbf{(a)} Average ideological estimation and \textbf{(b)} acceptance of statements, divided by statement pool (left/center/right panels) and respondent party affiliation (color). Error bars indicate 95\% confidence intervals (clustering errors by individual in Stata). We see the same overall ``flat" and ``linear" trends in each panel.
    \textbf{Comparing bar pairs:} The right bar of each pair corresponds to those viewing the ``identity-marked'' version; we see little, if any, impact of this information in most cases.  
    }
     \label{fig:multibars}
\end{figure*}

\begin{figure*}[h!]
\centering
\ \ \ \includegraphics[width=1.1\textwidth]{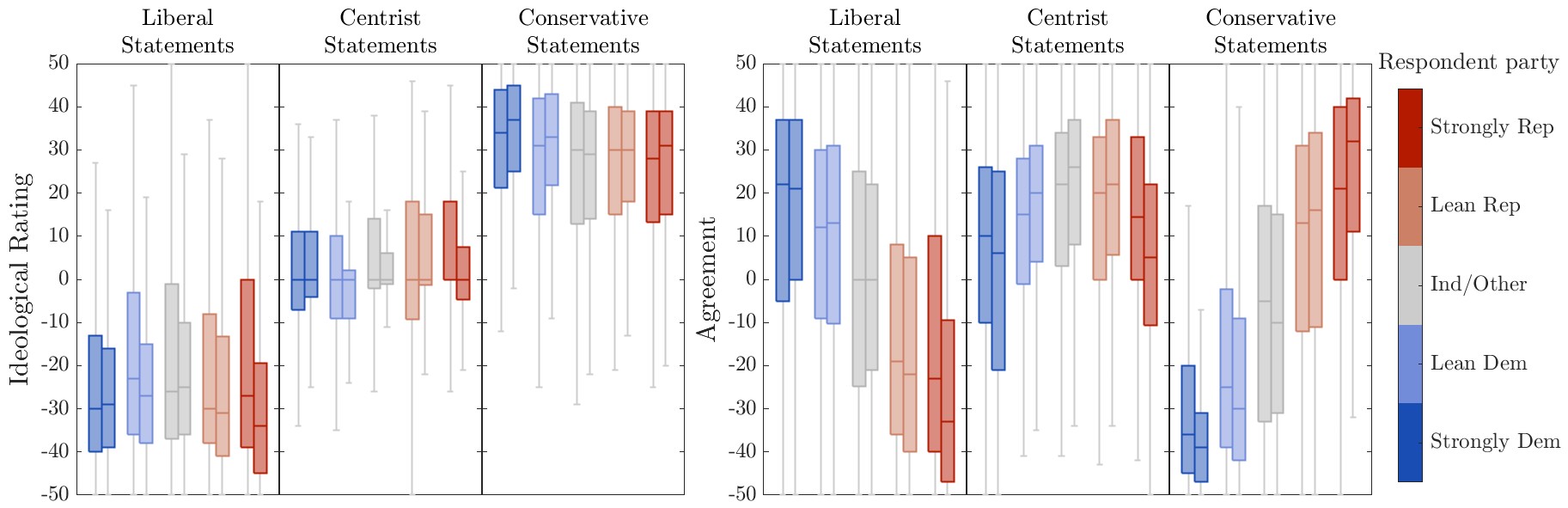}\\
\caption{\textbf{Box-plot view of data in Fig \ref{fig:multiscatters}, by party.} Box-plot view of the distributions of responses in Fig \ref{fig:multiscatters} broken down by political party affiliation and marked/unmarked condition. We can see the same predominant patterns: ``flatness" in the left panels signifying non-bias of observers in external ideological assessments, and quasi-linear self-favoring trends on the right for agreement with those statement pools. 
}
\label{fig:alternate_multibars}
\end{figure*}

Figs \ref{fig:13_major_selfplace_boxplots} through \ref{fig:alternate_multibars} present alternative breakdowns of response distributions by political party affiliation, rather than general ideology. Due to the strong relationship between ideology and party, the resulting patterns are very similar.\\

Fig \ref{fig:multibars} shows the mean response values for each party of respondent and each panel of Fig \ref{fig:multiscatters}. The left and right bars of each pair correspond to the unmarked and marked treatment conditions, respectively. Standard errors were calculated with clustering by individual in Stata. The one mean-difference that is significant at the $0.05$ level according to this analysis is strong Republicans' agreement (or rather, degree of disagreement) when viewing Democrat content, which was more negative/stronger disagreement for marked statements ($p=0.03$)---although this was not replicated in the other sample (see Fig \ref{fig:mv_multibars} in Appendix \ref{sec:mv_replication}). It is possible that additional differences between marked and unmarked statements could become more detectable at a larger sample size, underscoring the need for further work in this area.

Fig \ref{fig:alternate_multibars} displays more information about the entire distribution of responses underlying Fig \ref{fig:multibars}, rather than just the means and their standard errors.

\subsection{Individual Bias in External Ideology Estimation}
\begin{figure*}[htp] 
    \centering
    \begin{overpic}[width=.54\textwidth]{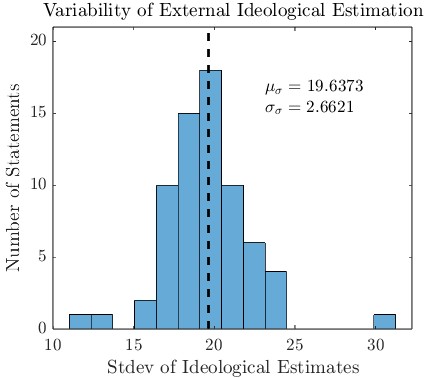}
    \put(14,78){\large{\textbf{a}}}
    \end{overpic}
    \begin{overpic}[width=.54\textwidth]{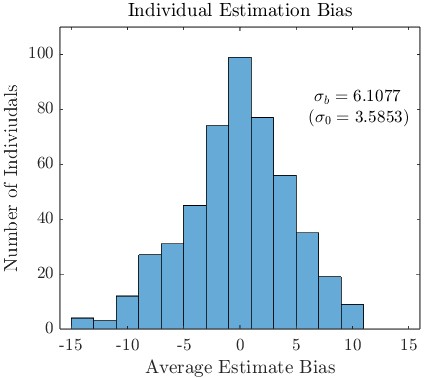}
    \put(16,76){\large{\textbf{b}}}
    \end{overpic}
    \caption{\textbf{External Ideology Assessment: Variability by Issue, and Individual Bias.} \textbf{(a)} Histogram of standard deviations for ideological assessment of the 68 political opinion statements. The upper outlier was a statement sympathetic to the USSR, which each side believed to be ideologically aligned with the other. \textbf{(b)} Average estimation bias by individual (ideology estimate - average ideology estimate, averaged across statements). The observed spread of this net bias is $\sigma_b = 6.11$, to be compared to the expected independent/null value $\sigma_0 = 3.59$; the mismatch indicates that there is some slight individual systematic bias in ideological estimation.}
     \label{fig:assessment_stdev_hist}
\end{figure*}

In Section \ref{sec:shared_external_ideo} in the main text we showed that estimated ideology distributions themselves aren't biased by ideology (or, as Fig \ref{fig:multibars}a shows, by party). However, the spread of ideological estimation around the mean value is considerable, as demonstrated by Fig \ref{fig:assessment_stdev_hist}a, which displays the standard deviation of ideological position estimates for each of the 68 statements (average per-statement standard deviation $\mu_\sigma = 19.64$). Furthermore, the selection of ideology from each corresponding distribution isn't entirely independent, as we can see by plotting individual average deviation/``bias" in Fig \ref{fig:assessment_stdev_hist}b---as a rough estimate, if individuals independently assigned ideologies with $\sigma=19.64$ for their 30 statements, we might expect their average ``bias" to have spread $\sigma_0 = 19.64/\sqrt{30} = 3.59$, and in Fig \ref{fig:assessment_stdev_hist}b the observed spread is $\sigma_b = 6.11$, indicating that individuals exhibit correlated deviations.

\setcounter{figure}{0} 
\renewcommand\thefigure{B\arabic{figure}}    

\section{Replication Figures} \label{sec:mv_replication}
Here we include copies of all figures in the main text (and a few alternate figures from Appendix \ref{sec:alternate_additional_figures}), utilizing instead the combined data from N = 166 US-resident Mechanical Turk Masters and N = 130 volunteers (of which 90 were undergraduates at a large midwestern university). These data were gathered first as a pilot study, and (while not a nationally representative sample) share nearly all the major effects and non-effects the Prolific sample exhibited. We make note of the few differences below.

\begin{figure*}[h!]
    \centering
    \begin{overpic}
        [width=.54\textwidth]{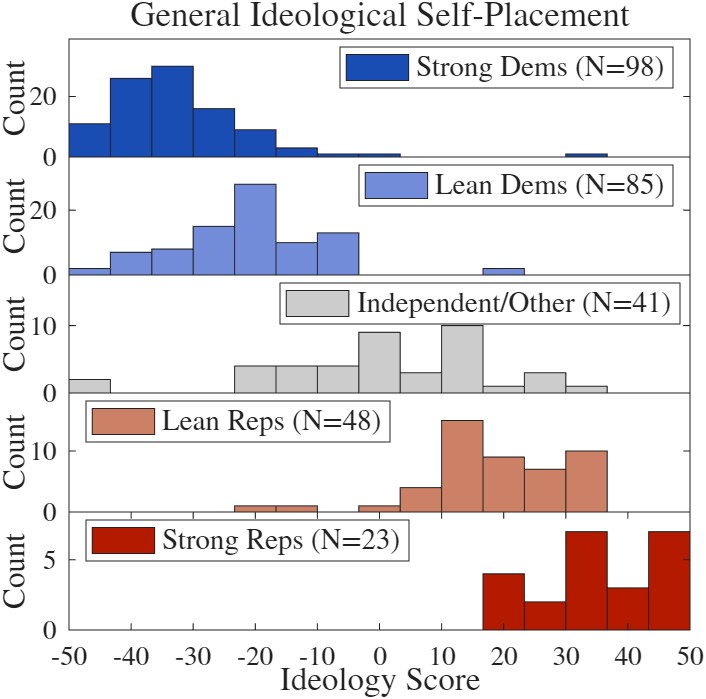}
        \put(8,95){\large{\textbf{a}}}
    \end{overpic}
    \begin{overpic}
        [width=.54\textwidth]{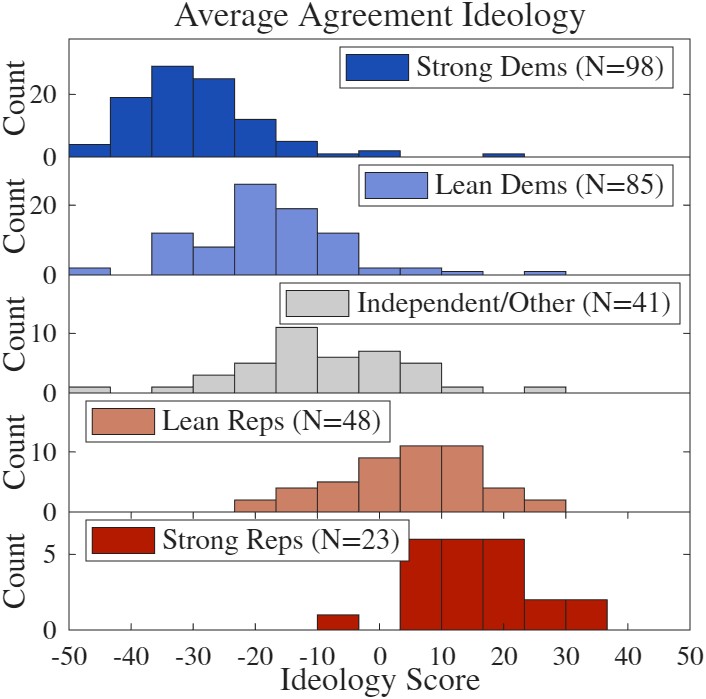}
        \put(8,95){\large{\textbf{b}}}
    \end{overpic}
    \\
    \vspace{2mm}
    \begin{overpic}
        [width=.54\textwidth]{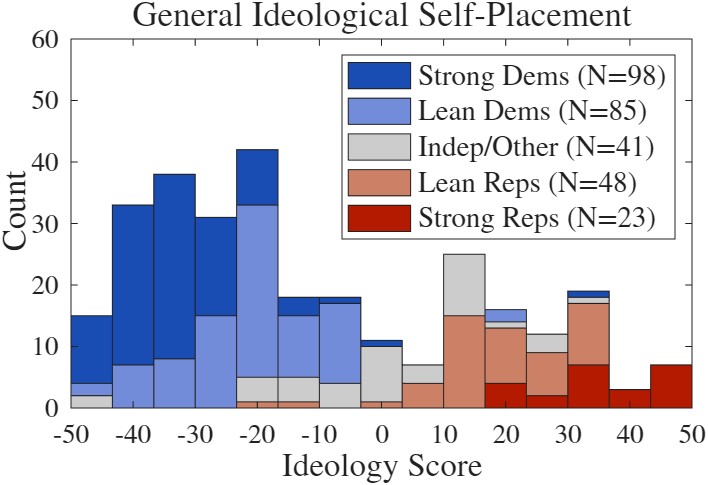}
        \put(8,64){\large{\textbf{c}}}
    \end{overpic}
    \begin{overpic}
        [width=.54\textwidth]{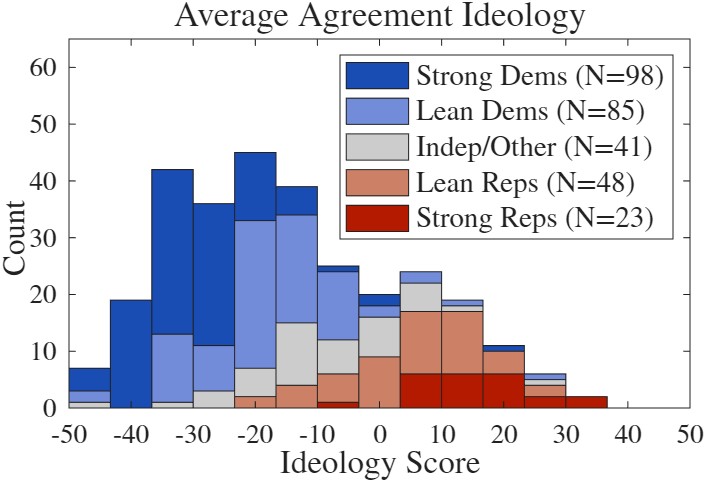}
        \put(8,64){\large{\textbf{d}}}
    \end{overpic}
    \caption{Version of Fig \ref{fig:ideo_hists}
    displaying data from volunteers and Mechanical Turk Masters. The same general pattern (well-distinguished party distributions, symmetrically spread for self-report and skewed liberal for agreement-average) is repeated. However, the large spike of Independents choosing exactly zero general ideology is not present, perhaps indicating a more nuanced/less ``na\"ive" respondent pool.  This shows the liberal skew in voluntary respondents, which was present in both sub-samples. Even after restricting Mechanical Turk respondents to self-identified US conservatives, responses stalled before anything close to parity was achieved.}
    \label{fig:mv_ideo_hists}
\end{figure*}

\begin{figure*}[h!] 
    \centering
    \begin{overpic}[width=1.1\textwidth]{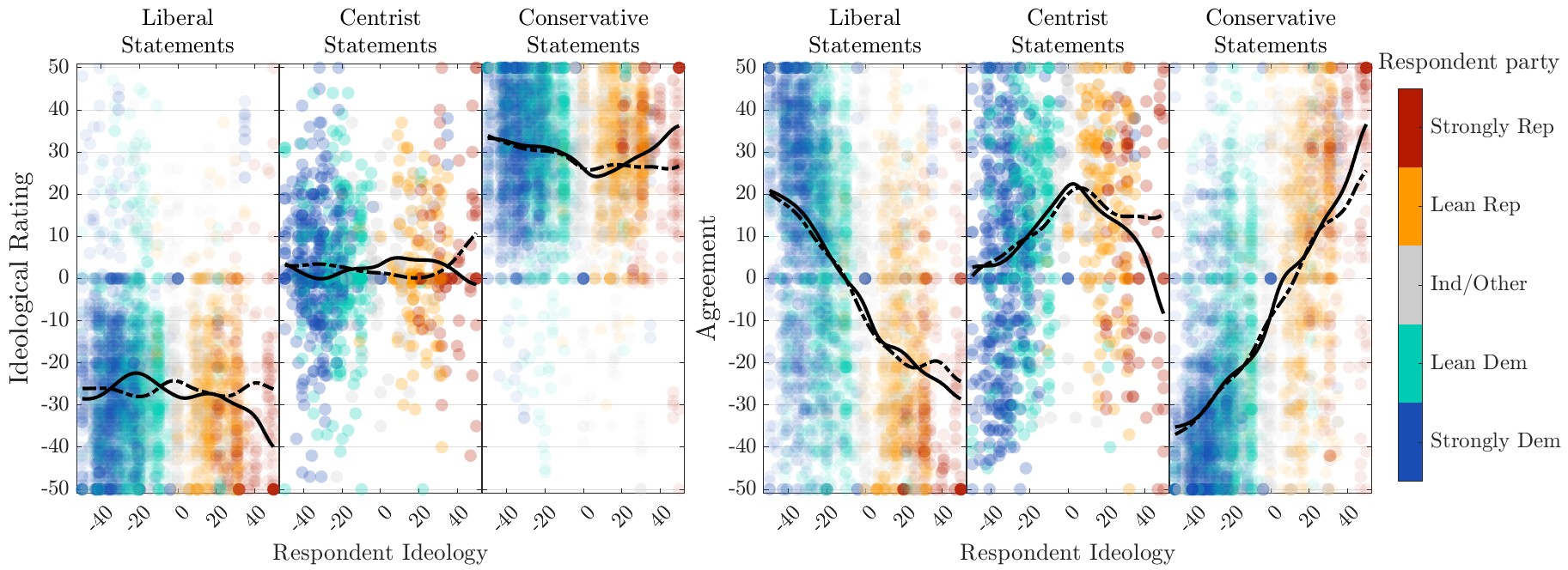}
    \put(5,33.5){\large{\textbf{a}}}
    \put(48,33.5){\large{\textbf{b}}}
    \end{overpic}
    \caption{Version of Fig \ref{fig:multiscatters} 
    displaying data from volunteers and Mechanical Turk Masters. This sample does not replicate any marked-unmarked condition differences between the trend-lines; in fact, several deviations are reversed. However, note that the extreme conservative side of these graphs are of very low certainty due to the small sample size there.
     }
     \label{fig:mv_multiscatters}
\end{figure*}

\begin{figure*}[h!]
    \centering
    \includegraphics[width=1.1\textwidth]{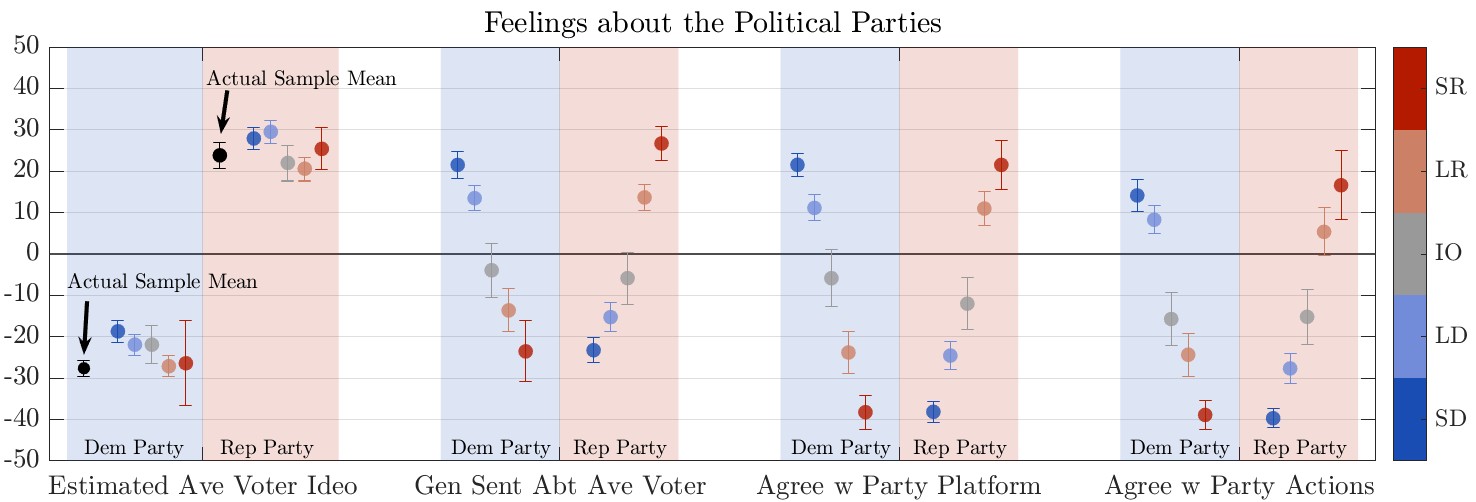}
    \caption{Version of Fig \ref{fig:party feelings} displaying data from volunteers and Mechanical Turk Masters. We see the same general patterns as Fig \ref{fig:party feelings}, but with near-perfect symmetry replacing the Republican upshift seen in that other dataset---here, both sides display the patterns Democrats did there.}
    \label{fig:mv_party feelings}
\end{figure*}

\begin{figure*}[h!] 
    \centering
    \begin{overpic}[width=1.1\textwidth]{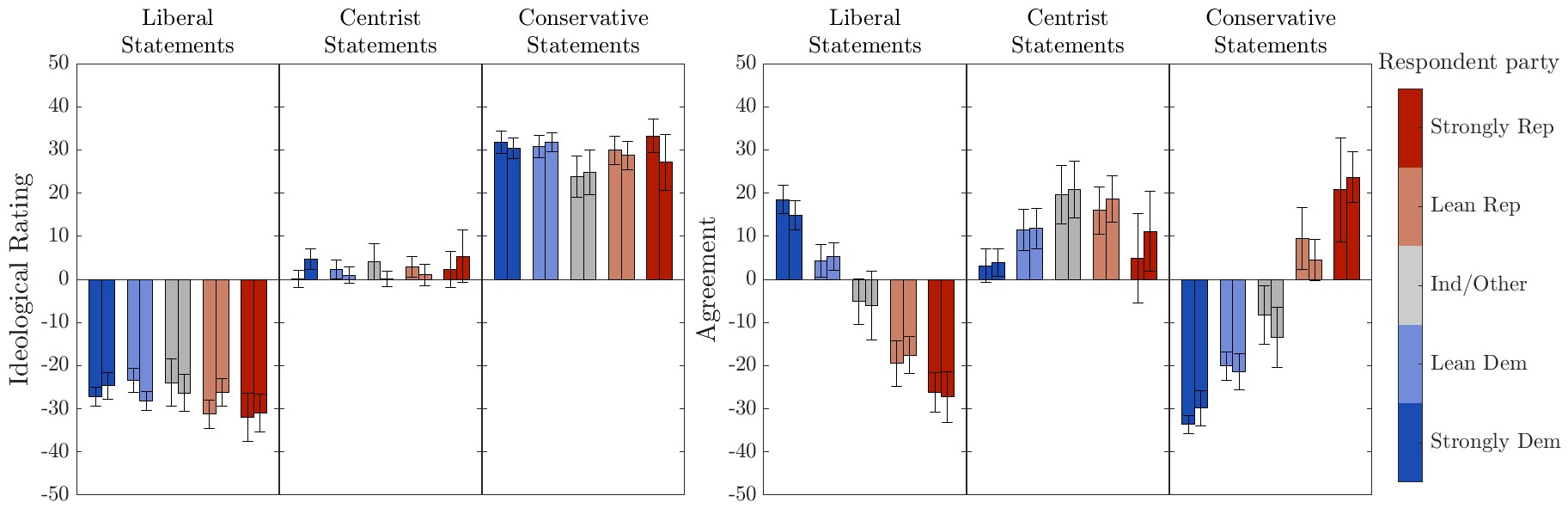}
    \put(5.5,26.5){\large{\textbf{a}}}
    \put(49.5,26){\large{\textbf{b}}}
    \end{overpic}
    \caption{Version of Fig \ref{fig:multibars} displaying data from volunteers ($N_V=130$) combined with  Mechanical Turk Masters ($N_M = 166$), with $95\%$ confidence intervals clustered by individual in Stata. We see essentially no signature of any treatment effect (comparing left and right members of each pair of bars). Many even exhibit a slight sign of the \textit{opposite} effect---i.e., the marked condition (right bar) has more \textit{moderate} ideology estimates, and more neutral agreement---though these opposite effects lack significance.
     }
    \label{fig:mv_multibars}
\end{figure*}

\medskip\noindent\textit{Sample Differences.} \label{sec:sample_differences}
This pilot sample (volunteers and Mechanical Turk Masters) skewed significantly liberal and Democratic overall (as seen in Figs \ref{fig:mv_ideo_hists}c and d), and notably did not show  the minor subgroup-specific treatment effects (see Fig \ref{fig:mv_multiscatters}, comparing trend curves, and \ref{fig:mv_multibars}a, comparing bar pairs). Nor did they share the asymmetric feelings about the political parties that the Prolific respondents did (see Fig \ref{fig:mv_party feelings}). This highlights the considerable degree of variability between subsets of the population---for instance, Prolific estimates that their respondents are more ``na\"ive" than Mechanical Turk workers, which may explain the increased susceptibility to speaker-identity framing bias among strong Republicans in the Prolific sample.

\begin{figure*} [h!]
    \centering
    \begin{overpic}
        [width=1.1\textwidth]{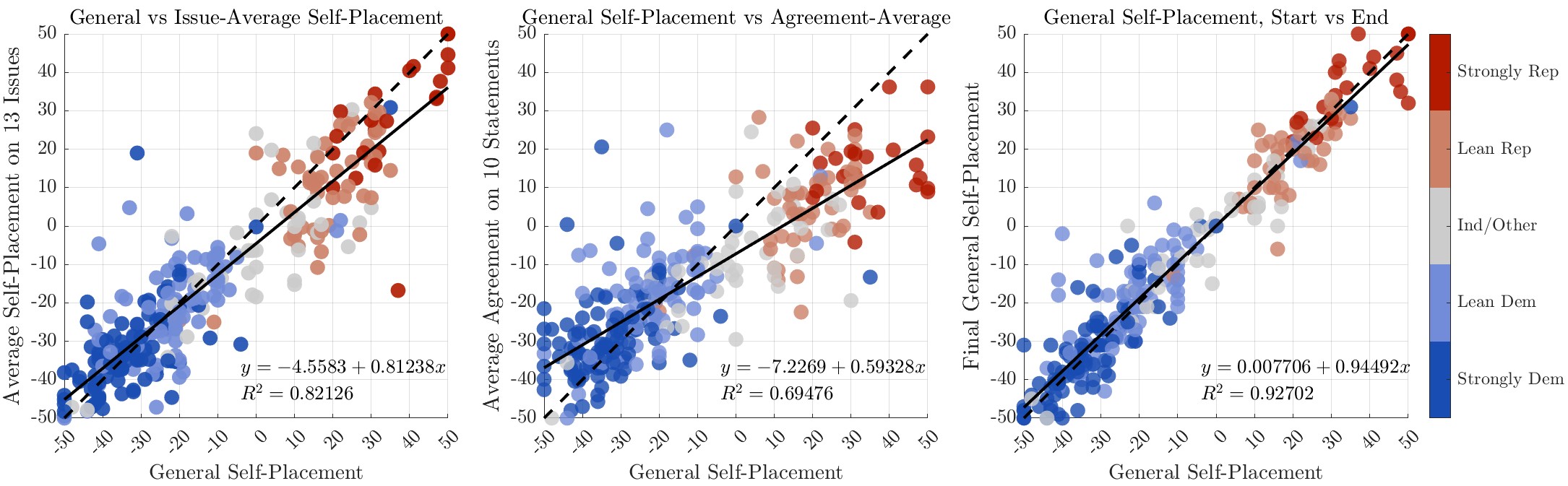}
        \put(4.5,27){\large{\textbf{a}}}
        \put(35,26.8){\large{\textbf{b}}}
        \put(65.7,27){\large{\textbf{c}}}
    \end{overpic}
    \caption{Version of Fig \ref{fig:ideo_scatter} displaying data from volunteers and Mechanical Turk Masters. The same overall patterns---good alignment with average issue-ideology, weaker but still considerable alignment with agreement-average ideology, and strong self-consistency over time---are repeated.}
    \label{fig:mv_ideo_scatter}
\end{figure*}

\begin{figure*}[h!] 
    \centering
    \includegraphics[width= 0.3\textwidth]{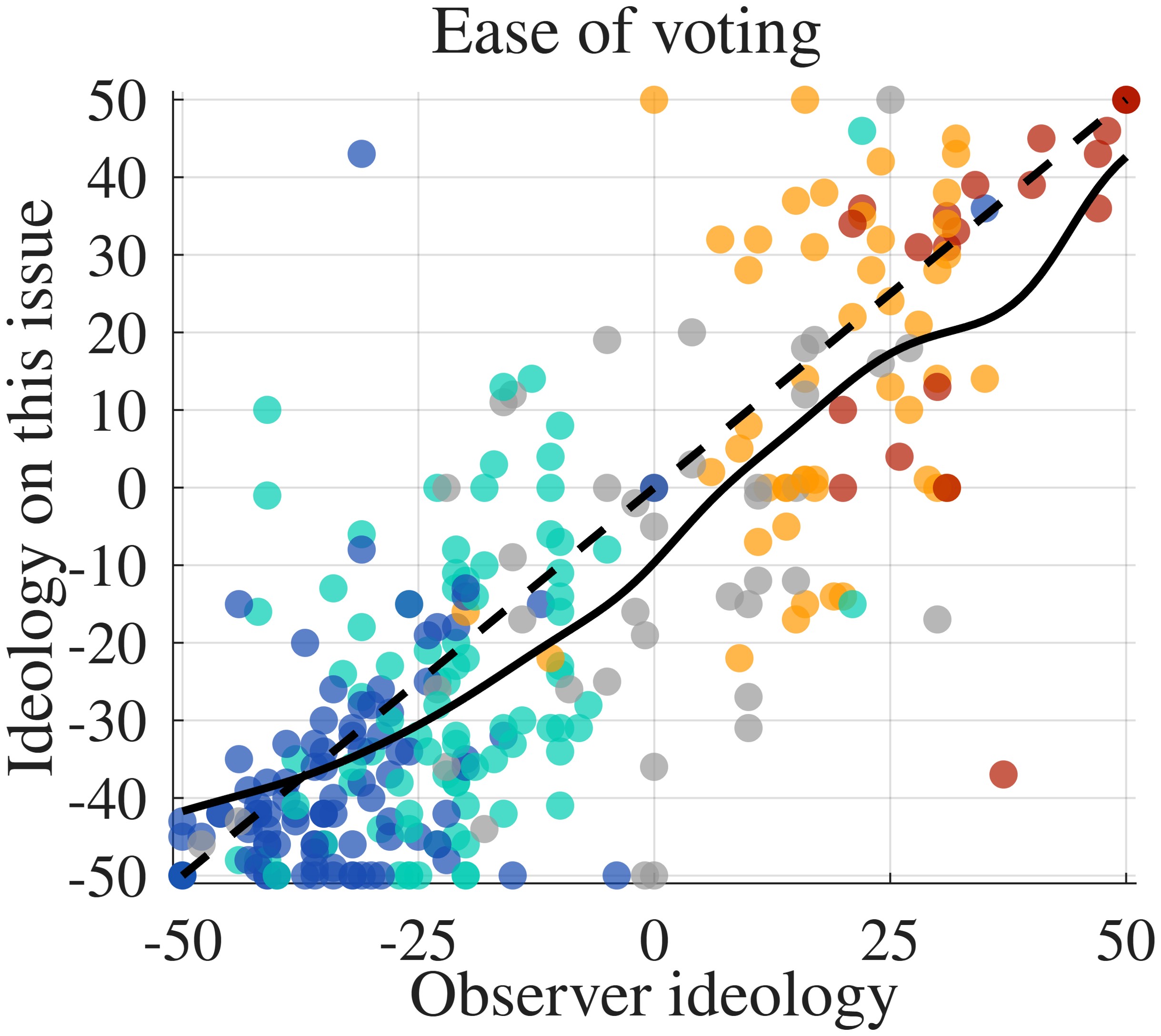}
    \raisebox{0.1\height}{\includegraphics[width= 0.095\textwidth]{jpgs/colorbar_short.jpg}}
    \\
    \includegraphics[width= 0.26\textwidth]{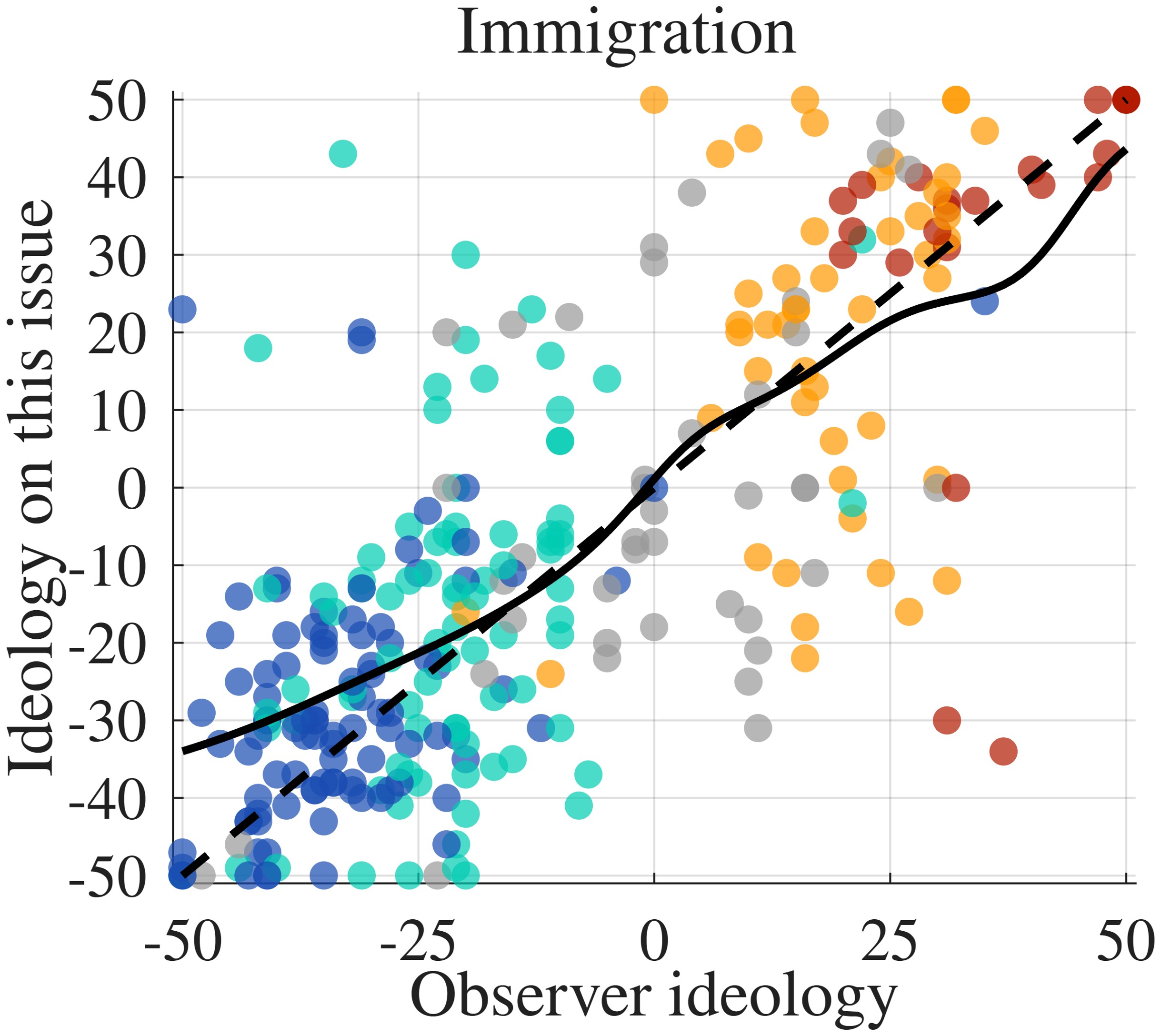}
    \includegraphics[width= 0.26\textwidth]{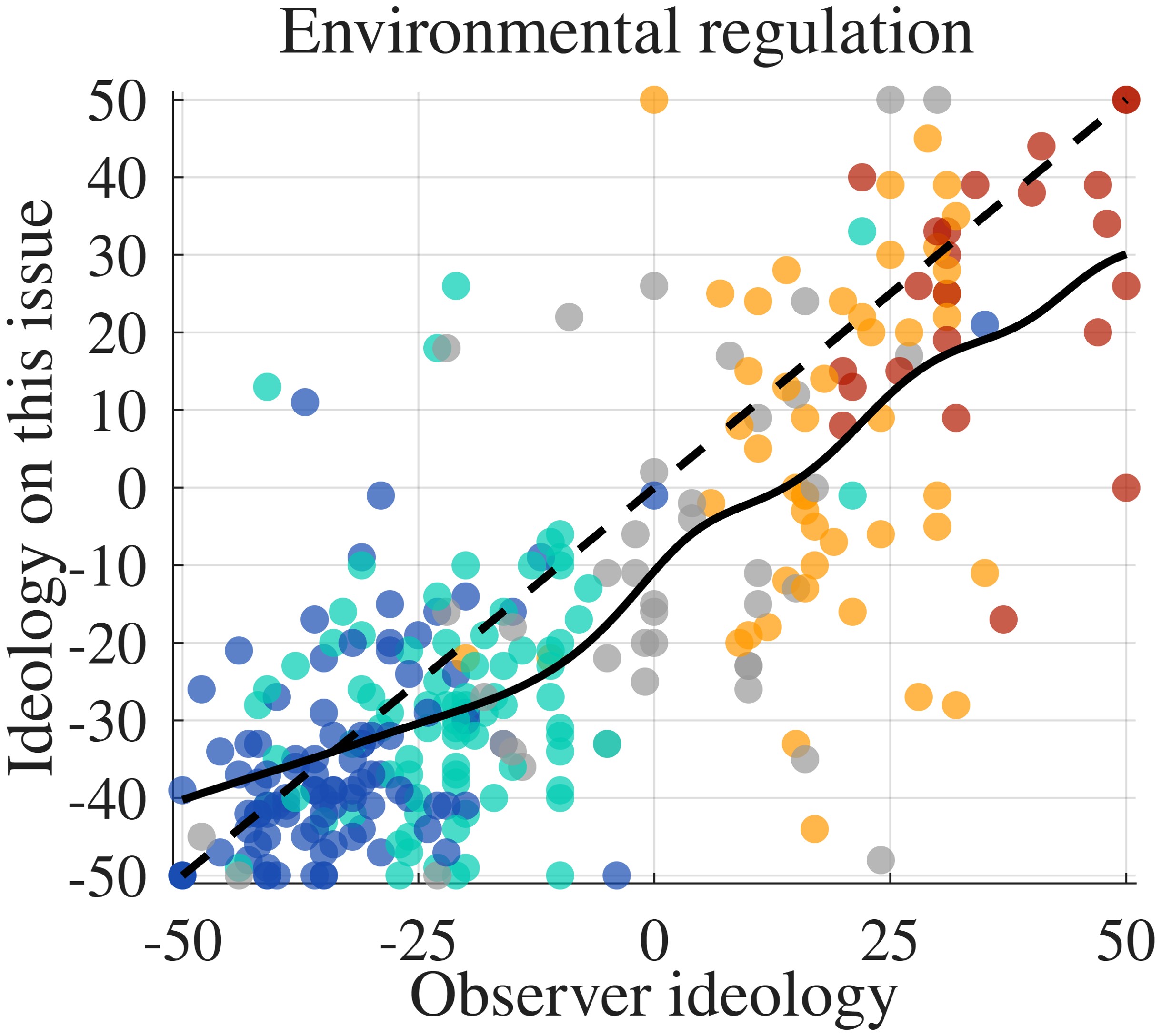}
    \includegraphics[width= 0.26\textwidth]{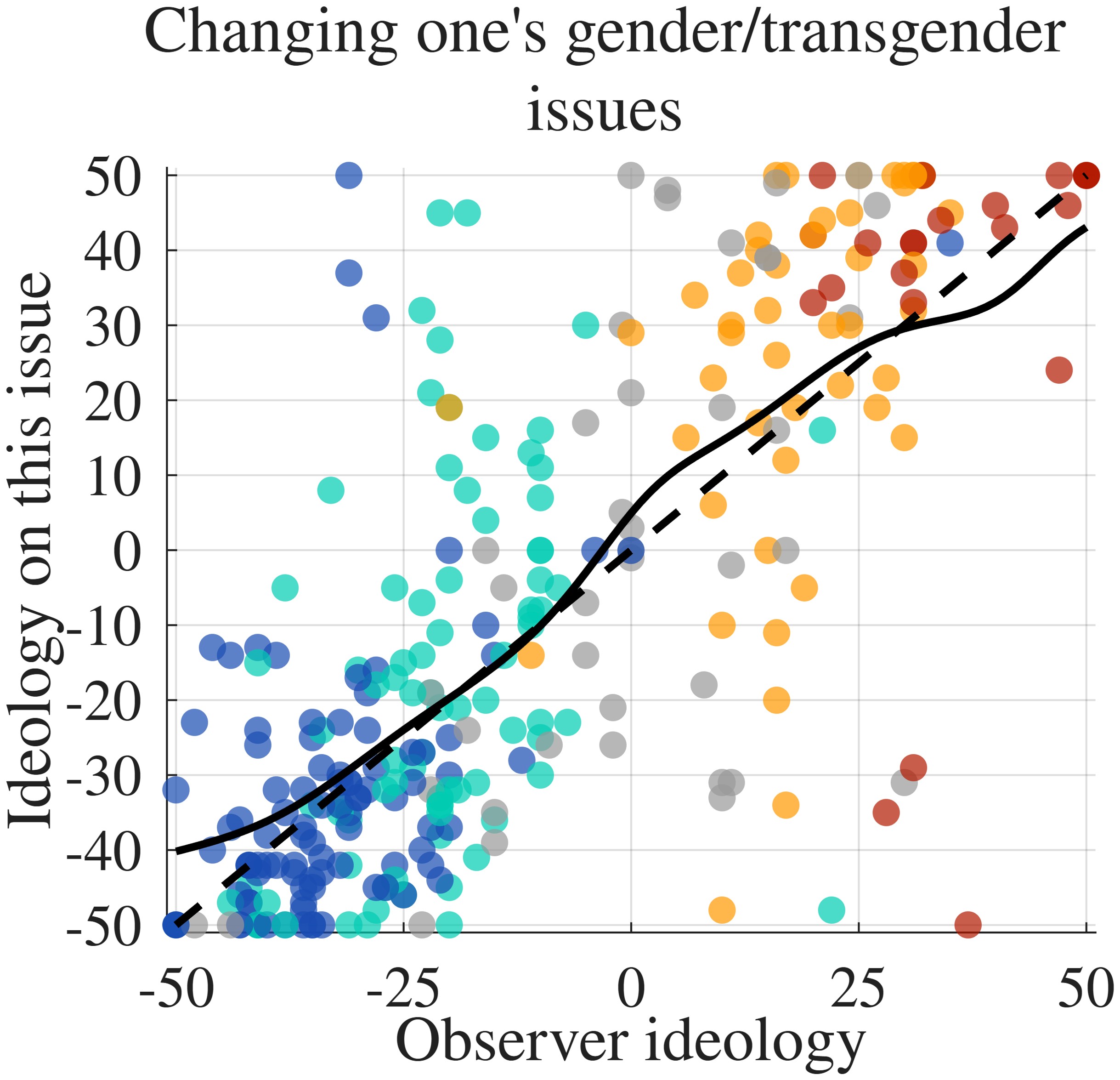}
    \includegraphics[width= 0.26\textwidth]{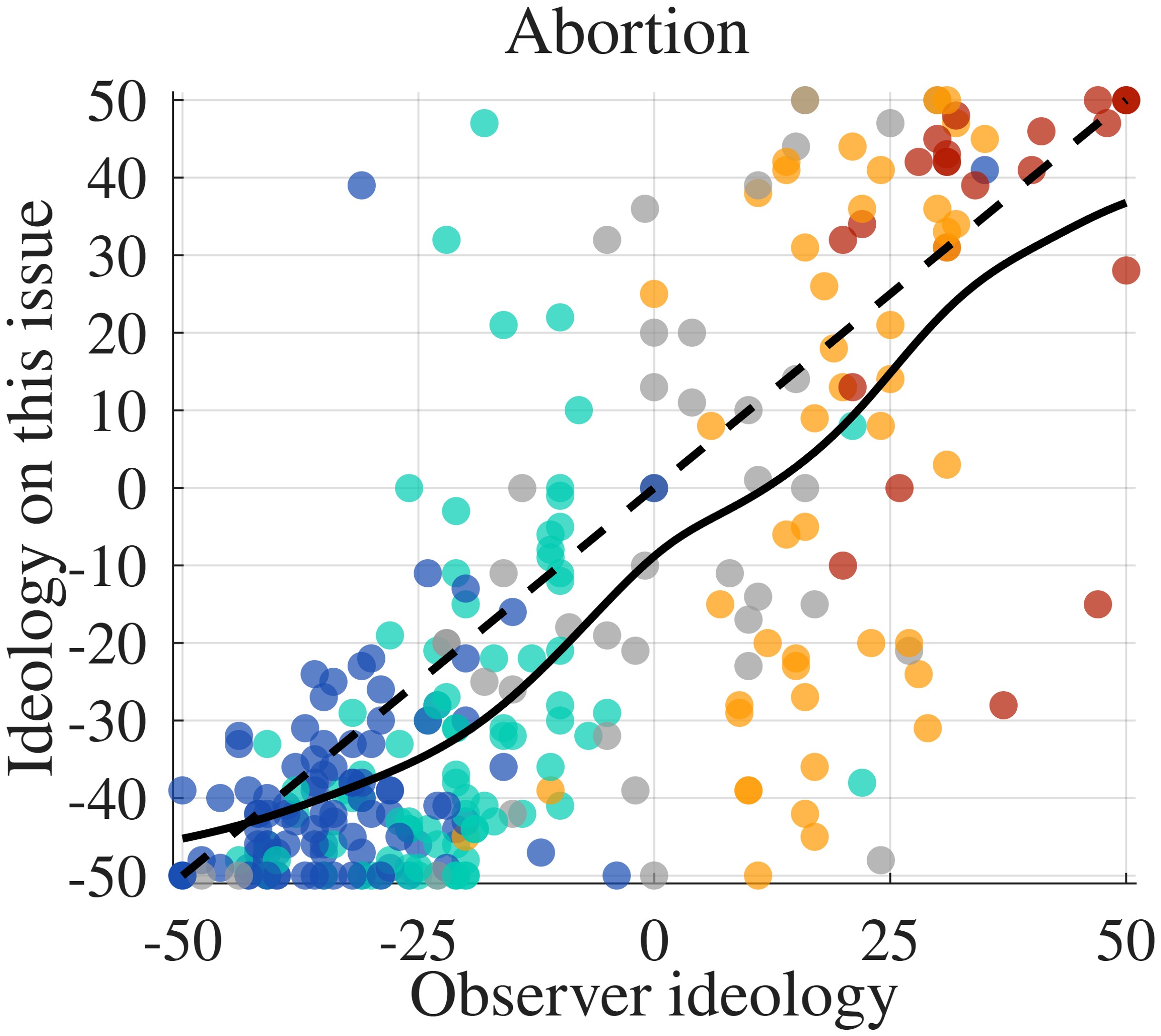}
    \\
    \vspace{2mm}
    \includegraphics[width= 0.26\textwidth]{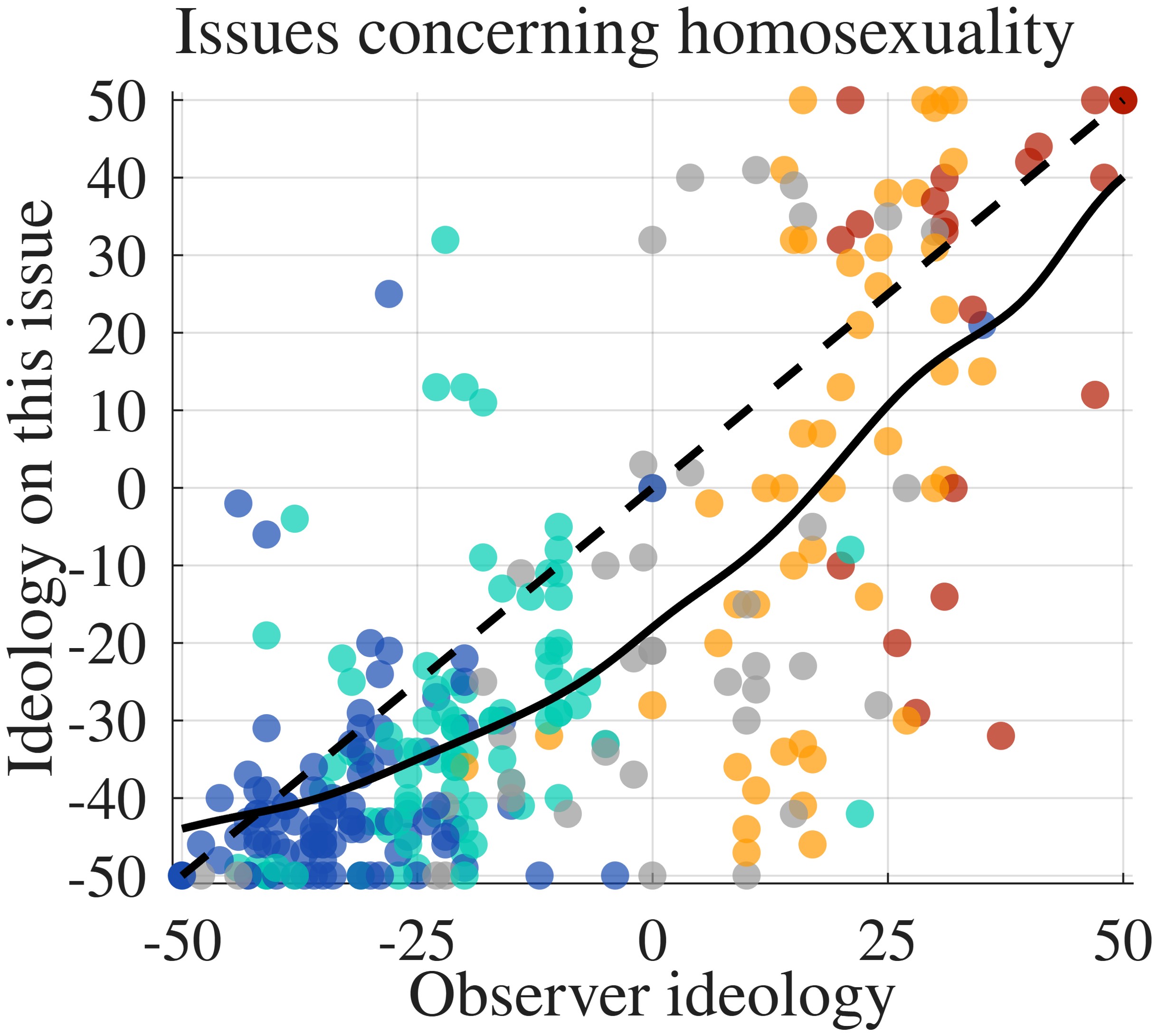}
    \includegraphics[width= 0.26\textwidth]{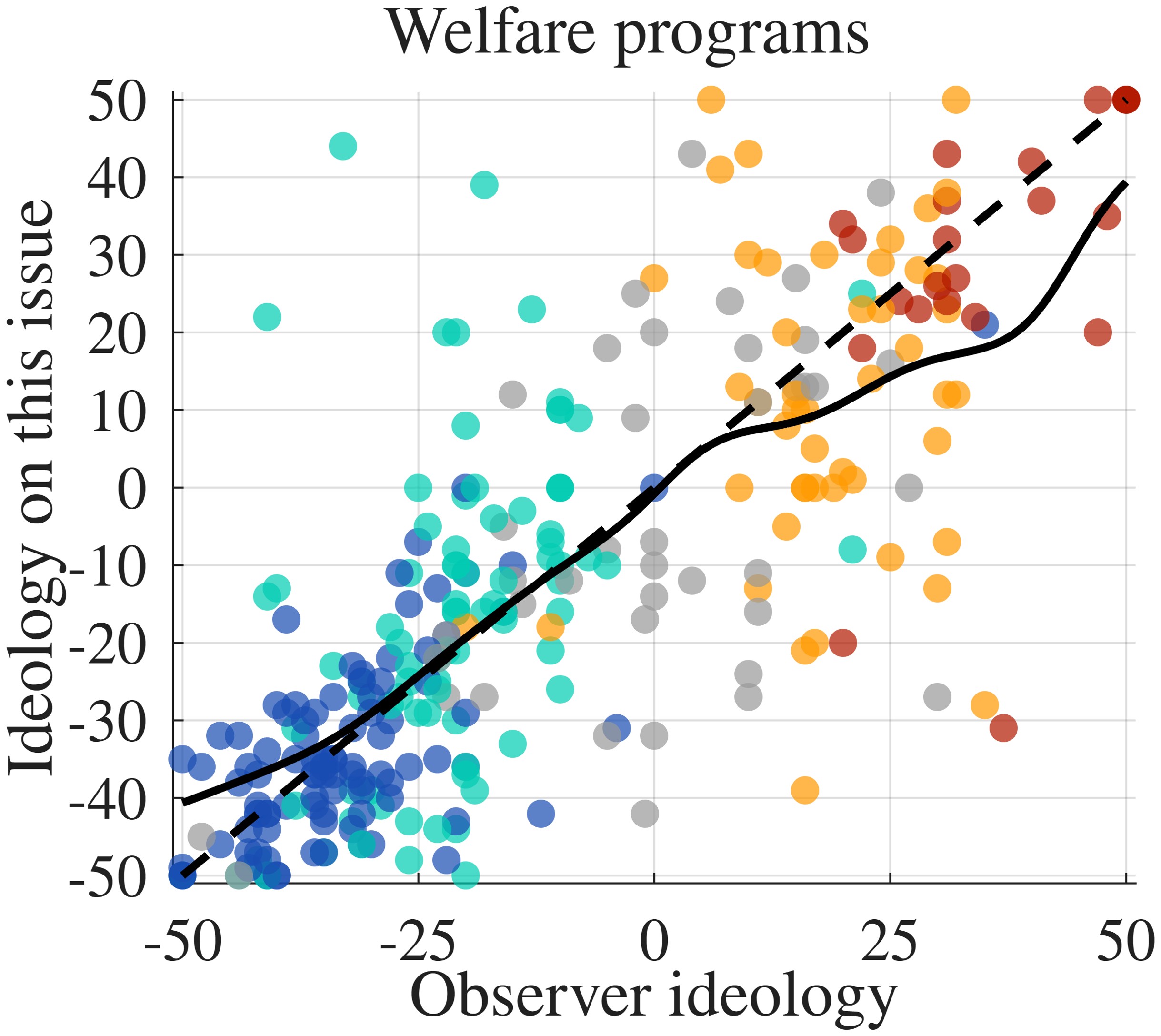}
    \includegraphics[width= 0.26\textwidth]{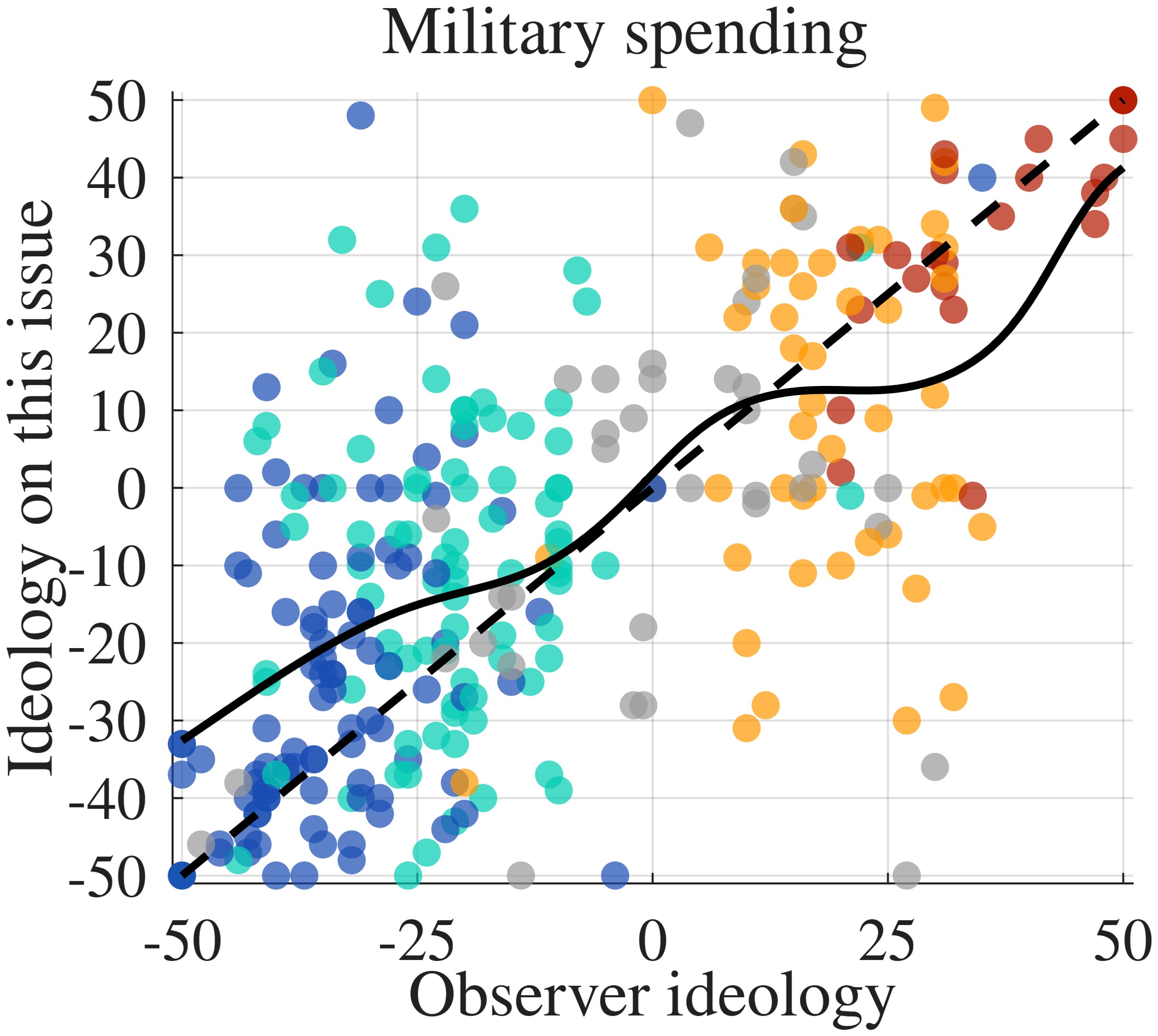}
    \includegraphics[width= 0.26\textwidth]{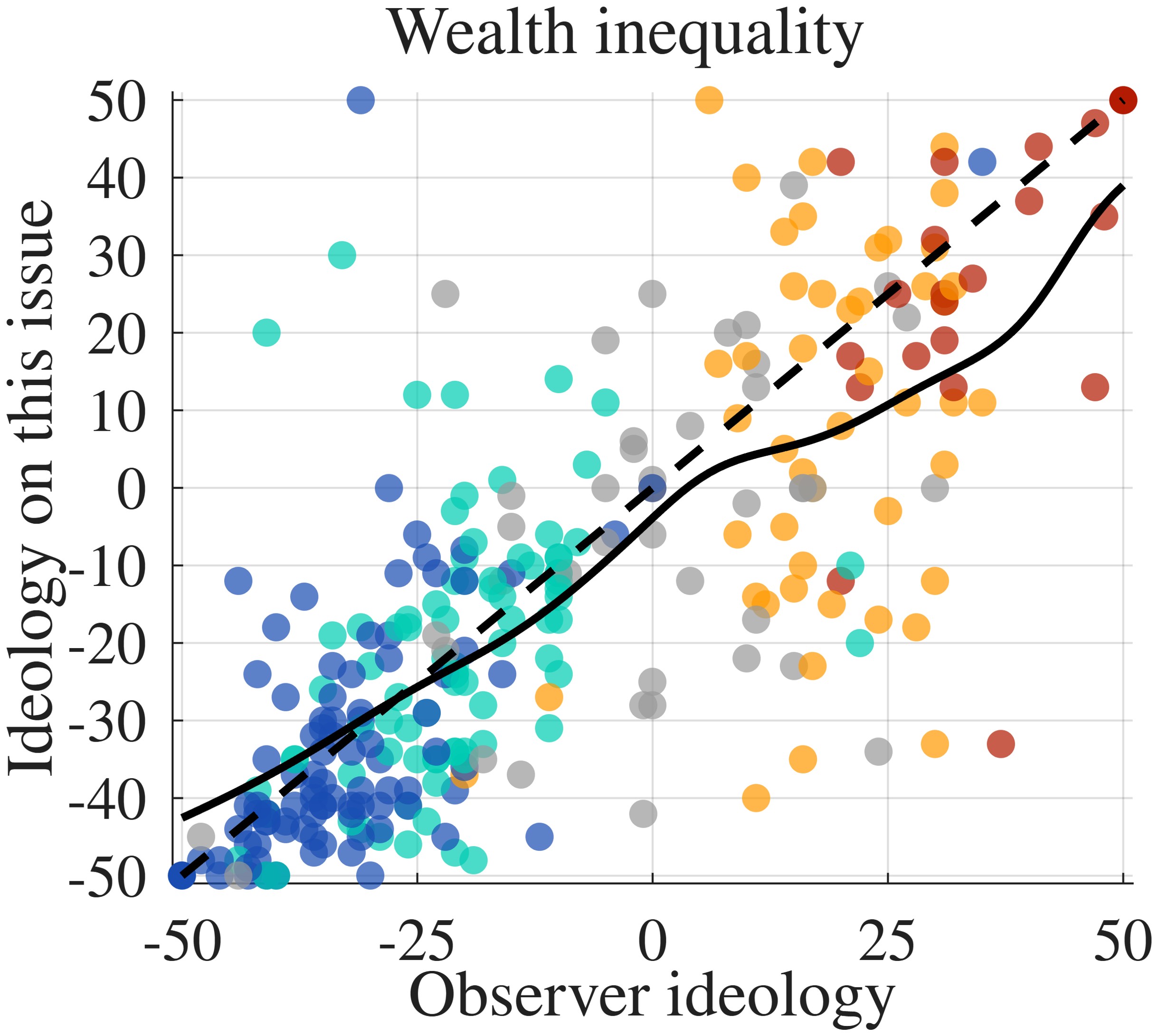}
    \\
    \vspace{2mm}
    \includegraphics[width= 0.26\textwidth]{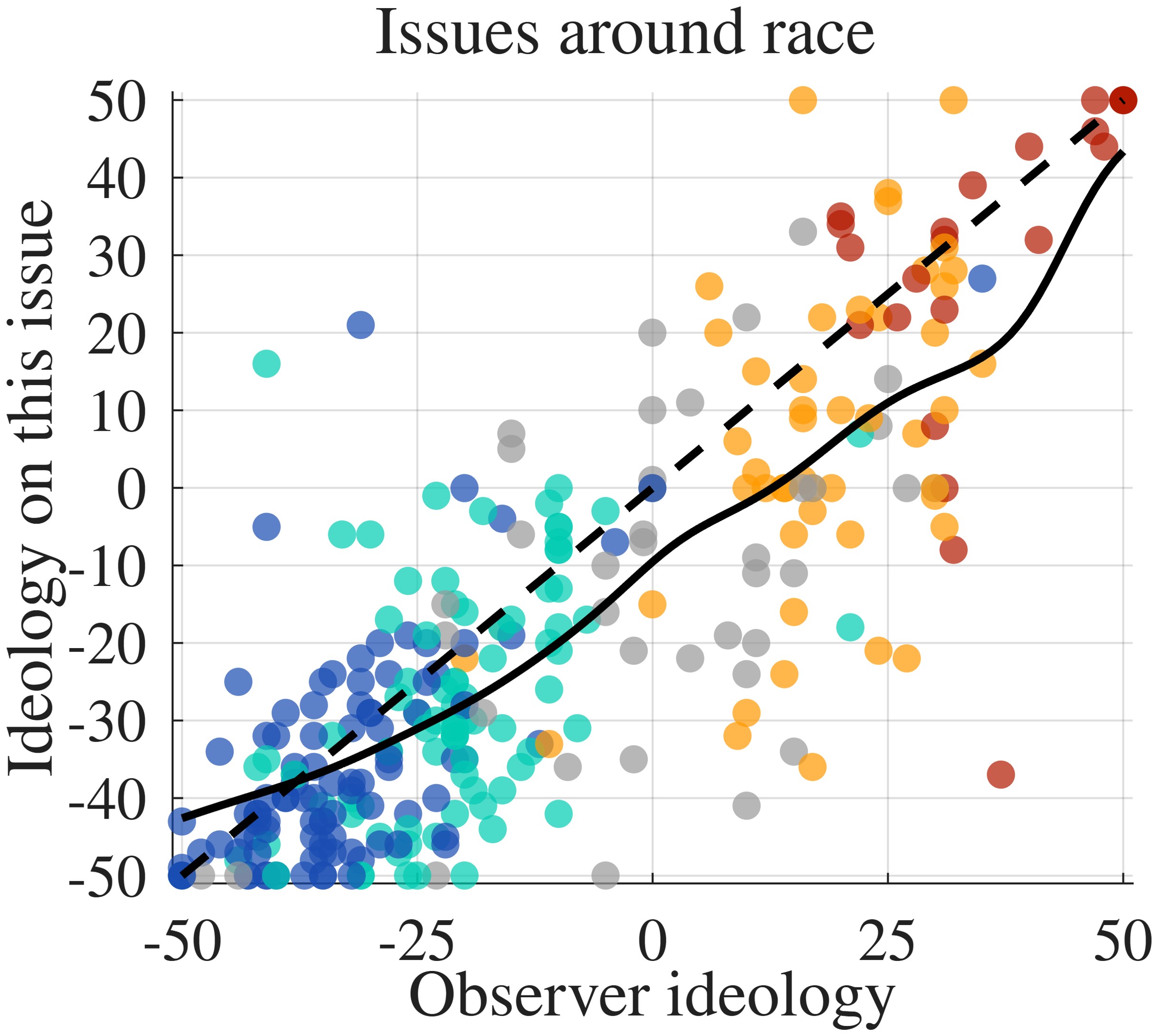}
    \includegraphics[width= 0.26\textwidth]{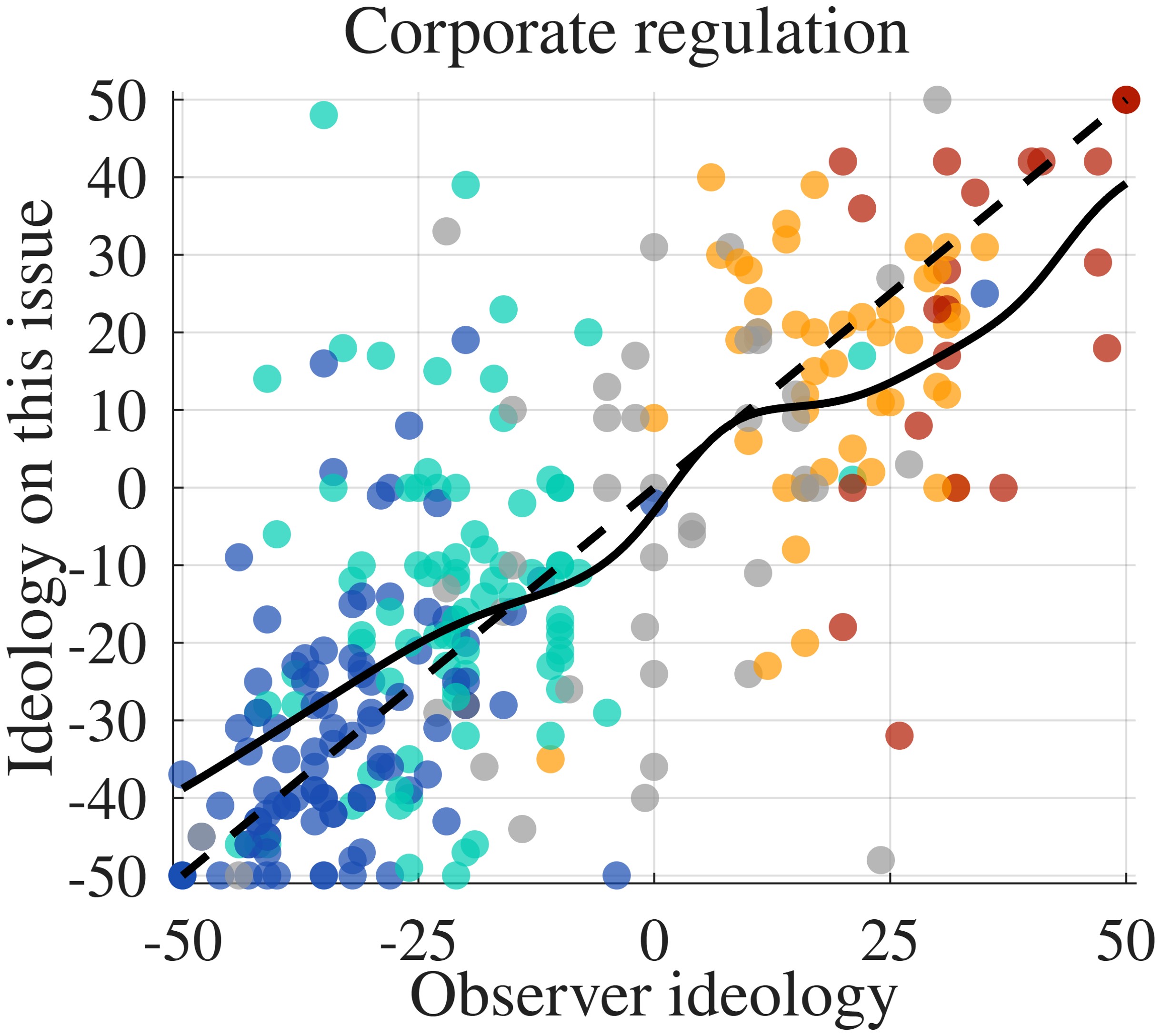}
    \includegraphics[width= 0.26\textwidth]{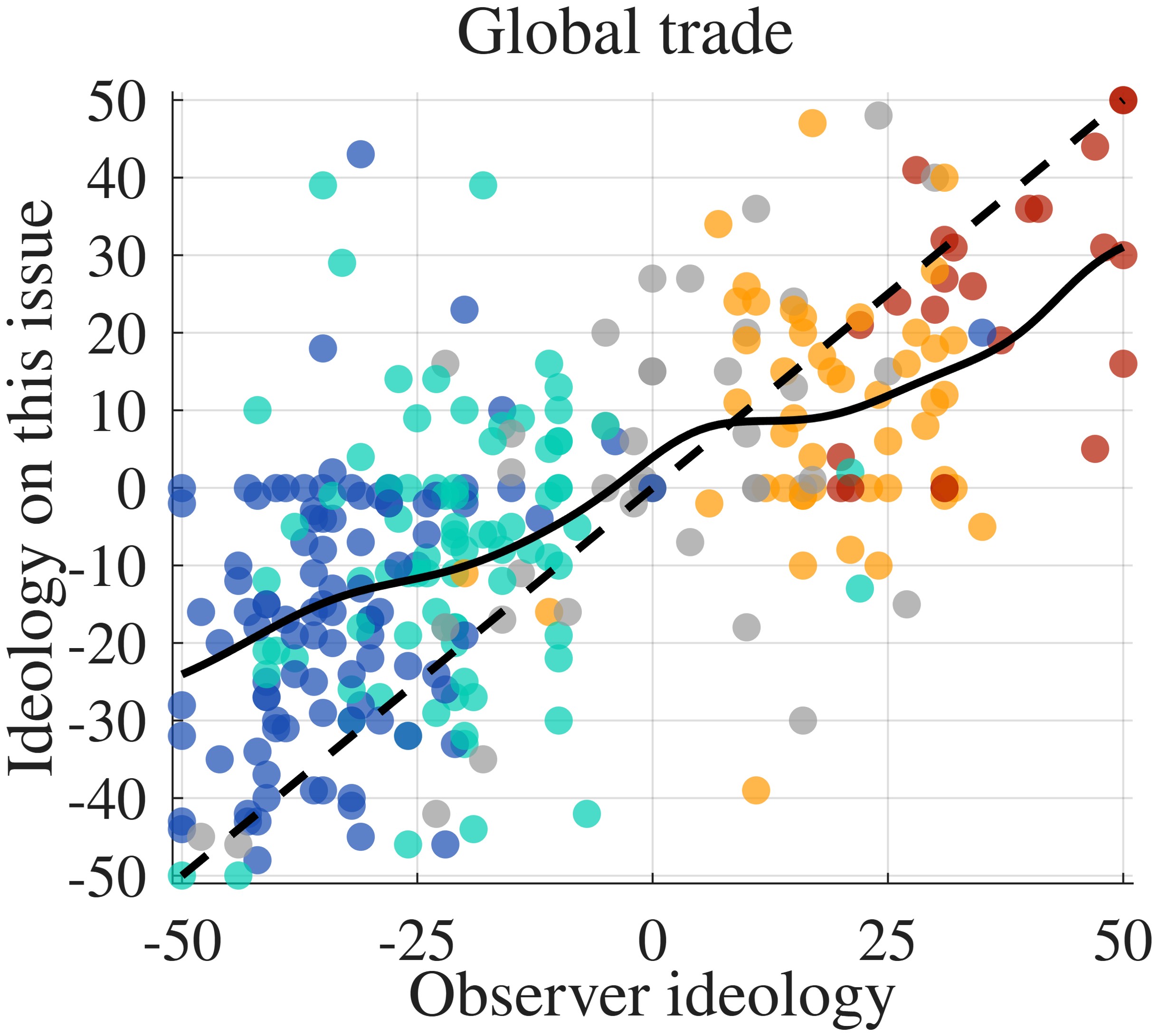}
    \includegraphics[width= 0.26\textwidth]{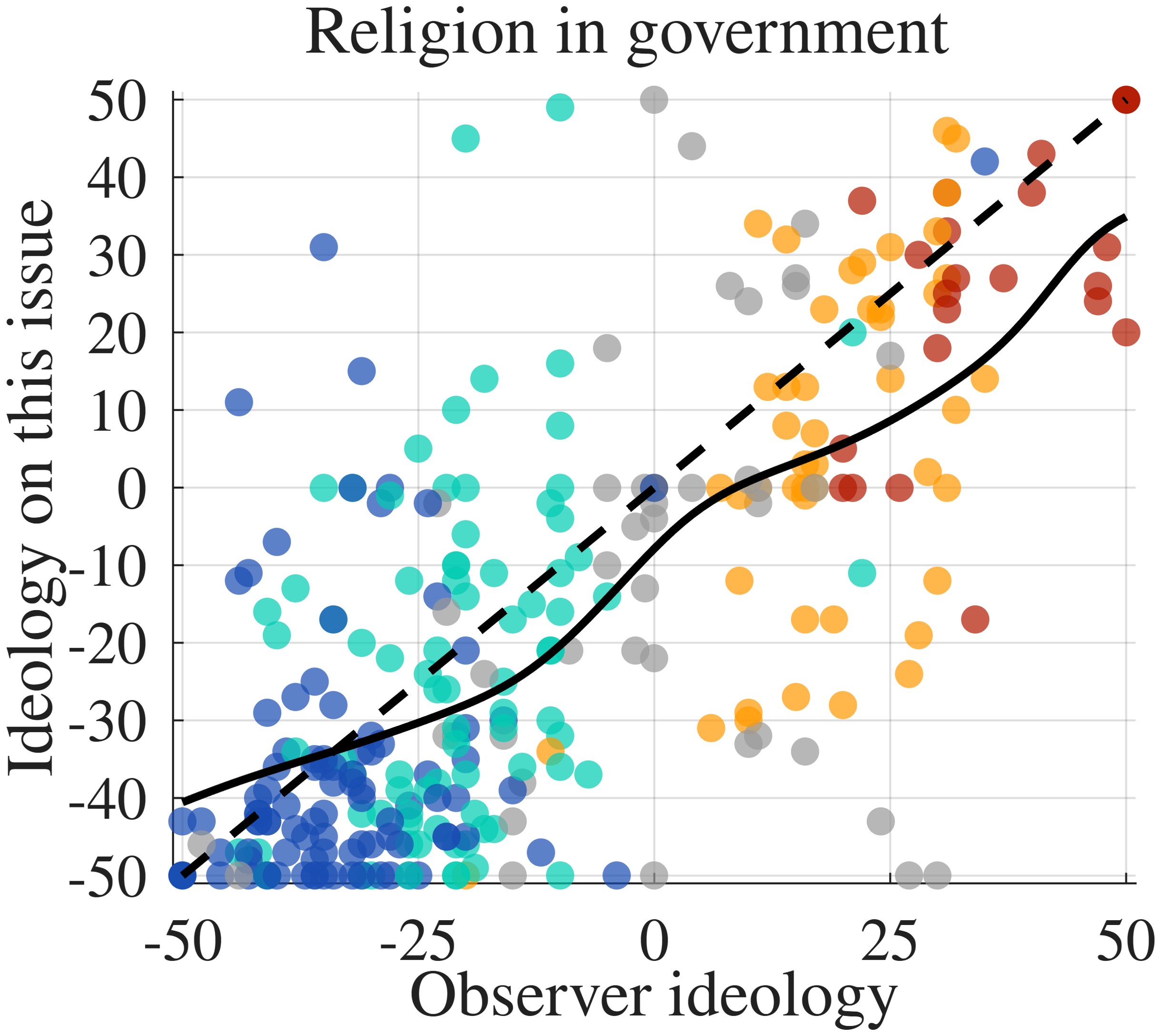}
    \caption{Version of Fig \ref{fig:13_major_selfplace_scatters}
    for the volunteer and Mechanical Turk sample (using the same issue ordering as the Prolific sample for ease of comparison, rather than its own decreasing-polarization ordering).
    }
    \label{fig:mv_13_major_selfplace_scatters}
\end{figure*}

\begin{figure*}[h!] 
    \centering
    
    \includegraphics[width=0.26\textwidth]{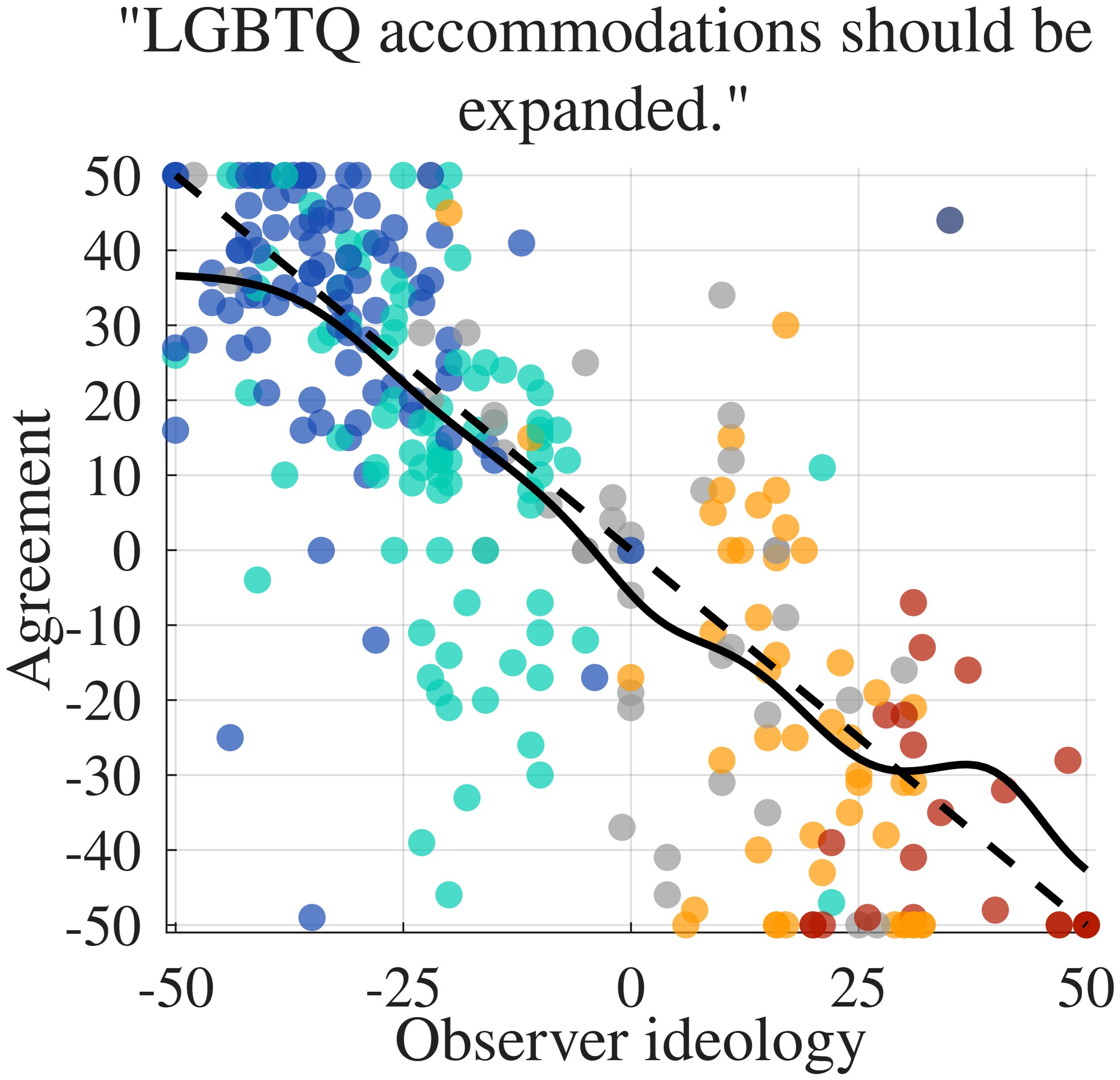}
    \includegraphics[width=0.26\textwidth]{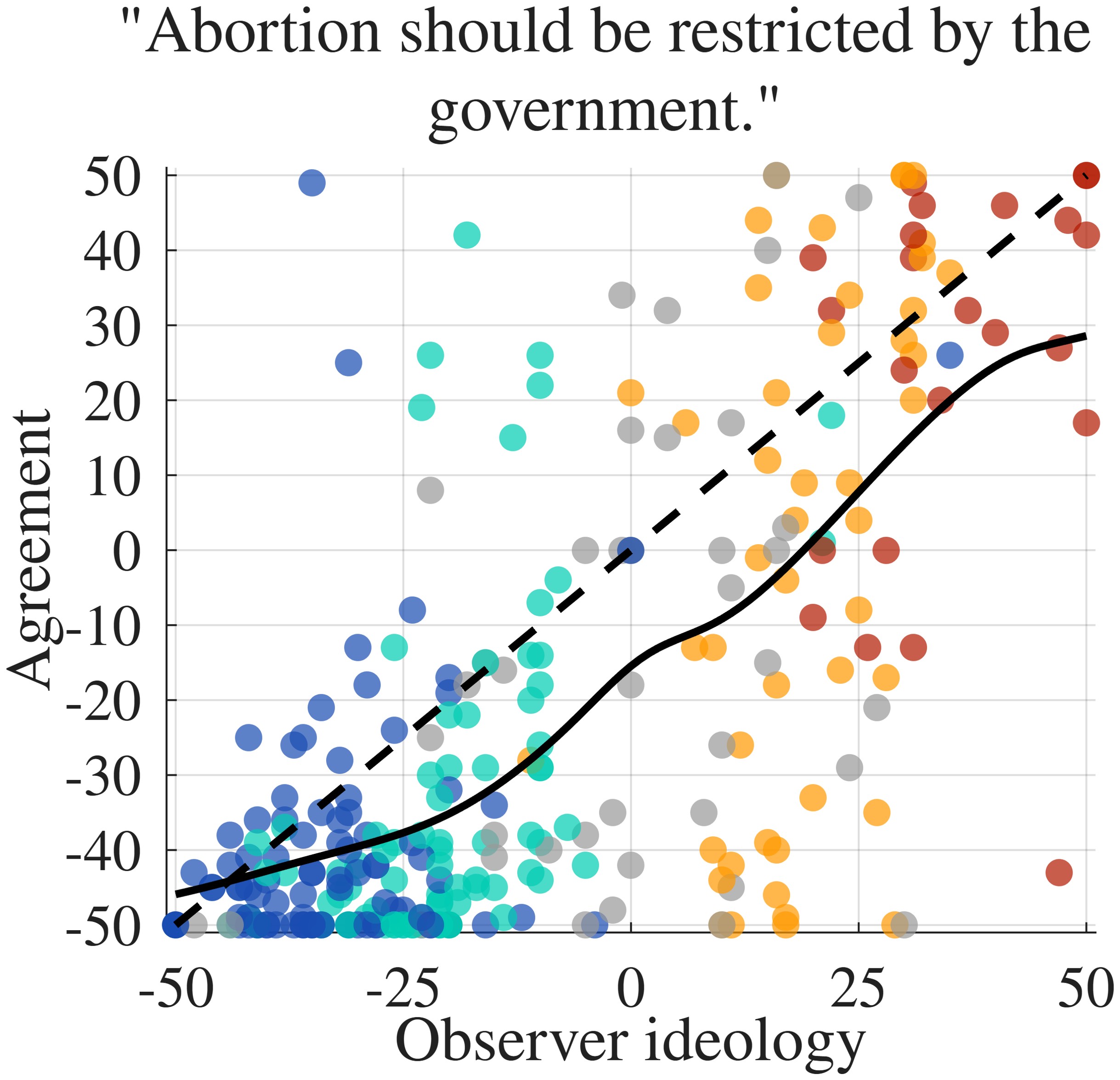}
    \includegraphics[width=0.26\textwidth]{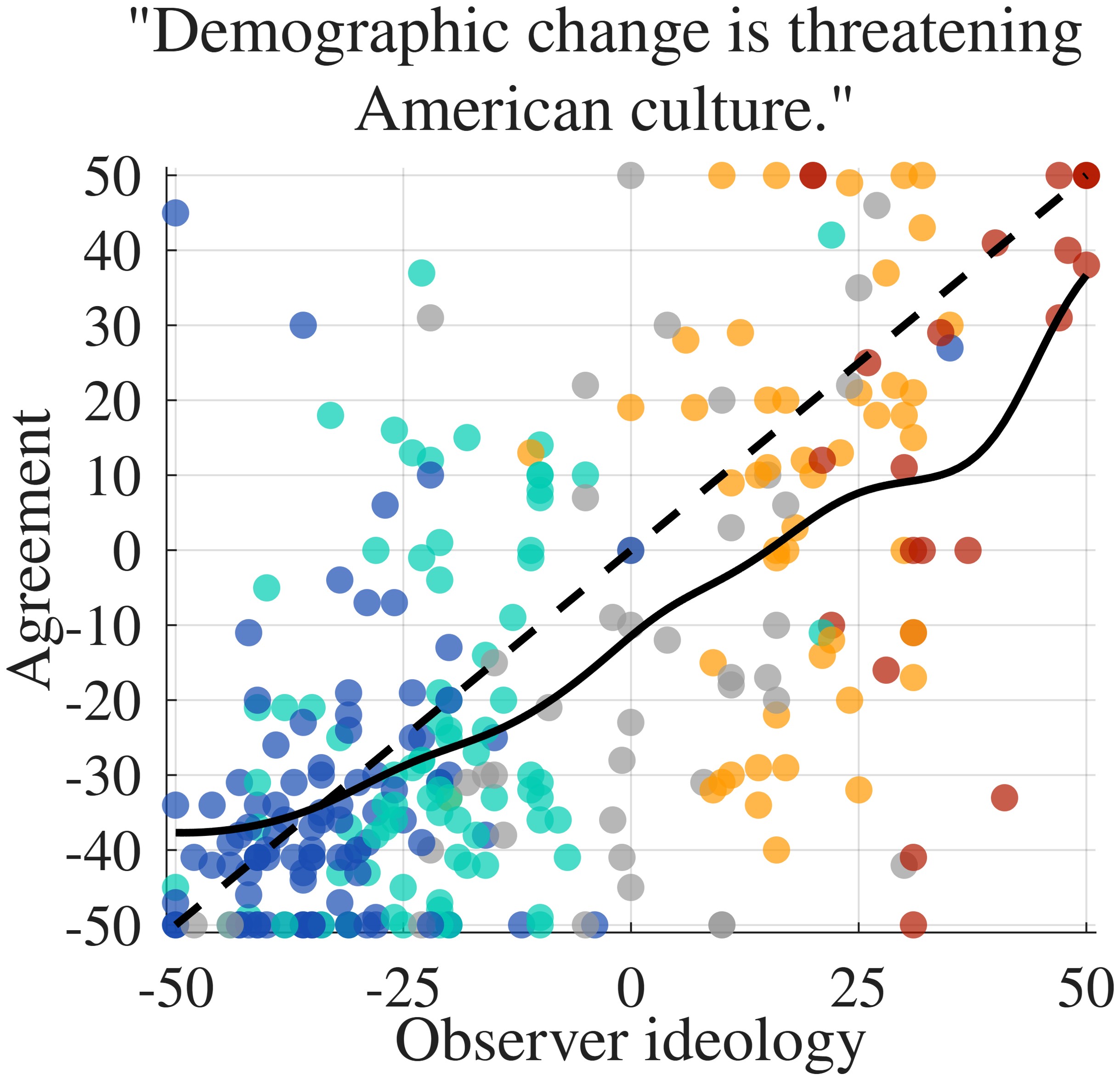}
    \raisebox{0.05\height}{\includegraphics[width= 0.09\textwidth]{jpgs/colorbar_short.jpg}}
    \\
    \vspace{2mm}
    \includegraphics[width=0.26\textwidth]{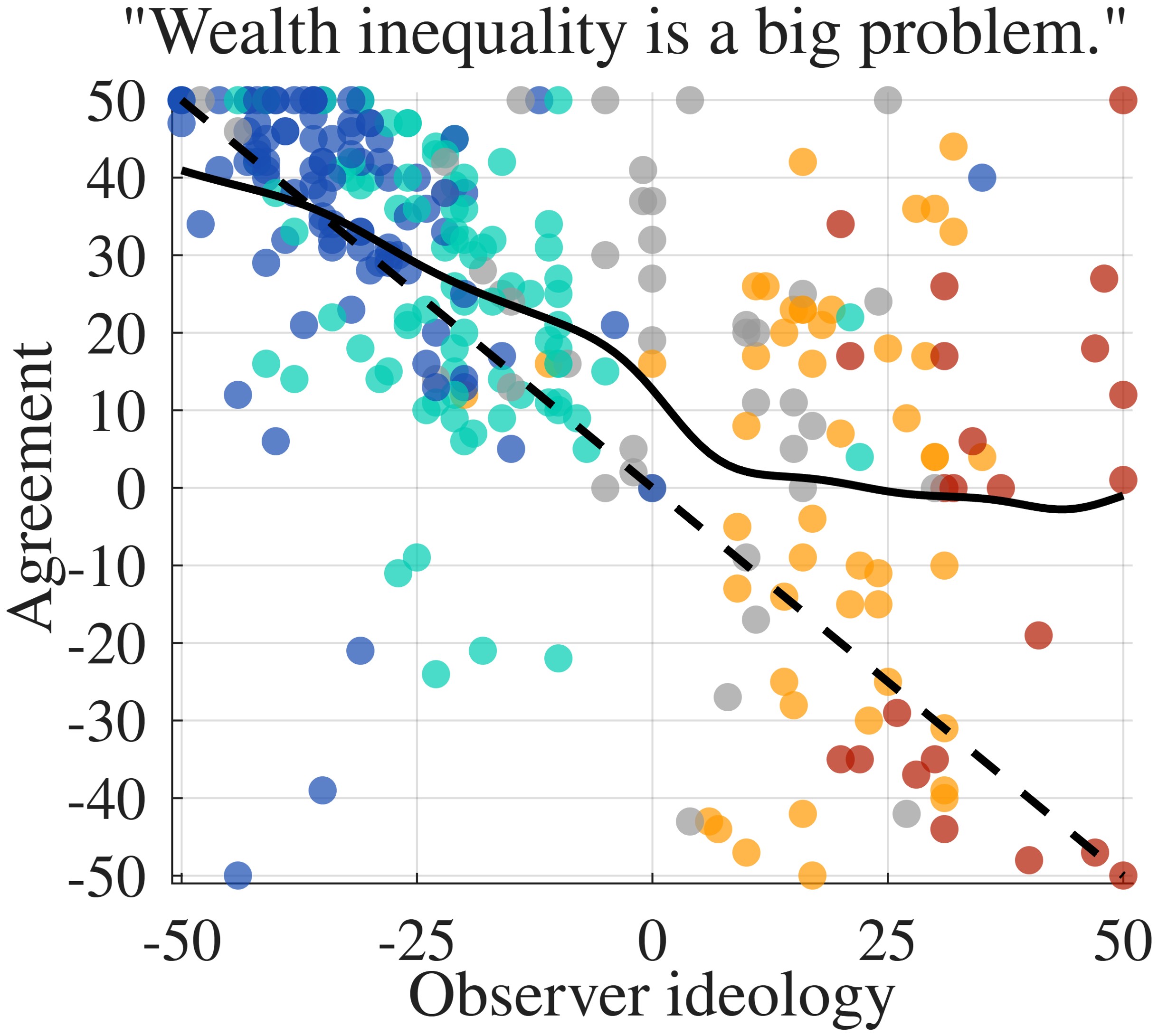}
    \includegraphics[width=0.26\textwidth]{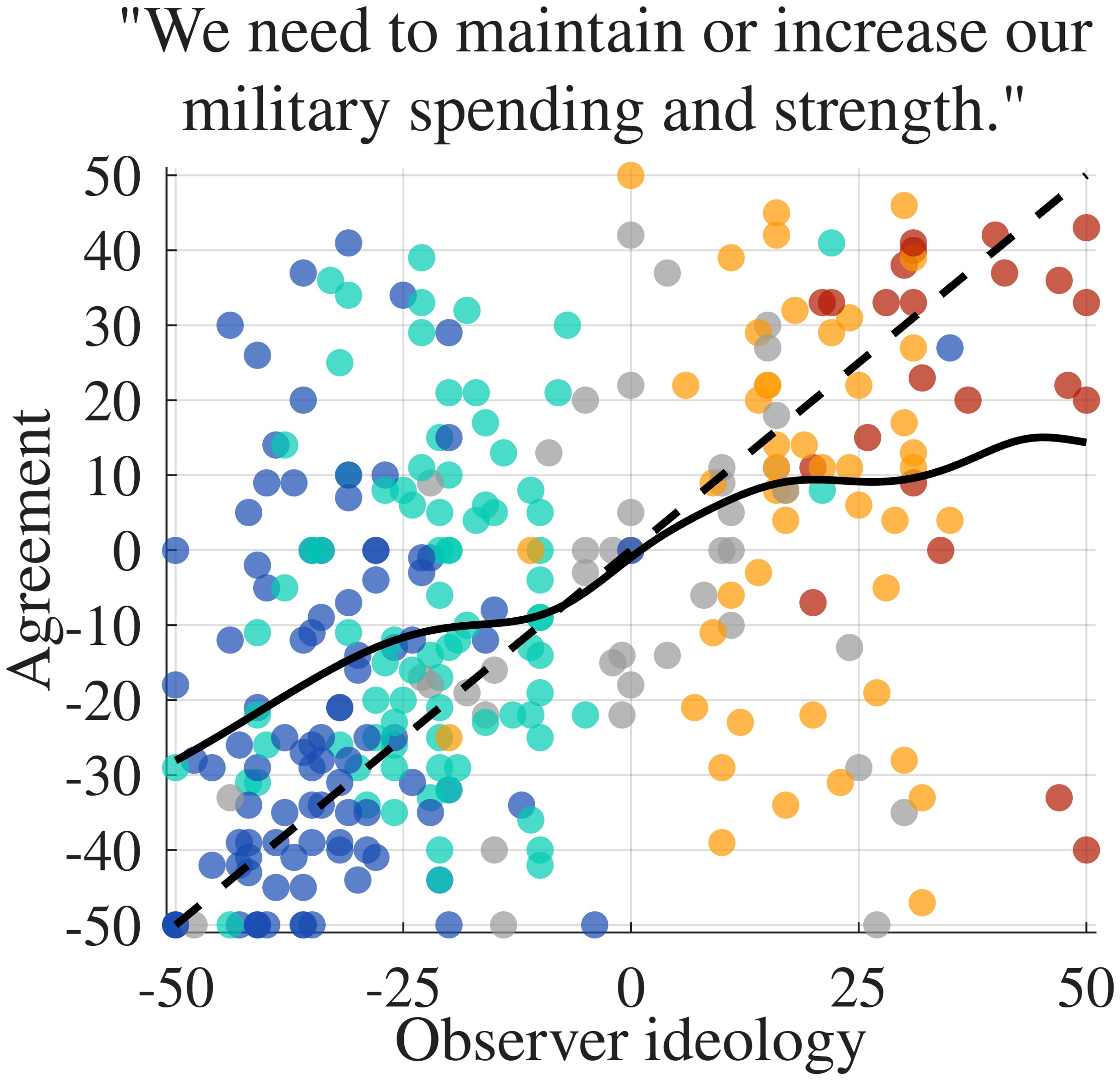}
    \includegraphics[width=0.26\textwidth]{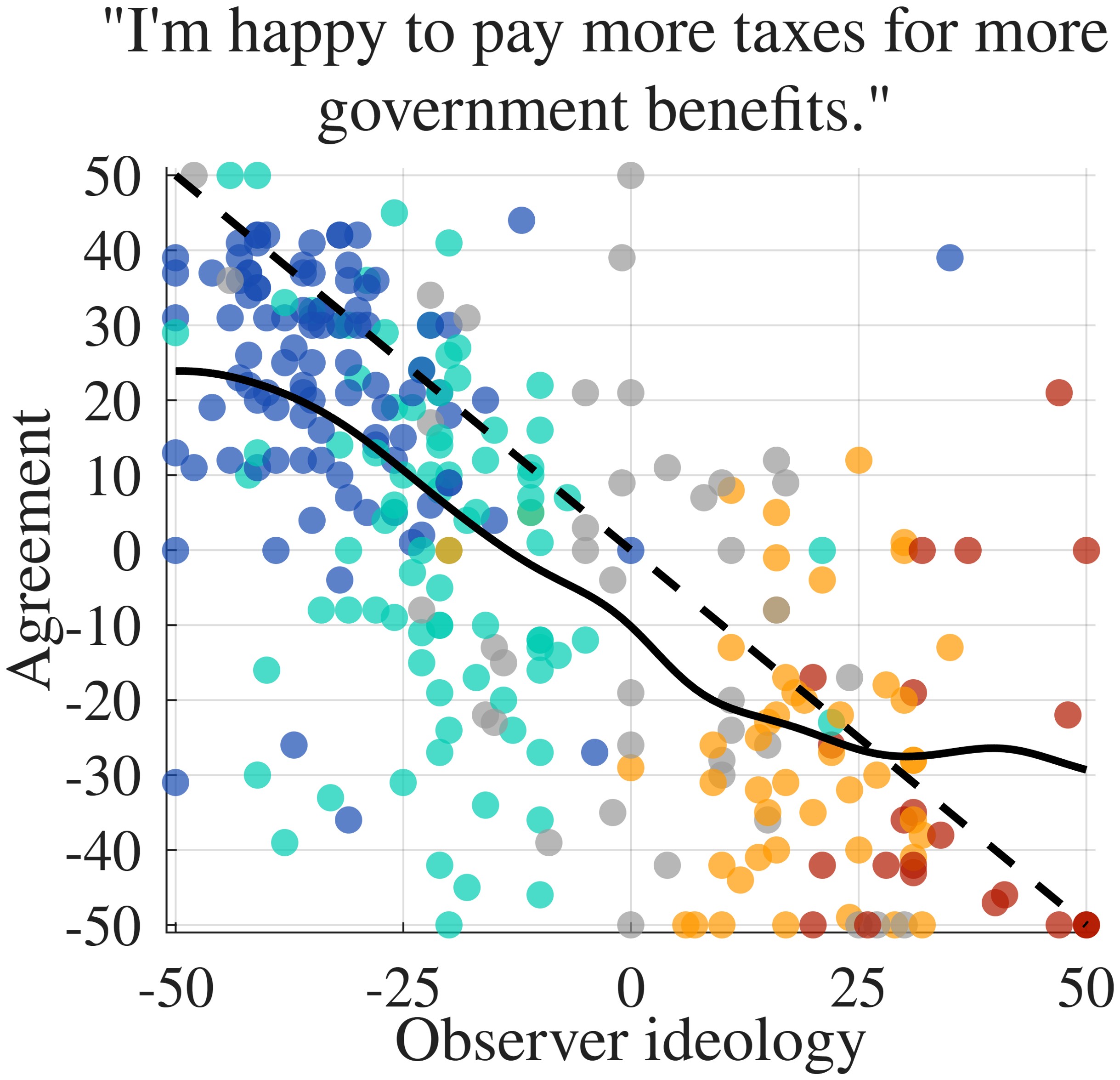}
    \includegraphics[width=0.26\textwidth]{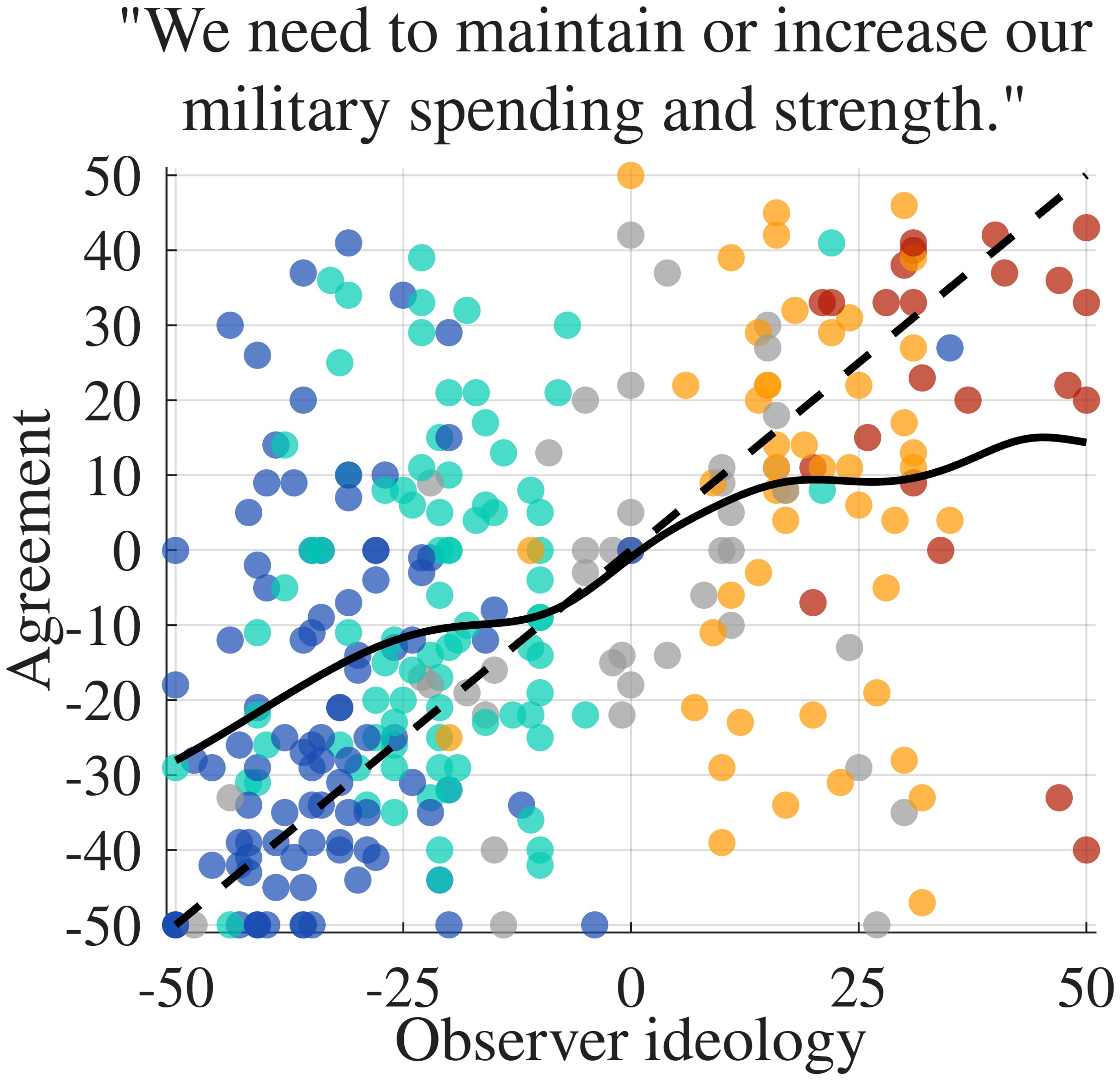}
    \\
    \vspace{2mm}
    \includegraphics[width=0.26\textwidth]{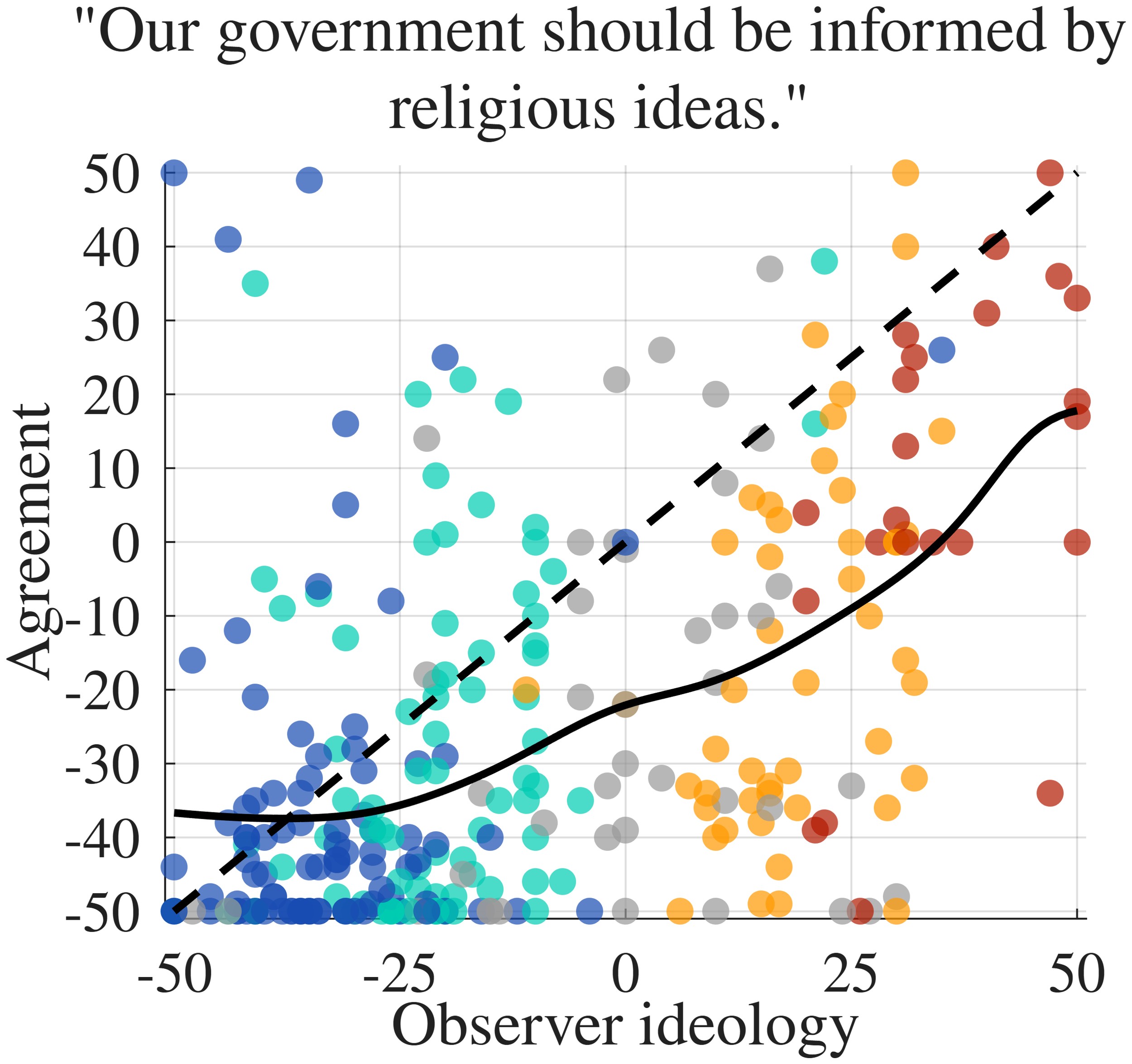}
    \includegraphics[width=0.26\textwidth]{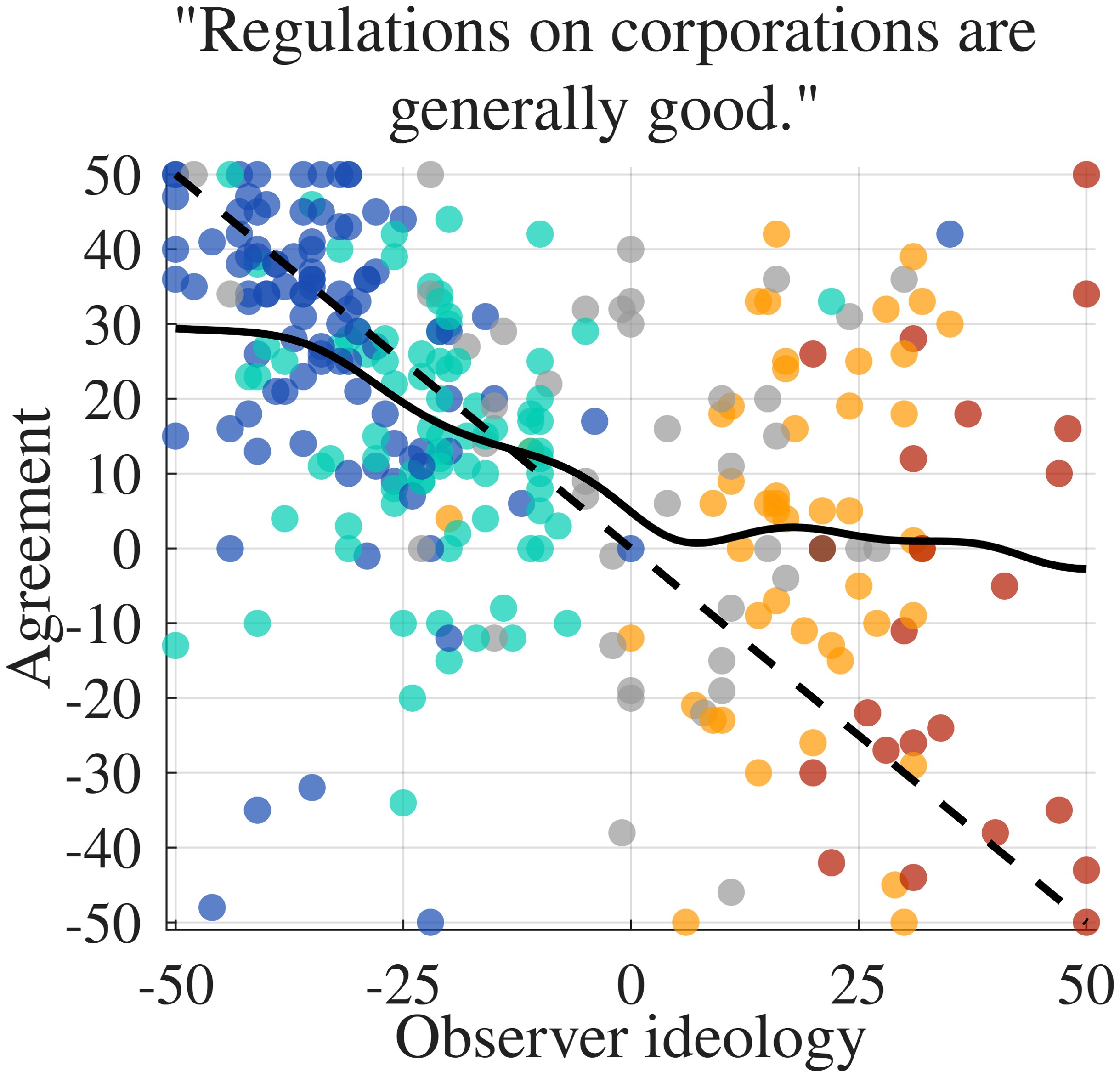}
    \includegraphics[width=0.26\textwidth]{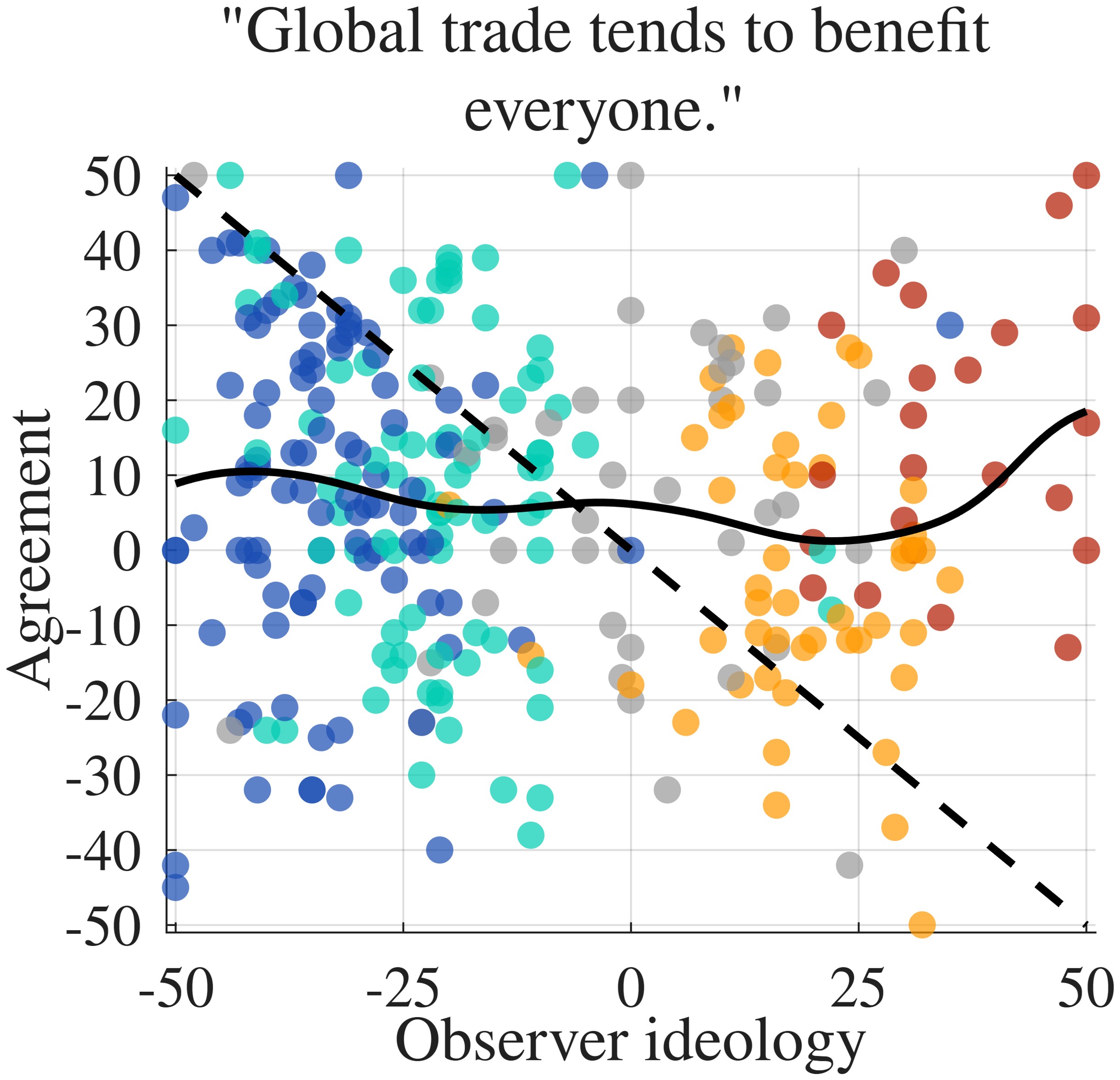}
    \caption{Version of Fig \ref{fig:10_major_agreement_scatters}
    displaying data from volunteers and Mechanical Turk Masters (using the same issue ordering as the Prolific sample for ease of comparison, rather than its own decreasing-polarization ordering).
    }
\label{fig:mv_10_major_agreement_scatters}
\end{figure*}

\begin{figure*}[h!]
    \centering
    \includegraphics[width=.54\textwidth]{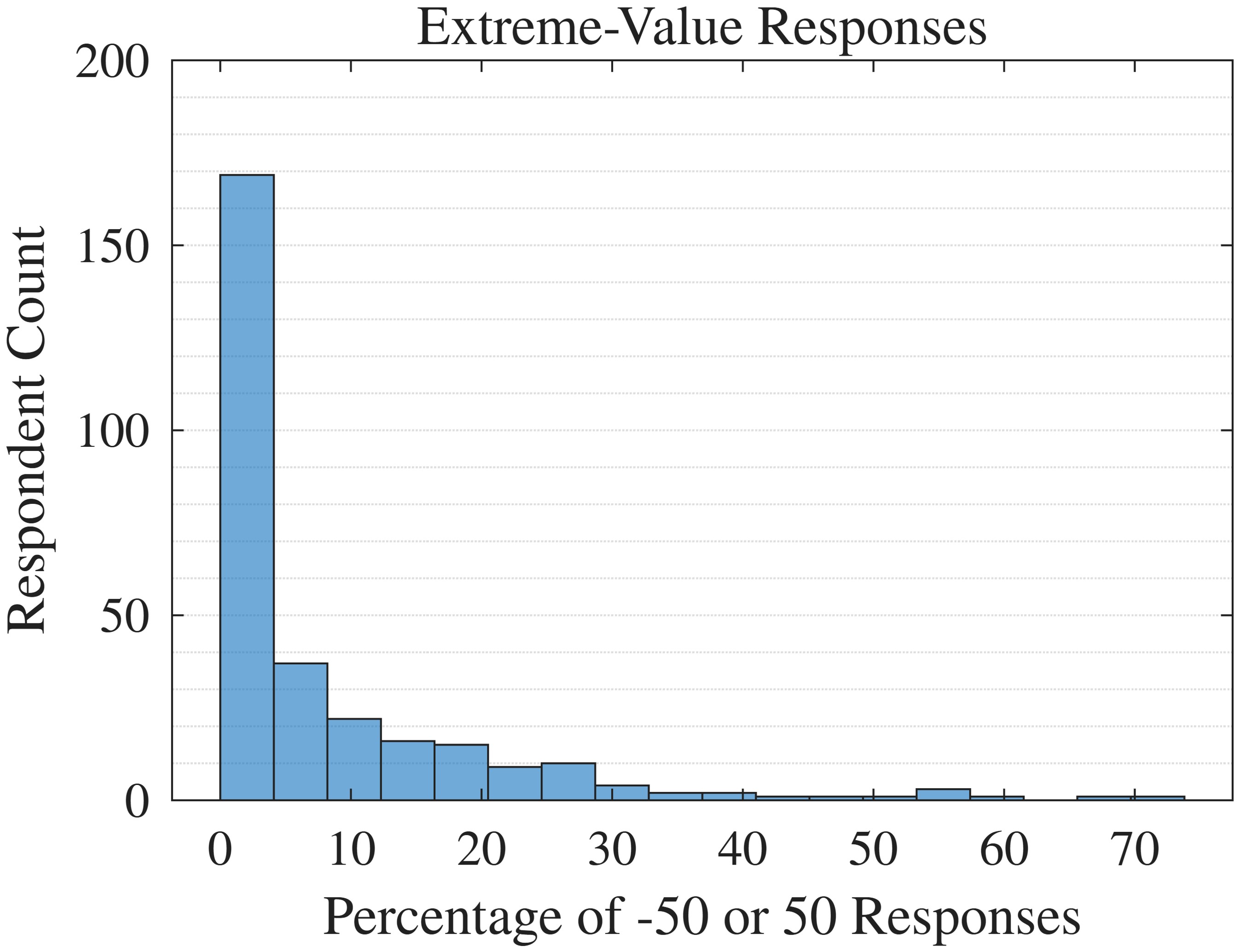}
    \includegraphics[width=.54\textwidth]{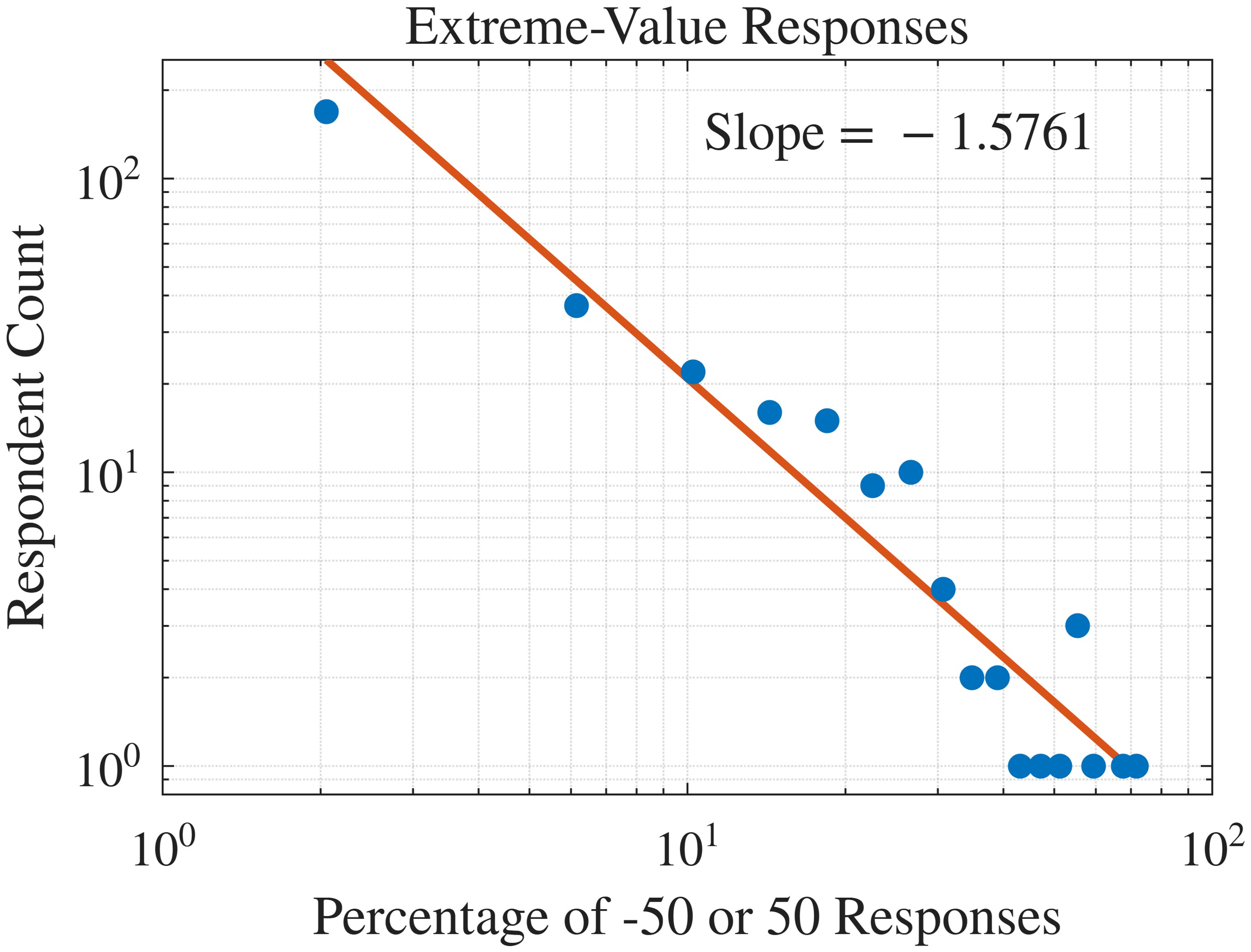}
    \caption{Version of Fig \ref{fig:extreme_responses} displaying data from volunteers and Mechanical Turk Masters. We see a similar apparent power law with exponent of about $-1.58$.}
    \label{fig:mv_extreme responses}
\end{figure*}

\begin{figure}[h!]
    \centering
    \includegraphics[width=0.54\columnwidth]{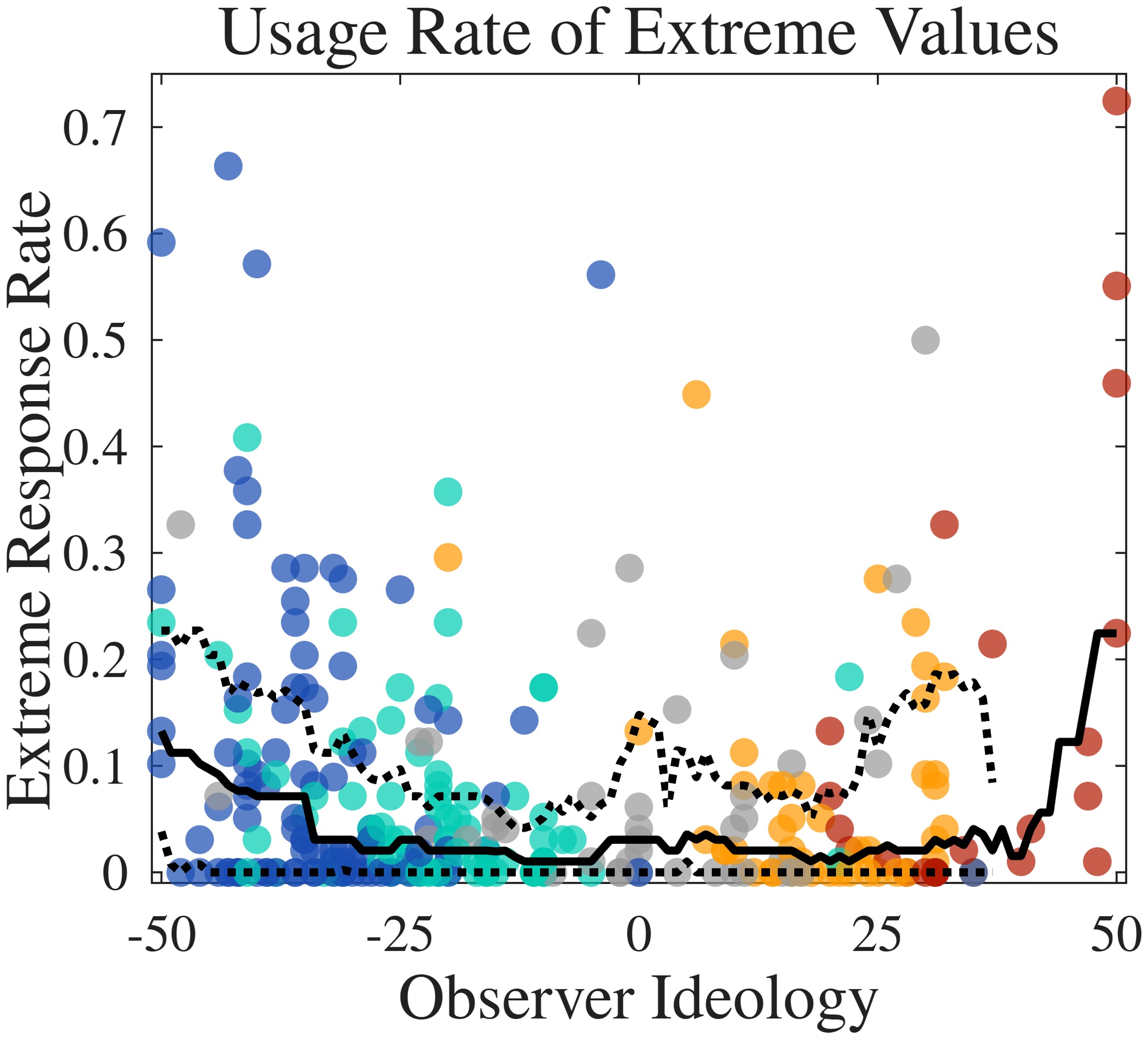}
    \caption{Version of Fig \ref{fig:extreme_response_scatter} displaying data from volunteers and Mechanical Turk Masters. We see a similar signature of extreme ideologues using extreme values more.}
    \label{fig:mv_extreme_response_scatter}
\end{figure}

\begin{figure*}[h!] 
    \centering
    \begin{overpic}[width=.54\textwidth]{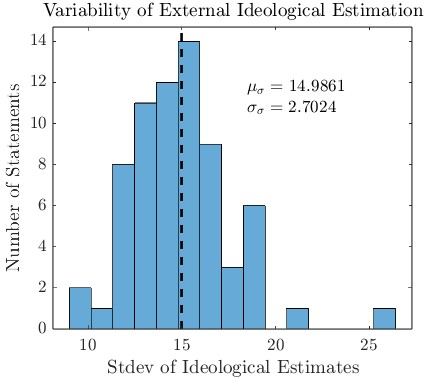}
    \put(14,78){\large{\textbf{a}}}
    \end{overpic}
    \begin{overpic}[width=.54\textwidth]{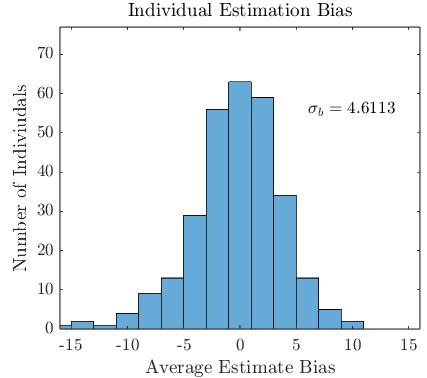}
    \put(16,76){\large{\textbf{b}}}
    \end{overpic}
    \caption{Version of Fig \ref{fig:assessment_stdev_hist} displaying data from volunteers and Mechanical Turk Masters. These respondents showed tighter ideological estimate distributions ($\mu_\sigma = 14.99$ instead of $19.64$), but had a similar signature of individual bias (i.e.~correlation of estimate deviations), since $\sigma_b = 4.61 > 2.74 = \sigma_0$.}
     \label{fig:mv_assessment_stdev_hist}
\end{figure*}

\clearpage
\setcounter{figure}{0} 
\renewcommand\thefigure{C\arabic{figure}}    

\section{Sentiment} \label{sec:sentiment}
\begin{figure*}[h!]
\centering
\includegraphics[width=.54\textwidth]{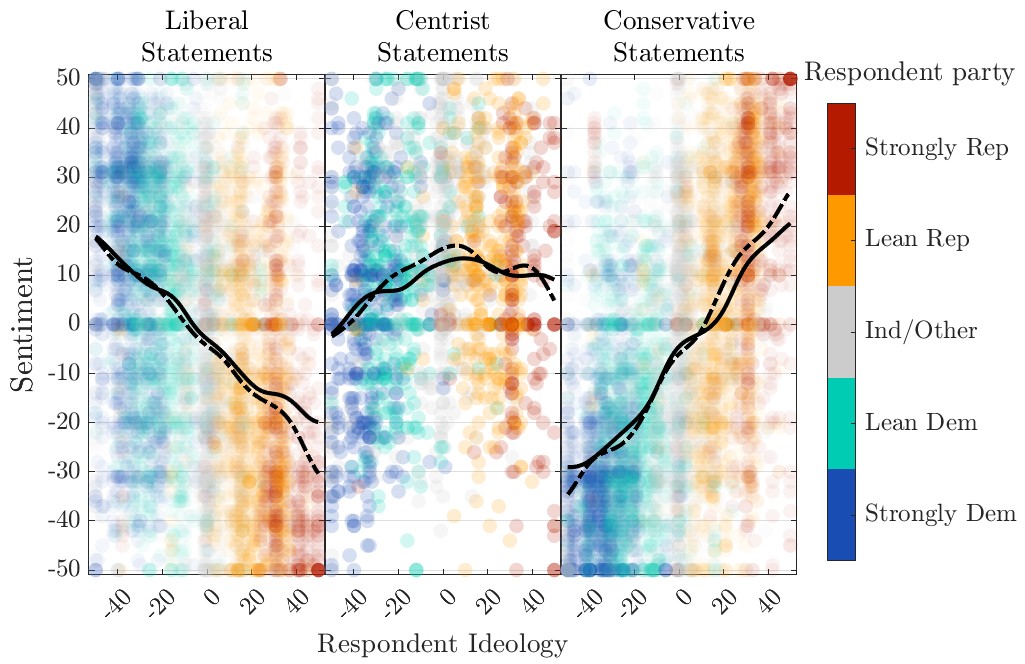}
\includegraphics[width=.54\textwidth]{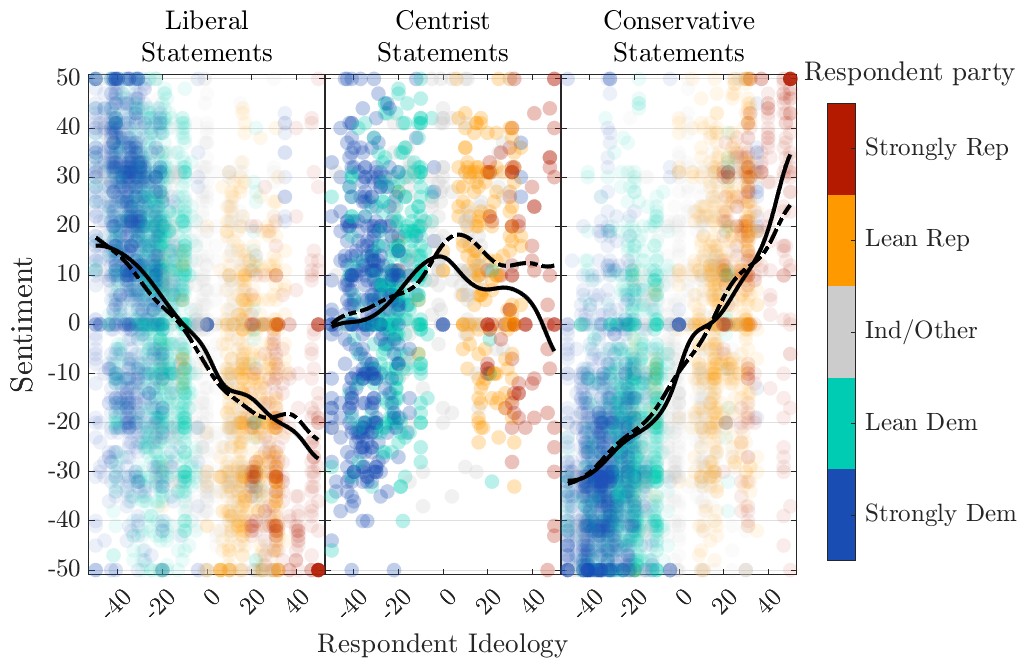}
\caption{Average positive/negative sentiment towards each statement pool among Prolific (\textbf{left}) and Volunteers/Mechanical Turk Masters (\textbf{right}), with Gaussian moving-average trend curves as in Fig \ref{fig:multiscatters}. Essentially the same pattern as Fig \ref{fig:multiscatters}b is seen, but at slightly lower amplitude.}
\label{fig:multiscatter_sentiment}
\end{figure*}

\begin{figure*}[h!]
\centering
\includegraphics[width=.54\textwidth]{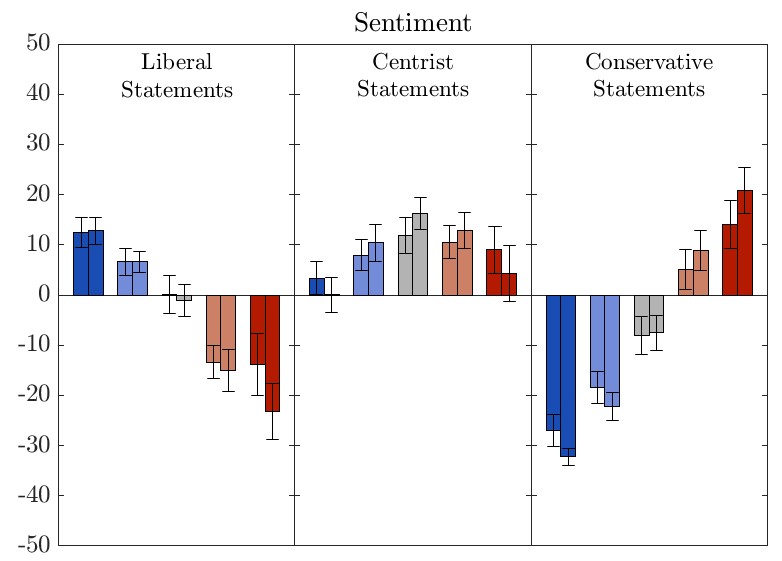}
\includegraphics[width=.54\textwidth]{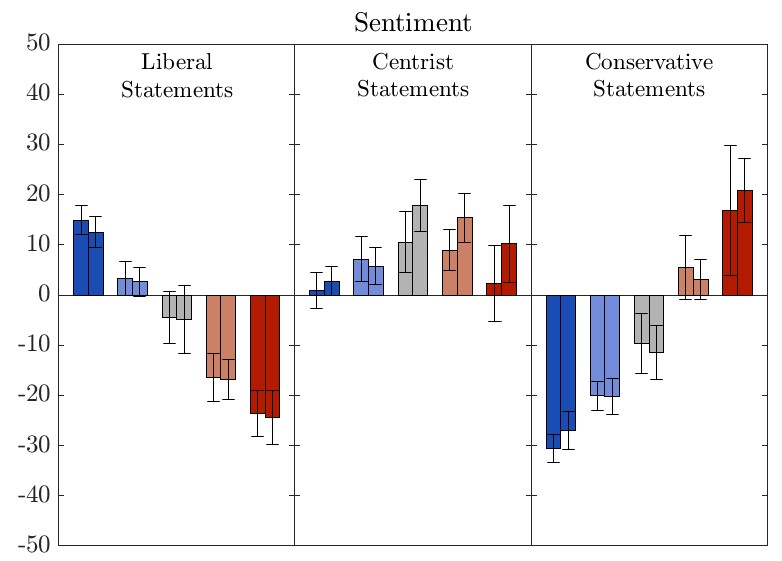}
\caption{Average positive/negative sentiment towards each statement pool among Prolific (\textbf{left}) and Volunteers/Mechanical Turk Masters (\textbf{right}), with $95\%$ confidence intervals clustered by individual in Stata. Essentially the same pattern as Fig \ref{fig:multibars}b is seen, but at slightly lower amplitude.}
\label{fig:multibar_sentiment}
\end{figure*}

We include Figs \ref{fig:multiscatter_sentiment}and \ref{fig:multibar_sentiment}, which present the emotional-sentiment versions of overall-agreement figures (Figs \ref{fig:multiscatters}b, \ref{fig:mv_multiscatters}b, \ref{fig:multibars}b and \ref{fig:mv_multibars}b). The same patterns are generally displayed, but at slightly lower amplitude than each corresponding agreement-based figure.

\end{document}